\pdfoutput=1

\documentclass[10pt]{article}
\usepackage[T1]{fontenc} 
\usepackage[left=3.0cm,right=3.0cm,top=3.0cm,bottom=3.0cm]{geometry} 
\usepackage{floatrow,svg,amsmath,amssymb,blindtext,natbib,subfig,caption,bm,bbm,tikz,setspace,mathabx,mathtools,amsthm,eurosym}

\fontfamily{pcr}\selectfont
\usepackage[hang,flushmargin]{footmisc}
\usetikzlibrary{shapes.geometric}
\usetikzlibrary{automata,positioning}
\usetikzlibrary{fit}

\makeatletter
\newcommand{\subalign}[1]{%
  \vcenter{%
    \Let@ \restore@math@cr \default@tag
    \baselineskip\fontdimen10 \scriptfont\tw@
    \advance\baselineskip\fontdimen12 \scriptfont\tw@
    \lineskip\thr@@\fontdimen8 \scriptfont\thr@@
    \lineskiplimit\lineskip
    \ialign{\hfil$\m@th\scriptstyle##$&$\m@th\scriptstyle{}##$\hfil\crcr
      #1\crcr
    }%
  }%
}

\makeatother

\renewcommand{\P}{\mathcal{P}}
\renewcommand{\L}{\mathcal{L}}
\newcommand{\N}{\mathcal{N}}
\newcommand{\Nb}{\mathbb{N}}
\newcommand{\E}{\mathcal{E}}
\newcommand{\C}{\mathcal{C}}
\newcommand{\G}{\mathcal{G}}

\newcommand{\one}{\mathbbm{1}}

\begin{document}
\begin{center}
  \LARGE
  \text{}\vspace{1cm}\\
  Reciprocity in Interbank Markets\footnote{The author is extremely grateful to Thomas Lux for helpful comments and suggestions. This research was supported through high-performance computing resources available at Kiel University Computing Center.}\vspace{0.5cm}\\

  \large
  Lutz Honvehlmann\medskip\\
  \normalsize
  \textit{Kiel University}\\
  \textit{Wilhelm-Seelig-Platz 1, 24118 Kiel, Germany}\\
  \textit{honvehlmann@economics.uni-kiel.de}\vspace{0.5cm}\\

  \large
  December 13, 2024\vspace{1cm}\\
  \normalsize
\end{center}

\begin{center}
    \textbf{Abstract}
\end{center}
Weighted reciprocity between two agents can be defined as the minimum of sending and receiving value in their bilateral relationship.
In financial networks, such reciprocity characterizes the importance of individual banks as both liquidity absorber and provider, a feature typically attributed to large, intermediating dealer banks.
In this paper we develop an exponential random graph model that can account for reciprocal links of each node simultaneously on the topological as well as on the weighted level.
We provide an exact expression for the normalizing constant and thus a closed-form solution for the graph probability distribution.
Applying this statistical null model to Italian interbank data, we find that before the great financial crisis (i) banks displayed significantly more weighted reciprocity compared to what the lower-order network features (size and volume distributions) would predict (ii) with a disappearance of this deviation once the early periods of the crisis set in, (iii) a trend which can be attributed in particular to smaller banks (dis)engaging in bilateral high-value trading relationships.
Moreover, we show that neglecting reciprocal links and weights can lead to spurious findings of triadic relationships.
As the hierarchical structure in the network is found to be compatible with its transitive but not with its intransitive triadic sub-graphs, the interbank market seems to be well-characterized by a hierarchical core-periphery structure enhanced by non-hierarchical reciprocal trading relationships.

\bigskip


\newpage
\tableofcontents
\newpage

\newpage
\section{Introduction}

Reciprocity is one of the most basic concepts in bi- and multilateral relationships.
Here, the term relationship is meant to be understood as a broad term which may encompass, for instance, technological, biological or social dimensions. In this context, reciprocity describes the tendency of such relationships to be mutual as opposed to one relationship partner only taking with the other one only giving.
Concentrating on the social aspect of reciprocated favors, the economic and behavioral literature typically aims at pointing out reasons for cooperative human behavior.
Standard explanations for mutually positive behavior besides purely selfish interests consist, e.g.,
of Altruism \citep{simon1993altruismand}, Morality \citep{sugden1984reciprocitythe} or Gift Exchange \citep{akerlof1982laborcontracts} around a fairness norm.

\medskip

The experimental literature in turn aims at testing such theories by developing games in which selfish rational individuals do not have an individual incentive to cooperate although reciprocal behavior yields higher payoffs in total (e.g., trust or gift exchange games).
Using such games in a one-shot setting, \cite{fehr1993doesfairness}, for instance, find experimental support for the gift exchange theory on labor markets in which employers offer above-equilibrium wages and employees reciprocate with above-equilibrium effort.
The authors also view this productivity-enhancing reciprocal behavior as a possible explanation for the macroeconomically observed downward nominal wage rigidity on the labor market.
\cite{fehr1999wagerigidity}, however, highlight that the effect largely disappears with complete contracts as only incomplete contracts allow employees to shirk their duties which in turn would be anticipated by employers.
In the absence of complete contracts but over a longer time horizon, (negative) reciprocity may also fulfill a reputational role \citep{milgrom1988economictheories} by alleviating possible trust issues via repeated transactions and the possibility of punishment.
The effect of anticipated future interactions as a cause for reciprocal behavior has also been experimentally verified by \cite{leider2009directedaltruism} in the context of social networks.

\medskip

We take a more abstract perspective in this paper using the theory of networks in which we denote a relationship between two nodes $A$ and $B$ on a directed graph\footnote{
  We will use the terms graph and network (as well as edge and link) interchangeably. Unless explicitly stated otherwise, graphs will be considered directed as our core concept of reciprocity is uninformative in undirected graphs since every undirected link can be equivalently interpreted as reciprocal.
}
by the term \textit{reciprocal} (in a binary sense) whenever a link from $A$ to $B$ implies the converse link from $B$ to $A$.
Reciprocity defined in this way has been shown, e.g., to greatly increase the existence probability of a \textit{giant strongly connected component} \citep{boguna2005generalizedpercolation}, an infinitely large directed sub-graph of the entire network in which each node can be reached from any other node in the component by following paths of directed links.
In epidemiology, \cite{newman2003randomgraphs}, for instance, illustrate by means of percolation theory that the existence of a giant component is closely related to the concept of transmissibility of diseases and epidemic thresholds such as reproduction values in SIR-type models \citep{kermack1927acontribution}.
Empirically, giant components are also reflected in the large reciprocity values of the network of articles on Wikipedia \citep{zlatic2006wikipediascollaborative}, or on a broader scale in the network of hyperlinks that constructs the World Wide Web \citep{serrano2007decodingthe} which enables efficient navigation among different websites due to the existence of return links on destination websites.
\cite{serrano2007decodingthe} furthermore point out that, besides the global role of generating a tightly connected center, reciprocity may also play an important role on the local level, by facilitating the establishment of persistent and tightly knit smaller communities.

\medskip

Graph theory has long had a focus on the topological nature of graphs induced by the set of edges, i.e., a set of two-node subsets (or tuples in directed graphs). Many types of networks however are naturally equipped with valued relationships, e.g., social network messages, electric circuits or trade webs.
Equipping the graph with a weight function from the set of edges to a set of weights then allows us to not only focus on the topological backbone of the network, but to also account for the valued nature shared by many networks of interest.
In such weighted networks we can still apply the reciprocity concept as previously defined because the underlying topology remains unchanged.
Nevertheless, \cite{squartini2013reciprocityof} develop a reciprocity notion that takes into account the actual weights rather than the mere existence of a mutual link.
\textit{Reciprocity} (in a weighted sense) is then defined as the minimum value of the two reciprocated link weights between $A$ and $B$.
This notion is now able to quantify the amount of reciprocity in weighted relationships while also providing a means to identify asymmetric relationships within established mutual links.
Using this concept of weighted reciprocity, \cite{squartini2013reciprocityof} find, e.g., social networks to be the most reciprocal and food webs to be the least reciprocal networks in a wide variety of datasets, confirming the intuition that social networks are fairly community-driven while food webs tend to follow a more hierarchical predator-prey structure.

\medskip

In general, a (binary) network can be fully characterized by its nodes and edges, the latter representing a direct relationship between two nodes.
Nodes however can also be indirectly connected via other nodes and thus still be \textit{reachable} even without a direct relationship.
The concept of a \textit{link}, as an edge connecting a pair of nodes, then naturally extends to the concept of a \textit{path}, as a sequence of distinct nodes in which each consecutive pair is connected by an edge itself \citep{newman2018networks}.
This close relationship between direct and indirect connections has led network theory to the development of null models \citep{solomonoff1951connectivityof, erdos1959onrandom, gilbert1959randomgraphs}. These are stochastic models of networks that preserve certain lower-order features so as to test whether these are able to generate other higher-order features.
The classical Erd\"{o}s-Renyi random graph preserves the total number of links in the network by imposing a single connection probability, $p$, across links.
The resulting distribution of edge counts per node, the so-called degree distribution, is however generally too homogeneous to explain the scale-free nature that is found in many empirical complex networks \citep{albert2000errorand}.
To account for this \textit{robust-yet-fragile} nature on random vs. targeted removal of nodes in such heterogeneous graphs, proper null models thus need to replicate the heavily skewed degree distributions, which in turn necessitates heterogenous connection probabilities across pairs of nodes.

\medskip

\cite{bender1978theasymptotic}, and later \cite{molloy1995acritical}, then generalized the Erd\"{o}s-Renyi random graph on undirected graphs into the \textit{Configuration Model} that preserves not only the total edge count of the network in general but also the local edge count of each node in particular.
This in turn leads to heterogeneous connection probabilities dependent on the specific edge counts of any two nodes.
\cite{newman2001randomgraphs} apply generating functions to analyze directed networks while \cite{park2004statisticalmechanics} apply the principle of maximum entropy to cast this model into the family of exponential random graphs.
Enforcing such distributional constraints on average allowed them to naturally extend the baseline model not only to directed but also to weighted networks in closed analytical forms.\footnote{Weighted networks are in fact introduced by solving the isomorphic problem of multiple unweighted networks which share the same set of nodes but generally differ in the placement of edges among pairs of nodes and networks.}
Based on homogeneous versions of these results for directed, yet unweighted, networks as well as for weighted, yet undirected, networks, \cite{garlaschelli2009generalizedbosefermi} combined these concepts into a traditional random graph that controls not only for the global edge count but also for the total sum of weights in the network.
As expected, however, this homogeneous, two-parameter model once again generates fairly thin-tailed distributions across nodal edge counts and weight sums.
This is why \cite{mastrandrea2014enhancedreconstruction} re-integrate heterogeneity into the model by controlling for these two factors (edge count and weight sums) on the node level.
They find that those stochastic models of weighted networks which do not account for unweighted properties, such as the nodal distribution of edge counts, are unable to generate networks with a realistic topological structure.
With a similar focus, \cite{squartini2011analyticalmaximum} provide a general overview on the maximum entropy derivation and maximum likelihood estimation while \cite{squartini2015unbiasedsampling} develop unbiased sampling schemes of these models.

\medskip

The framework of exponential random graphs also allows for the integration of reciprocity into closed form graph probability distributions \citep{garlaschelli2006multispeciesgrandcanonical}. Reciprocity necessarily needs directionality. In this directed but unweighted context, the problem then translates into breaking down the counts of outward and inward links further into those of reciprocal links and integrating over all possible network configurations. As reciprocal connections are closed paths between two nodes, a model that controls for these patterns seems a suitable candidate model for the investigation of higher-order properties such as closed paths of length three, or general triadic motifs \citep{milo2002networkmotifs}. In a follow-up study, \cite{squartini2013earlywarningsignals} apply this reciprocal configuration model to the Dutch interbank network consisting of binary links among banks and find strong changes of these sub-network structures prior to and within the financial crisis of 2007-2009.
Restricting the analysis to the topological backbone of the network by projecting weights onto binary edges can be a useful starting point but discards all information incorporated in the distribution of edge weights.
As this problem also applies to unweighted reciprocity, \cite{squartini2013reciprocityof} instead concentrate on the weighted notion of reciprocity and derive an exponential random graph that controls for the latter but not the former type of reciprocity on the node level.
While their model is now able to generate weighted networks, it severely lacks the capability to generate sparse network structures as alluded to by \cite{mastrandrea2014enhancedreconstruction} in the simpler context of plain edge counts and corresponding weights.

\medskip

In this paper, we try to bridge this gap in the literature by integrating \textit{both} unweighted \textit{and} weighted reciprocity into a single family of exponential random graphs.
In particular, we derive the closed form graph probability distribution over the universe of all directed, non-negative integer-valued networks controlling for both types of reciprocity distributions across nodes.
We then derive the resulting first-order conditions that identify the parameters of the model and furthermore decompose the distribution to derive an efficient sampling procedure.
Our \textit{Reciprocal Enhanced Configuration Model} (RECM) not only nests the purely unweighted \citep{garlaschelli2006multispeciesgrandcanonical, squartini2013earlywarningsignals} or weighted \citep{squartini2013reciprocityof} reciprocal configuration models but also the enhanced configuration model by \cite{mastrandrea2014enhancedreconstruction} which was able to simultaneously control for first-order unweighted and weighted degrees but unable to replicate the corresponding reciprocity sequences.
The model of \cite{garlaschelli2006multispeciesgrandcanonical} controls for such mutual links on the extensive margin, i.e. whether a (reciprocal) link exists or not, but does not account for the intensive margin, i.e. the value of an existing link.
On the other hand, the model of \cite{squartini2013reciprocityof} performs well on the intensive margin, but generally fails on the extensive margin as it generates mostly dense networks.

\medskip

As our model preserves both extensive and intensive margin of reciprocal as well as unilateral connections, we provide a statistical model that now enables the analysis of networks in which links are weighted but not every link among any two nodes necessarily has to exist, a case which is oftentimes (e.g. in social or economic networks) the most natural setting for modeling relationships among different entities.
Since the model reduces to the aforementioned models in the special cases of sparse but unweighted or weighted but dense networks, our model can generally be applied for the investigation of reciprocity as long as the underlying data represents simple, directed graphs with non-negative integer weighted values (which includes the binary set of weights for unweighted networks as special case).

\medskip

Simultaneously controlling for unweighted and weighted reciprocity furthermore enables us to not only investigate triadic binary motifs \citep{milo2002networkmotifs} but also to look at their weighted equivalents.
Even if a triangle of connected nodes exist, such an observation may be less meaningful if the underlying weight provision is extremely heterogeneous (e.g. if a subset of the links within the triad just provides a minimal amount of weight).
Such connected triangles may themselves be incomplete sub-graphs as not every edge among the three nodes has to exist (e.g. a chain of nodes in which the two end-nodes do not entertain a direct edge among themselves, or a cycle of nodes consisting of just three unilateral links).
An analysis of weighted triads within the framework of a weighted probabilistic network model unable to generate incomplete networks has thus remained a difficult endeavor.
Our model however is able to generate incomplete network structures and controls for first- and second-order network properties so that an analysis not only of binary but also of weighted triads is possible.
To this end, we also generalize the triadic motifs from \cite{milo2002networkmotifs} to weighted versions which nest the respective unweighted versions when applied to unweighted networks.
Both versions of triadic motifs in empirical networks can then be filtered for lower order (weighted) degree and reciprocity effects using our probabilistic network model to assess whether their fluctuations are just random outcomes of lower-order structural properties of the network.

\medskip

In an application, we finally estimate these different models on network data of quarterly trading volume among banks on the Italian electronic market for interbank deposits.
We concentrate on time series of the main global network properties such as trading volume and trading partners with a focus on (a-)symmetries in monetary flows.
Afterwards, we take a look at the meso-scale properties by dividing banks into large core and smaller peripheral banks to see which group of banks drives the results in the aggregate.
In order to see how these results on the macro- and meso-scale relate to the micro-scale, we further analyze the networks in smaller constellations in the form of triadic motifs.
To this end, we first define in section two the basic concepts from network theory.
Section three then introduces the principle of maximum entropy and derives the different graph probability distributions.
In section four we provide a short literature overview on interbank networks.
Here we recall theoretical and empirical results as well as stylized facts and institutional details of such decentralized markets.
After describing the underlying dataset, section five presents the descriptive results in combination with predictions from the various null models. Section six concludes.

\newpage
\section{Network Characteristics}
We will analyze reciprocity within the framework of network theory. For a thorough introduction into the topic of modern network theory, \cite{newman2018networks} provides an excellent overview.
We nevertheless recapitulate in this chapter the basic concepts that are necessary to follow the train of thought in later chapters on probabilistic networks and distributions.

\medskip

As a first step we define finite sets of nodes $\N := \{1,2,...,n\}$ and directed edges $\E := \{(i,j) : i, j \in \N, i \neq j \}$ comprising $d := |\E| \leq n(n-1)$ links without self-loops. If there is a direct link $i \Rightarrow j$ between two nodes $i,j \in \N$ it will thus be an element $(i,j) \in \E$ of the set of edges of the network. Note that a direct connection $i \Rightarrow j$ makes no statement about the reverse direction, i.e., about a possible counter-directed link, $i \Leftarrow j$, or a lack thereof, $i \not\Leftarrow j$.

\medskip

We will furthermore equip our graphs with a set of weights which we will assume to be the positive integers $\Nb_+ = \{1,2,3, ...\}$.\footnote{
  We could equally well allow for weights to be continuous over the positive real numbers.
  The theoretical results in this paper would then have appropriately redefined functional forms, the basic intuition however remains the same.
  We therefore opt for discrete weights for ease of exposition and the fact that the weights in our application (monetary units) are better characterized by countability.
}
Defining a weighting function $w: \E \longrightarrow \Nb_+$, we can the express our graph as a triple $\G := (\N, \E, w)$ or a weighted $n \times n$ adjacency matrix $W := \{w_{ij}\}$ with non-negative elements $w_{ij} \in \Nb$ and zero diagonal.

\subsection{Elementary Statistics}

The previous definitions render the configuration space of networks infinitely large but still countable.
Nevertheless, the complexity that such a large ensemble of networks brings with it easily justifies the use of simplifying statistics.
Even if we were dealing just with unweighted networks we would still have $2^{n(n-1)}$ possible network configurations.
Since the topology, i.e., the unweighted representation of the network given by edge set $\E$, provides additional information on top of the weighted perspective, we briefly define the indicator function
\begin{align}
  \one (w_{ij}) :=
  \begin{cases}
    1, &\text{if } w_{ij} > 0\\
    0, &\text{otherwise}
  \end{cases}
\end{align}
which allows us to recover the unweighted adjacency matrix $A := \{a_{ij}\} = \{ \one (w_{ij})\}$ with $a_{ij} \in \{0,1\}$ and $a_{ii} = 0$ from our weighted starting point $W$.

\medskip

This also leads us to the definition of our most elementary network statistic, the number of links in the network
\begin{align}
  d := \displaystyle\sum_{1 \leq i, j \leq n} a_{ij}
\end{align}
using linear algebraic instead of graph-theoretic terms. Normalizing this property by the amount of possible links $m := n(n-1)$ in an $n$-node network yields the \textit{density}, $D := d / m$, which is a measure of completeness or interconnectedness of the underlying network or, equivalently, the edge-wise linking probability if all edges were equally likely to occur.

\medskip

Taking weights into account, we can also define the total amount of weight exchanged in the network by
\begin{align}
  s := \displaystyle\sum_{1 \leq i, j \leq n} w_{ij}
\end{align}
summing over all elements of the weighted adjacency matrix $W$. Normalizing this property, though this time not by the maximum but by the actual number of links $d$, we can define the \textit{weighted density}, $S := s / d$, as a measure of average weight distributed over all $m$ realized edges of the network.\footnote{
  Had we defined a set of weights with a finite upper bound, one could have used that bound as alternative normalization factor so that the interpretation would be similar to the unweighted case.
}

\medskip

Another (less reductionist) avenue of complexity reduction is to create summary statistics on the node level.
The neighborhood of node $i$ consists of all neighbors to which it has an outgoing link, $N_i^{\Rightarrow} := \{j \in \N: (i,j) \in \E\}$, or to which it has an incoming link, $N_i^{\Leftarrow} := \{j \in \N: (j,i) \in \E\}$, which can thus be denoted by $N_i := N_i^{\Rightarrow} \cup N_i^{\Leftarrow}$.
In undirected networks, which we are only implicitly dealing with in this paper, an evident elementary statistic is the number of links connected to a node, or simply \textit{degree}.
An analogous concept also exists in directed networks as \textit{total degree}, $d_i := |N_i|$, of node $i$. The directedness of links however also gives rise to a useful decomposition into the number of outgoing as well incoming links
\begin{align}
  d_i^{\Rightarrow} &:= \sum_{1 \leq j \leq n} a_{ij}\\
  d_i^{\Leftarrow} &:= \sum_{1 \leq j \leq n} a_{ji}
\end{align}
called \textit{out-degree} and \textit{in-degree} of node $i$ respectively. Total degree, $d_i = d_i^{\Rightarrow} + d_i^{\Leftarrow}$, in a network without self-loops thus follows from the simple sum of both of its directed subtypes.
Similarly, there are weighted counterparts called \textit{out-strength} and \textit{in-strength} of node $i$
\begin{align}
  s_i^{\Rightarrow} &:= \sum_{1 \leq j \leq n} w_{ij}\\
  s_i^{\Leftarrow} &:= \sum_{1 \leq j \leq n} w_{ji}
\end{align}
which give the total weight that node $i$ either sends to or receives from its neighborhood respectively.
This is why oftentimes they are also labeled as weighted out- and in-degree in the literature.
Obviously, we can recover the total amount of weight exchanged in the entire network, $s = \sum_{1 \leq i,j \leq n} w_{ij}$, by summing over all individual out- or in-strengths just like the total number of links, $d = \sum_{1 \leq i,j \leq n} a_{ij}$, was recoverable by summing over all individual out- or in-degrees.

\subsection{Network Reciprocity}
Having defined the neighborhood of node $i$ as $N_i = N_i^{\Rightarrow} \cup N_I^{\Leftarrow}$ with out-neighborhood $N_i^{\Rightarrow} = \{ j \in \N: (i,j) \in  \E \}$ and in-neighborhood $N_i^{\Leftarrow} = \{ j \in \N : (j,i) \in \E \}$ allowed us to summarize out-, in- and total degrees in terms of cardinalities of these sets.
It is important to note that in many empirical applications out- and in-neighborhoods are generally not disjoint for a sizable subset of nodes.
This common neighborhood $N_i^{\leftrightarrow} := N_i^{\Rightarrow} \cap N_i^{\Leftarrow}$ that collects all neighbors to which $i$ has a simultaneous sending and receiving relationship can in turn be used to isolate all its neighbors to which it either has exclusively sending relationships, $N_i^{\rightarrow} := N_i^{\Rightarrow} \setminus N_i^{\leftrightarrow}$, or exclusively receiving relationships, $N_i^{\leftarrow} := N_i^{\Leftarrow} \setminus N_i^{\leftrightarrow}$.

\medskip

In particular, what becomes apparent from the definition of the common or reciprocal neighborhood, $N_i^{\leftrightarrow} = \{ j \in \N : (i,j) \in \E \land (j,i) \in \E \}$, is that we need to shift focus from single edges, $(i,j)$ or $i \Rightarrow j$, to dyadic edges, $\{(i,j), (j,i)\}$ or $i \leftrightarrow j$.
While we identify the former with an element of the adjacency matrix $a_{ij} = 1$, we identify the latter with $a_{ij} = 1 \,\land\, a_{ji} = 1$ or simply as a product of single, counter-directional, binary edges (eq. \ref{eq:def_aijleftrightarrow}).
\begin{align}
  a_{ij}^{\leftrightarrow} &:= a_{ij} a_{ji}
  \label{eq:def_aijleftrightarrow}
  \\
  a_{ij}^{\rightarrow} &:= a_{ij} (1 - a_{ji})
  \label{eq:def_aijrightarrow}
  \\
  a_{ij}^{\leftarrow} &:= (1 - a_{ij}) a_{ji}
  \label{eq:def_aijleftarrow}
\end{align}
If the dyadic edge only contains an outgoing connection but lacks the incoming counterpart, $\{(i,j)\}$ or $i \rightarrow j$, we identify it algebraically with eq. (\ref{eq:def_aijrightarrow}) while the opposite case, an incoming but no outgoing connection, $\{(j,i)\}$ or $i \leftarrow j$, is represented by eq. (\ref{eq:def_aijleftarrow}) or $a_{ji}^{\rightarrow}$ by symmetry.
We immediately verify that all three cases are mutually exclusive and decompose the single edges of a dyad into a linear combination of unilateral and reciprocal connections:
\begin{align}
  a_{ij} &= a_{ij}^{\rightarrow} + a_{ij}^{\leftrightarrow}\\
  a_{ji} &= a_{ij}^{\leftarrow} + a_{ij}^{\leftrightarrow}
\end{align}
We furthermore define the residual lack of reciprocal or unilateral connection by the product $a_{ij}^{\not\leftrightarrow} := (1 - a_{ij}) (1 - a_{ji})$ so that $a_{ij} = a_{ji} = 0 \Leftrightarrow a_{ij}^{\not\leftrightarrow} = 1$.
These concepts and definitions have a long tradition in sociology \citep{moreno1934whoshall, moreno1938statisticsof, holland1981anexponential} and have recently also gained traction in the field of statistical physics \citep{garlaschelli2004patternsof, garlaschelli2006multispeciesgrandcanonical} as modeling device for chemical potentials.
In particular, they allow us to formulate a decomposed version of out- and in-degree as sizes of the previously defined neighborhoods, $N_i^{\leftrightarrow}$, $N_i^{\rightarrow}$ and $N_i^{\leftarrow}$,
\begin{align}
  d_i^{\leftrightarrow} &:= \sum_{1 \leq j \leq n} a_{ji}^{\leftrightarrow}
  \label{eq:def_dileftrightarrow}\\
  d_i^{\rightarrow} &:= \sum_{1 \leq j \leq n} a_{ij}^{\rightarrow}
  \label{eq:def_dirightarrow}\\
  d_i^{\leftarrow} &:= \sum_{1 \leq j \leq n} a_{ji}^{\leftarrow}
  \label{eq:def_dileftarrow}
\end{align}
as \textit{reciprocated degree}, \textit{non-reciprocated out-degree} and \textit{non-reciprocated in-degree} respectively.
From the definitions of reciprocated dyads and by linearity, it follows immediately that the original out-degree, $d_i^{\Rightarrow} = d_i^{\rightarrow} + d_{i}^{\leftrightarrow}$, and in-degree, $d_i^{\Leftarrow} = d_i^{\leftarrow} + d_{i}^{\leftrightarrow}$, are simple combinations of the non-reciprocated and reciprocated degrees.

\medskip

As opposed to local, or node-level, network variables we may also just globally count reciprocal links
\begin{align}
  d^{\leftrightarrow} := \sum_{1 \leq i,j \leq n} a_{ij}^{\leftrightarrow}
\end{align}
or look at its normalized counterpart, the \textit{reciprocity ratio} $D^{\leftrightarrow} := d^{\leftrightarrow} / d$, i.e., the fraction of reciprocal to actual links in the network.

\medskip

While these unweighted measures of reciprocity followed uniquely from elementary set-theoretic principles, a weighted notion of reciprocity, similar to the notion of weighted density, leaves multiple options for a definition.
The literature has tried to use concepts around the idea of symmetry like imbalance $w_{ij} - w_{ji}$ \citep{serrano2007patternsof}, edge bias $w_{ij} / (w_{ij} + w_{ji})$ \citep{kovanen2010reciprocityof}, coherence $\sqrt{4w_{ij} w_{ji}} / (w_{ij} + w_{ji})$ \citep{akoglu2012quantifyingreciprocity}\footnote{
  \cite{akoglu2012quantifyingreciprocity} however also define a measure called ratio, $\min\{w_{ij}, w_{ji}\} / \max\{w_{ij}, w_{ji}\}$, which is close to our favorite measure of weighted reciprocity but still not applicable in our upcoming chapter on reciprocity-preserving network ensembles due to its non-factorizable denominator.
}
or just sets an arbitrary threshold to binarize the network and apply the unweighted reciprocity measure.
\cite{squartini2013reciprocityof} however came up with a simple and intuitive definition of weight-reciprocated edges (eq. \ref{eq:def_wijleftrightarrow})
\begin{align}
  w_{ij}^{\leftrightarrow} &:= \min\{w_{ij}, w_{ji}\}
  \label{eq:def_wijleftrightarrow}\\
  w_{ij}^{\rightarrow} &:= w_{ij} - w_{ij}^{\leftrightarrow}
  \label{eq:def_wijrightarrow}\\
  w_{ij}^{\leftarrow} &:= w_{ji} - w_{ij}^{\leftrightarrow}
  \label{eq:def_wijleftarrow}
\end{align}
which at the same time allows for weight-unreciprocated out-edges (eq. \ref{eq:def_wijrightarrow}) and in-edges (eq. \ref{eq:def_wijleftarrow}) respectively.
The authors list several advantages (e.g., scale invariance) of this definition in their paper, the major advantage we briefly highlight however is the consistency vis-\`a-vis the definition of unweighted reciprocity:
if $w_{ij}, w_{ji} \in \{0,1\}$ are unweighted, the definitions recover exactly equations (\ref{eq:def_aijleftrightarrow}) - (\ref{eq:def_aijleftarrow}) while also being well-defined for the nil-case $w_{ij} = w_{ji} = 0$.
Although the latter case is exhaustively covered in eq. (\ref{eq:def_wijleftrightarrow}), we explicitly define $w_{ij}^{\not\leftrightarrow} := a_{ij}^{\not\leftrightarrow} = [1 - \one (w_{ij})] [1 - \one (w_{ij})]$ for notational convenience.
Note however that for $w_{ij}, w_{ji} \in \mathbb{N}$, unweighted and weighted notions of reciprocity generally yield different results as they are measuring different aspects of networked relationships.
This measure of reciprocity may be interpreted as a form of exchanged flow of trust and thus bears some resemblance to a measure of indirect trust flow, $\min \{w_{ij}, w_{jk}\}$, of \cite{karlan2009trustand} that emphasizes the idea of a trust chain facilitating the establishment of a direct relationship $w_{ik}$ of two otherwise unconnected nodes.

\medskip

As it was the case for unweighted reciprocity, the weight-reciprocated and weight-unreciprocated edges also map back into their underlying weighted edges as
\begin{align}
  w_{ij} &= w_{ij}^{\rightarrow} + w_{ij}^{\leftrightarrow}
  \label{eq:id_wij}
  \\
  w_{ji} &= w_{ij}^{\leftarrow} + w_{ij}^{\leftrightarrow}
  \label{eq:id_wji}
\end{align}
Weighted reciprocity therefore is able to capture (a-)symmetries in the weight distribution of dyads and hence contains more detailed information on the nature of the reciprocal relationship beyond the binary indicator that we presented before. Aggregating weight-reciprocated edges on the node level then yields
\begin{align}
  s_i^{\leftrightarrow} &:= \sum_{1 \leq j \leq n} w_{ij}^{\leftrightarrow}
  \label{eq:def_sileftrightarrow}\\
  s_i^{\rightarrow} &:= \sum_{1 \leq j \leq n} w_{ij}^{\rightarrow}
  \label{eq:def_sirightarrow}\\
  s_i^{\leftarrow} &:= \sum_{1 \leq j \leq n} w_{ij}^{\leftarrow}
  \label{eq:def_sileftarrow}
\end{align}
\textit{reciprocated strength}, \textit{non-reciprocated out-strength} and \textit{non-reciprocated in-strength} as local measures on the node-level while the main network-wide global measure of weighted reciprocity reads
\begin{align}
  s^{\leftrightarrow} := \sum_{1 \leq i,j \leq n} s_{ij}^{\leftrightarrow}
\end{align}
with $S^{\leftrightarrow} := s^{\leftrightarrow} / d^{\leftrightarrow}$ yielding the average reciprocated weight in such a relationship.

\medskip

We close this section with one particular use-case in favor of weighted measures: the analysis of time-aggregated networks.
Many empirical networks are (close to) continuous processes with possibly very short activation times of links.
In order to reveal a signal (i.e., true relationship structures) in the noise (i.e., random activations on short time-scales) of such high-frequency settings, the literature (as well as this paper) often resorts to aggregating networks over multiple periods, thereby filtering noise and making such signals identifiable.
At the same time, however, one loses information due to aggregation.

\medskip

In the case of unweighted reciprocity this is particularly concerning, as the statistic will not be able to pick up any further signal once the edges under investigation have been reciprocated.
If the relationship becomes strongly asymmetrical, i.e., if just one node continues to provide, then any unweighted measure would be unable to identify such a one-sided relationship and, what is worse, falsely consider it a perfectly reciprocal relationship.
A weighted measure in turn would be able to correctly measure such a relationship by treating repeated links as weights. Thus, time aggregation naturally leads to weighted networks and consequently needs models that are able (i) to go beyond simple binary graphs while (ii) taking into account that not every link possibly had been activated throughout the process.\footnote{The treatment of time-varying networks is beyond the scope of this paper. For a general treatment on temporal networks, we refer to \cite{holme2012temporalnetworks} who also touch upon the topic of time aggregation.}

\newpage
\section{Network Ensembles}
We have established our universe of graphs $\G$ as the set of any directed, non-negative integer-weighted $n$-node networks without self-loops, and recalled elementary statistics on those networks. These elementary statistics were motivated by definitions of first-order (or direct) neighborhoods. It is immediately clear that higher-order (or indirect) neighborhoods, i.e., neighbors of neighbors of neighbors..., may capture other important aspects of the given network. While direct and indirect neighborhoods could in principle be unrelated for specific topologies, there is usually at least some form of dependence among and within these types of neighborhoods.
\medskip

To illustrate the point, we resort to a simplified family of \cite{bonacich1987powerand} centrality measures $c(A,\beta) = \sum_{k=0}^{\infty} \beta^k A^{k+1} I = A I + \beta A^2 I + \beta^2 A^3 I + ... $, for a given directed network adjacency matrix $A$, an $n$-dimensional vector $I = (1,1,...,1)^T$ of ones, as well as parameter $0 < \beta < 1/\lambda$ driving higher-order influences (with $\lambda_1$ being the dominant eigenvalue of network adjacency matrix $A$), and consider its two edge cases: If we let $\beta \rightarrow 0$, we get $c = AI$ and therefore recover a vector of out-degrees over all nodes capturing only their direct neighborhoods.
In any other case, however, the elements of the vector $c$ would contain centralities that take into account both the first-order as well as all higher-order walks (i.e., sequences of edges) and thus the direct but also all indirect neighborhoods \citep{benzi2015onthe}.
Larger values of $\beta$ imply less discounting of higher-order walks which may also be interpreted as higher probability of further propagating a shock from a receiving node to any of its neighbors.
Hence, given the other edge case of $\beta \rightarrow 1/\lambda$ approaching its upper bound, the influence of higher-order walks becomes maximal while still admitting a finite solution for the centrality measure.
Due to the upper bound being the reciprocal dominant eigenvalue of the network, this special case also goes by the name \textit{eigenvector centrality}.

\medskip

If one is now interested in the impact of strategic positioning of certain nodes in the network using such a family of centrality measures, as a consequence, one needs a way to disentangle the degree-based size effect from a pure, i.e. higher-order, centrality effect.
This is where probabilistic networks come into play.
By constructing a probability distribution that preserves the degrees of a given network but randomizes all other aspects, it is possible to disentangle such an effect.
Centrality values for a given configuration are properly weighted by the respective graph probabilities from the model so that the empirical counterparts can be statistically assessed by comparing them with the model outcome.
If the centrality values of the empirical network match in distribution those of the network ensemble, then one may infer that higher-order neighborhoods did not add more than noise to the statistics of the individual nodes.

\medskip

In directed networks, this centrality measure can analogously be defined for incoming connections, $c(A^T,\beta)$, so that $\beta = 0$ leads to a vector of in-degrees $c = A^T I$.
If both centralities are contrasted with their direct degree-based counterparts, the result of \cite{boguna2005generalizedpercolation} that reciprocal connections percolate the network poses the question whether one should not also control for node degrees in the adjacency matrix of reciprocal connections, $A^{\leftrightarrow} := \{a_{ij}^{\leftrightarrow}\}$, in order not to mistake direct reciprocal factors for higher-order neighborhood effects.
To put it in context, note that the percolation threshold in undirected networks is given by the classical \cite{molloy1995acritical} criterion $\sum_d d (d-2) P(d) > 0$ guaranteeing the existence of a giant connected component, which in turn is always fulfilled in degree distributions $P(d)$ for which the second moment does not exist.
For directed networks in which the directionality of edges and paths matters, this result generalizes directly into an existence condition for the giant \textit{strongly} connected component in directed networks $\sum_{d^{\Rightarrow}, d^{\Leftarrow}} d^\Rightarrow (d^\Leftarrow - 1) P(d^\Rightarrow, d^\Leftarrow) > 0$ based on the joint distribution of in- and out-degrees \citep{dorogovtsev2001giantstrongly}.

\medskip

\cite{boguna2005generalizedpercolation} combine both approaches and explicitly model edge types as either uni- or bidirectional to recover the two previous criteria as limiting cases.
Their detailed approach however allowed them to furthermore investigate the impact of reciprocal edges on the percolation threshold which they find to always exist so that reciprocity can be considered a "percolation catalyst".
This effect is particularly strong for scale-free degree distributions with scaling coefficient less than three, i.e., with infinite variance.
In this case, the threshold is in fact zero so that an infinitesimally small amount of reciprocity is sufficient for the existence of a giant strongly connected component in which all nodes can reach and be reached via directed paths.
Going back to the notion of eigenvector centrality from before, the existence of a giant strongly connected component is even a necessary condition in any incompletely connected network for eigenvector centrality to be well-defined according to the Perron-Frobenius theorem as otherwise the underlying matrix of the network would not be irreducible.
Due to this large non-local impact of reciprocity, we think that a probabilistic network model that investigates higher-order network properties should control for such an elementary network effect.

\subsection{Maximum Entropy Distributions}
We approach the general problem of finding appropriate network ensembles by resorting to the principle of maximum entropy \citep{jaynes1957informationtheory1} in constructing probability distributions that maximize entropy as a measure of randomness subject to average constraints we impose on the configurations.
The microcanonical approach to entropy maximization preserves the specified constraints exactly, whereas the canonical approach preserves them on average but still assigns low (high) graph probabilities to configurations which are less (more) in line with these constraints \citep{squartini2011analyticalmaximum}.
While the former is more precise for well-specified problems, the slightest amount of measurement error on just one of the constraints however leads to a statistical ensemble in which the true network is of measure zero.
In contrast, the canonical approach is robust in this respect while also admitting solutions in closed form for specific functional forms of the constraints.
We therefore rely on the canonical approach throughout this paper.

\medskip

To this end, let us briefly recall the general formalism based on \cite{park2004statisticalmechanics}. Let $P: \G \rightarrow [0,1]$ be a probability (counting) measure on our universe of graphs, then the appearance of a low probability network $P(G) \rightarrow 0$ has a high informational value $\ln \left[ 1 / P(G) \right]$ so that we can maximize expected surprisal as $\max_{P \in \mathbb{P}} \mathbb{E}^P [ - \ln P(G)]$ to find a maximum entropy probability measure out of the universe of possible measures $\mathbb{P}$. Unconstrained optimization, e.g., would lead to a discrete uniform distribution on graphs, $P(G) = 1 / |\G|$, which is by construction maximally uninformative. In order to preserve \textit{some} information we have to deal with constrained optimization, so let us state our objective function, the Shannon entropy, as
\begin{align}
  S(\G) = - \sum_{G \in \G} P(G) \ln P(G)
\end{align}
and the $q-$dimensional vector of constraints $x := (x_1,...,x_q)$, where $x_k := x_k(G)$ denotes real-valued statistics involving different nodes or edges in the network $G \in \G$. If we denote an empirical reference network by $G^*$ and by $x_k^* := x_k^*(G^*)$ its corresponding statistics, we can summarize all $k \in \{1,...,q\}$ constraints
\begin{align}
  x_k^* = \sum_{G \in \G} P(G) x_k(G)
\end{align}
that we impose \textit{on average} on the ensemble of networks $\G$. Maximizing the Shannon entropy subject to these network constraints necessitates furthermore the Kolmogorov axiom
\begin{align}
  \sum_{G \in \G} P(G) = 1
\end{align}
on valid probability measures so that this discrete problem can be conveniently summarized by a Lagrangian over all discrete probability mass functions with domain $\G$
\begin{align}
  \L(P) = S + \sum_{k=1}^{q} \gamma_k \left(x_k^* - \sum_{G \in \G} P(G) x_k(G) \right) + \delta \left( 1 - \sum_{G \in \G} P(G) \right)
\end{align}
and Lagrangian multipliers on constraints $\{\gamma_k\}$ and normalizing constant $\delta$ so that the resulting first-order conditions $\partial \L / \partial P(G) = \ln P(G) + 1 + \delta + \sum_{k} \gamma_k x_k(G) = 0$ yield a probability mass function of the exponential family
\begin{align}
  P(G) = \frac{ e^{- H(G)} }{Z}
\end{align}
where the \textit{Hamiltonian} $H(G) := \sum_k \gamma_k x_k(G) $ summarizes the network constraints and the \textit{Partition Function} $Z := \exp \left( 1 + \delta\right) = \sum_{G \in \G} \exp \left( - H(G) \right)$ acts as normalizing constant. The difficulty generally lies in the large configuration space $\G$ which makes an exact enumeration oftentimes infeasible, thus precluding a closed-form solution of $Z$. In the following we present linear dyadic independent models for which a closed-form solution exists and which furthermore allow for the design of an exact sampling scheme from the probability distribution of networks \citep{squartini2015unbiasedsampling}.

\subsection{Directed Enhanced Configuration Model (DECM)}
In the context of the maximum entropy graph probability distributions, \cite{mastrandrea2014enhancedreconstruction} and \cite{bianconi2009entropyof} based on \cite{garlaschelli2009generalizedbosefermi} derive the so-called \textit{Directed Enhanced Configuration Model} (DECM) in the context of directed weighted networks.
We rely on their formulation for discrete weights, but note that the continuous case \citep{gabrielli2019grandcanonical} bears large similarities.
The family of configuration models imposes constraints on the node-level which however directly translate into constraints on the global network-level.
While the older literature (e.g., \cite{molloy1995acritical}) concentrated on unweighted networks and kept fixed out- and in-degrees, the DECM additionally imposes constraints on the sequences of out- and in-strengths.

\medskip

In particular the authors formulate an irreducibility conjecture, showing that imposing \textit{only} the weighted information \textit{without} imposing also topological information (in the form of directed degrees) allocates most of the probability mass on dense networks so that network characteristics that involve topological information can generally not be reproduced.
The DECM therefore \textit{enhances} a purely weighted configuration model which preserves directed strength sequences with topological information in the form of directed degree sequences.

\medskip

Using the formalism of the previous section, the corresponding Hamiltonian then reads
\begin{align}
  H = \sum_{1 \leq i \leq n}
  \left(
    \kappa_i^{\Rightarrow} d_i^{\Rightarrow} +
    \kappa_i^{\Leftarrow} d_i^{\Leftarrow} +
    \lambda_i^{\Rightarrow} s_i^{\Rightarrow} +
    \lambda_i^{\Leftarrow} s_i^{\Leftarrow}
  \right)
\end{align}
with $4n$ Lagrangian multipliers on nodal out-degrees, $\{\kappa_i^{\Rightarrow}\}$, and in-degrees, $\{\kappa_i^{\Leftarrow}\}$, as well as on out-strengths, $\{\lambda_i^{\Rightarrow}\}$, and in-strengths $\{\lambda_i^{\Leftarrow}\}$. Redefining these multipliers with $k_i^{\Rightarrow} := \exp \{- \kappa_i^{\Rightarrow}\}$, $k_i^{\Leftarrow} := \exp \{- \kappa_i^{\Leftarrow}\}$, $l_i^{\Rightarrow} := \exp \{- \lambda_i^{\Rightarrow}\}$ and $l_i^{\Leftarrow} := \exp \{- \lambda_i^{\Leftarrow}\}$ the probability distribution over graphs $G$ identified by weighted $n \times n$ matrices $W = \{w_{ij}\}$ can be expressed with $P(G) := P(\{w_{ij}\} | \{k_i^{\Rightarrow}, k_i^{\Leftarrow}, l_i^{\Rightarrow}, l_i^{\Leftarrow}\})$ as
\begin{align}
  P(G)
  =
  \prod_{i,j: i \neq j}
  \frac{
    \left(
      k_i^{\Rightarrow} k_j^{\Leftarrow}
    \right)^{
      \one (w_{ij})
    }
    \left(
      l_i^{\Rightarrow} l_j^{\Leftarrow}
    \right)^{
      w_{ij}
    }
    \left(
      1 - l_i^{\Rightarrow} l_j^{\Leftarrow}
    \right)
  }{
    1 - l_i^{\Rightarrow} l_j^{\Leftarrow} + k_i^{\Rightarrow} k_j^{\Leftarrow} l_i^{\Rightarrow} l_j^{\Leftarrow}
  }
\end{align}

The parameters (or multipliers) can be estimated by maximizing the likelihood function or by solving the following system of non-linear first-order conditions over all nodes $i \in \{1,...,n\}$:

\begin{align}
  d_i^{\Rightarrow}
  &=
  \sum_{j: j \neq i}
  \frac{
    k_i^{\Rightarrow} k_j^{\Leftarrow} l_i^{\Rightarrow} l_j^{\Leftarrow}
  }{
    1 - l_i^{\Rightarrow} l_j^{\Leftarrow} + k_i^{\Rightarrow} k_j^{\Leftarrow} l_i^{\Rightarrow} l_j^{\Leftarrow}
  }
  \\
  d_i^{\Leftarrow}
  &=
  \sum_{j: j \neq i}
  \frac{
    k_j^{\Rightarrow} k_i^{\Leftarrow} l_j^{\Rightarrow} l_i^{\Leftarrow}
  }{
    1 - l_j^{\Rightarrow} l_i^{\Leftarrow} + k_j^{\Rightarrow} k_i^{\Leftarrow} l_j^{\Rightarrow} l_i^{\Leftarrow}
  }
  \\
  s_i^{\Rightarrow}
  &=
  \sum_{j: j \neq i}
  \frac{
    k_i^{\Rightarrow} k_j^{\Leftarrow} l_i^{\Rightarrow} l_j^{\Leftarrow}
  }{
    \left(
      1 - l_i^{\Rightarrow} l_j^{\Leftarrow}
    \right)
    \left(
      1 - l_i^{\Rightarrow} l_j^{\Leftarrow} + k_i^{\Rightarrow} k_j^{\Leftarrow} l_i^{\Rightarrow} l_j^{\Leftarrow}
    \right)
  }
  \\
  s_i^{\Leftarrow}
  &=
  \sum_{j: j \neq i}
  \frac{
    k_j^{\Rightarrow} k_i^{\Leftarrow} l_j^{\Rightarrow} l_i^{\Leftarrow}
  }{
    \left(
      1 - l_j^{\Rightarrow} l_i^{\Leftarrow}
    \right)
    \left(
      1 - l_j^{\Rightarrow} l_i^{\Leftarrow} + k_j^{\Rightarrow} k_i^{\Leftarrow} l_j^{\Rightarrow} l_i^{\Leftarrow}
    \right)
  }
\end{align}

\subsection{Reciprocal Weighted Configuration Model (RWCM)}
\cite{squartini2013reciprocityof} observe a sizeable amount of weighted reciprocity in social, technological and financial networks. Based on this observation they develop the \textit{Reciprocal Weighted Configuration Model} (RWCM) in the class of exponential random graphs.\footnote{
  The authors present several subclasses of the model. We refer to their most general version called Weighted Reciprocated Configuration Model.
}
This is at the same time a step forward and two steps backward compared to the \textit{Directed Enhanced Configuration Model} (DECM): one step forward in that they not only implicitly preserve out- and in-strength sequences but also explicitly keep fixed the weighted reciprocity structure of each node; two steps backward however, in that they do not impose restrictions on purely topological properties like directed degree sequences which makes the model ill-suited for most applications involving incomplete networks.

\medskip

Summarizing this information on weighted constraints in a Hamiltonian function
\begin{align}
  H = \sum_{1 \leq i \leq n}
  \left(
    \lambda_i^{\rightarrow} s_i^{\rightarrow}
    +
    \lambda_i^{\leftarrow} s_i^{\leftarrow}
    +
    \lambda_i^{\leftrightarrow} s_i^{\leftrightarrow}
  \right)
\end{align}
with $3n$ parameters on non-reciprocal out-strengths, $\{\lambda_i^{\rightarrow}\}$, non-reciprocal in-strengths, $\{\lambda_i^{\leftarrow}\}$, as well as on reciprocal strengths, $\{\lambda_i^{\leftrightarrow}\}$ which we can again redefine by $l_i^{\rightarrow} := \exp\{-\lambda_i^{\rightarrow}\}$, $l_i^{\leftarrow} := \exp\{-\lambda_i^{\leftarrow}\}$ and $l_i^{\leftrightarrow} := \exp\{-\lambda_i^{\leftrightarrow}\}$ so as to express the weighted reciprocity-preserving probability distribution $P(G) := P(\{w_{ij}\} | \{l_i^{\rightarrow}, l_i^{\leftarrow}, l_i^{\leftrightarrow}\})$ as
\begin{align}
  P(G)
  =
  \prod_{i, j: j > i}
  \frac{
    \left(l_i^{\rightarrow} l_j^{\leftarrow}\right)^{w_{ij}^{\rightarrow}}
    \left(l_i^{\leftarrow} l_j^{\rightarrow}\right)^{w_{ij}^{\leftarrow}}
    \left(l_i^{\leftrightarrow} l_j^{\leftrightarrow}\right)^{w_{ij}^{\leftrightarrow}}
    \left(1 - l_i^{\rightarrow} l_j^{\leftarrow}\right)
    \left(1 - l_i^{\leftarrow} l_j^{\rightarrow}\right)
    \left(1 - l_i^{\leftrightarrow} l_j^{\leftrightarrow}\right)
  }{
    1 - l_i^{\rightarrow} l_j^{\leftarrow} l_i^{\rightarrow} l_j^{\leftarrow}
  }
\end{align}
whose parameters can again be determined by maximizing the corresponding likelihood function or solving the following system of non-linear first-order conditions over all nodes $i \in \{1,...,n\}$:
\begin{align}
  s_i^{\rightarrow}
  &=
  \sum_{j: j \neq i}
  \frac{
    l_i^{\rightarrow} l_j^{\leftarrow}
    \left(
      1 - l_i^{\leftarrow} l_j^{\rightarrow}
    \right)
  }{
    \left(
      1 - l_i^{\rightarrow} l_j^{\leftarrow}
    \right)
    \left(
      1 - l_i^{\rightarrow} l_j^{\leftarrow} l_i^{\rightarrow} l_j^{\leftarrow}
    \right)
  }
  \\
  s_i^{\leftarrow}
  &=
  \sum_{j: j \neq i}
  \frac{
    l_i^{\leftarrow} l_j^{\rightarrow}
    \left(
      1 - l_i^{\rightarrow} l_j^{\leftarrow}
    \right)
  }{
    \left(
      1 - l_i^{\leftarrow} l_j^{\rightarrow}
    \right)
    \left(
    1 - l_i^{\rightarrow} l_j^{\leftarrow} l_i^{\rightarrow} l_j^{\leftarrow}
    \right)
  }
  \\
  s_i^{\leftrightarrow}
  &=
  \sum_{j: j \neq i}
  \frac{
    l_i^{\leftrightarrow} l_j^{\leftrightarrow}
  }{
    1 - l_i^{\leftrightarrow} l_j^{\leftrightarrow}
  }
\end{align}

\subsection{Reciprocal Enhanced Configuration Model (RECM)}
We are going to remedy in this section the shortcomings of the \textit{Reciprocal Weighted Configuration Model} (RWCM) by taking the irreducibility conjecture of \cite{mastrandrea2014enhancedreconstruction} seriously and \textit{enhancing} the model with information on the topological nature of the networks. In particular we will enhance the ensemble with reciprocal and non-reciprocal degrees presented in earlier sections which will allow a joint analysis of both concepts. Fortunately, ensembles for unweighted reciprocities have been developed for unweighted networks \citep{holland1981anexponential, park2004statisticalmechanics, garlaschelli2006multispeciesgrandcanonical}. By integrating these into the model from the previous section we will derive a generalized version which we will coin, in spirit of the existing models, the \textit{Reciprocal Enhanced Configuration Model} (RECM).

\medskip

We will once again start with formulating the Hamiltonian that captures the relevant binary as well as weighted, and unilateral as well as reciprocal constraints
\begin{align}
  H = \sum_{1 \leq i \leq n}
  \left(
    \kappa_i^{\rightarrow} d_i^{\rightarrow}
    +
    \kappa_i^{\leftarrow} d_i^{\leftarrow}
    +
    \kappa_i^{\leftrightarrow} d_i^{\leftrightarrow}
    +
    \lambda_i^{\rightarrow} s_i^{\rightarrow}
    +
    \lambda_i^{\leftarrow} s_i^{\leftarrow}
    +
    \lambda_i^{\leftrightarrow} s_i^{\leftrightarrow}
  \right)
\end{align}
with $6n$ parameters. Using the definitions on (non$+$)reciprocal degree (eqs. \ref{eq:def_dileftrightarrow} - \ref{eq:def_dileftarrow}) and strength (eqs. \ref{eq:def_sileftrightarrow} - \ref{eq:def_sileftarrow}) as well as the symmetries involved in unweighted (eqs. \ref{eq:def_aijleftrightarrow} - \ref{eq:def_aijleftarrow}) and weighted (eqs. \ref{eq:def_wijleftrightarrow} - \ref{eq:def_wijleftarrow}) links we can rewrite the Hamiltonian as a function of (non$+$)reciprocal edges:
\begin{align}
  \hspace{-2.5cm}
  H
  =
  \sum_{i,j: j>i}
  &
  \Big[
    \left(
      \kappa_i^{\rightarrow} + \kappa_j^{\leftarrow}
    \right)
    a_{ij}^{\rightarrow}
    +
    \left(
      \kappa_i^{\leftarrow} + \kappa_j^{\rightarrow}
    \right)
    a_{ij}^{\leftarrow}
    +
    \left(
      \kappa_i^{\leftrightarrow} + \kappa_j^{\leftrightarrow}
    \right)
    a_{ij}^{\leftrightarrow}
  \nonumber\\
  +
  \hspace{0.25cm}
  &
    \left(
      \lambda_i^{\rightarrow} + \lambda_j^{\leftarrow}
    \right)
    w_{ij}^{\rightarrow}
    +
    \left(
      \lambda_i^{\leftarrow} + \lambda_j^{\rightarrow}
    \right)
    w_{ij}^{\leftarrow}
    +
    \left(
      \lambda_i^{\leftrightarrow} + \lambda_j^{\leftrightarrow}
    \right)
    w_{ij}^{\leftrightarrow}
  \Big]
\end{align}
If we redefine parameters again by
$k_i^{\rightarrow} := \exp\{-\kappa_i^{\rightarrow}\}$,
$k_i^{\leftarrow} := \exp\{-\kappa_i^{\leftarrow}\}$,
$k_i^{\leftrightarrow} := \exp\{-\kappa_i^{\leftrightarrow}\}$,
$l_i^{\rightarrow} := \exp\{-\lambda_i^{\rightarrow}\}$,
$l_i^{\leftarrow} := \exp\{-\lambda_i^{\leftarrow}\}$,
$l_i^{\leftrightarrow} := \exp\{-\lambda_i^{\leftrightarrow}\}$
and assume dyad independence like in the previous models, the partition function $Z = \sum_{G \in \G} e^{-H}$ can be expressed as
\begin{align}
  Z =
  \prod_{i,j: j > i}
  \sum_{(w_{ij}^{\rightarrow}, w_{ij}^{\leftarrow}, w_{ij}^{\leftrightarrow})}
  \left( k_i^{\rightarrow} k_j^{\leftarrow} \right)^{
    a_{ij}^{\rightarrow}
  }
  \left( k_i^{\leftarrow} k_j^{\rightarrow} \right)^{
    a_{ij}^{\leftarrow}
  }
  \left( k_i^{\leftrightarrow} k_j^{\leftrightarrow} \right)^{
    a_{ij}^{\leftrightarrow}
  }
  \left( l_i^{\rightarrow} l_j^{\leftarrow} \right)^{
    w_{ij}^{\rightarrow}
  }
  \left( l_i^{\leftarrow} l_j^{\rightarrow} \right)^{
    w_{ij}^{\leftarrow}
  }
  \left( l_i^{\leftrightarrow} l_j^{\leftrightarrow} \right)^{
    w_{ij}^{\leftrightarrow}
  }
  \label{eq:def_Z_recm}
\end{align}
where the summation runs over all admissible triplets $(w_{ij}^{\rightarrow}, w_{ij}^{\leftarrow}, w_{ij}^{\leftrightarrow})$ from which we know that they are able to reconstruct the corresponding dyads $(w_{ij}, w_{ji}) \in \mathbb{N} \times \mathbb{N}$ based on eqs. (\ref{eq:id_wij}, \ref{eq:id_wji}). Denoting the edge-wise summands in eq. (\ref{eq:def_Z_recm}) by
\begin{align}
  Z_{ij}
  &:=
  \left( k_i^{\rightarrow} k_j^{\leftarrow} \right)^{
    \one \left(
      w_{ij}^{\rightarrow} + w_{ij}^{\leftrightarrow}
    \right)
    \left[
      1
      -
      \one \left(
        w_{ij}^{\leftarrow} + w_{ij}^{\leftrightarrow}
      \right)
    \right]
  }
  \nonumber\\
  &\,\,\times
  \left( k_i^{\leftarrow} k_j^{\rightarrow} \right)^{
    \left[
      1
      -
      \one \left(
        w_{ij}^{\rightarrow} + w_{ij}^{\leftrightarrow}
      \right)
    \right]
    \one \left(
      w_{ij}^{\leftarrow} + w_{ij}^{\leftrightarrow}
    \right)
  }
  \nonumber\\
  &\,\,\times
  \left( k_i^{\leftrightarrow} k_j^{\leftrightarrow} \right)^{
      \one \left(
        w_{ij}^{\rightarrow} + w_{ij}^{\leftrightarrow}
      \right)
      \one \left(
        w_{ij}^{\leftarrow} + w_{ij}^{\leftrightarrow}
      \right)
  }
  \nonumber\\
  &\,\,\times
  \left( l_i^{\rightarrow} l_j^{\leftarrow} \right)^{
    w_{ij}^{\rightarrow}
  }
  \left( l_i^{\leftarrow} l_j^{\rightarrow} \right)^{
    w_{ij}^{\leftarrow}
  }
  \left( l_i^{\leftrightarrow} l_j^{\leftrightarrow} \right)^{
    w_{ij}^{\leftrightarrow}
  }
\end{align}
we verify that this is indeed the case for summation over disjoint and exhaustive sets
\begin{align}
  Z =
  \prod_{i,j: j > i}
  \left(
    \sum_{(0, 0, \mathbb{N})}
    Z_{ij}
    +
    \sum_{(\mathbb{N}_+, 0, \mathbb{N})}
    Z_{ij}
    +
    \sum_{(0, \mathbb{N}_+, \mathbb{N})}
    Z_{ij}
  \right)
\end{align}
as $w_{ij}^{\rightarrow} > 0$ implies $w_{ij}^{\leftarrow} = 0$, while $w_{ij}^{\leftarrow} > 0$ implies $w_{ij}^{\rightarrow} = 0$, independent of $w_{ij}^{\leftrightarrow}$. Assuming $0 < l_i^{\rightarrow} l_j^{\leftarrow}, l_i^{\leftarrow} l_j^{\rightarrow}, l_i^{\leftrightarrow} l_j^{\leftrightarrow} < 1$ for all $i,j \in \{1,...,n\}$, like the previous models did, allows us to separately calculate the additive partition function components as
\begin{align}
  \sum_{w_{ij}^{\leftrightarrow} = 0}^{\infty}
  Z_{ij}
  \bigg|_{
    \substack{
      w_{ij}^{\rightarrow} = 0\\
      w_{ij}^{\leftarrow} = 0
    }
  }
  =
  1
  +
  k_i^{\leftrightarrow} k_j^{\leftrightarrow}
  \sum_{w_{ij}^{\leftrightarrow} = 1}^{\infty}
  \left(
    l_i^{\leftrightarrow} l_j^{\leftrightarrow}
  \right)^{
    w_{ij}^{\leftrightarrow}
  }
  =
  \frac{
    1 - l_i^{\leftrightarrow} l_j^{\leftrightarrow}
    +
    k_i^{\leftrightarrow} k_j^{\leftrightarrow} l_i^{\leftrightarrow} l_j^{\leftarrow}
  }{
    1 - l_i^{\leftrightarrow} l_j^{\leftrightarrow}
  }
  \label{eq:Zij_00N}
\end{align}
for purely weighted reciprocal relationships as well as for out-asymmetric weights
\begin{align}
  \sum_{
    \substack{
      w_{ij}^{\leftrightarrow} = 0\\
      w_{ij}^{\rightarrow} = 1
    }
  }^{\infty}
  Z_{ij}
  \bigg|_{
    w_{ij}^{\leftarrow} = 0
  }
  &=
  \sum_{w_{ij}^{\leftrightarrow} = 0}^{\infty}
  \left(
    k_i^{\rightarrow} k_j^{\leftarrow}
  \right)^{
    1 - \one \left( w_{ij}^{\leftrightarrow} \right)
  }
  \left(
    k_i^{\leftrightarrow} k_j^{\leftrightarrow}
  \right)^{
    \one \left( w_{ij}^{\leftrightarrow} \right)
  }
  \left(
    l_i^{\leftrightarrow} l_j^{\leftrightarrow}
  \right)^{
    w_{ij}^{\leftrightarrow}
  }
  \sum_{w_{ij}^{\rightarrow} = 1}^{\infty}
  \left(
    l_i^{\rightarrow} l_j^{\leftarrow}
  \right)^{
    w_{ij}^{\rightarrow}
  }
  \nonumber\\
  &=
  \frac{
    k_i^{\rightarrow} k_j^{\leftarrow}
    \left( 1 - l_i^{\leftrightarrow} l_j^{\leftrightarrow} \right)
    +
    k_i^{\leftrightarrow} k_j^{\leftrightarrow}
    l_i^{\leftrightarrow} l_j^{\leftrightarrow}
  }{
    1 - l_i^{\leftrightarrow} l_j^{\leftrightarrow}
  }
  \frac{
    l_i^{\rightarrow} l_j^{\leftarrow}
  }{
    1 - l_i^{\rightarrow} l_j^{\leftarrow}
  }
  \label{eq:Zij_N+0N}
\end{align}
and by symmetry for in-asymmetric weights
\begin{align}
  \sum_{
    \substack{
      w_{ij}^{\leftrightarrow} = 0\\
      w_{ij}^{\leftarrow} = 1
    }
  }^{\infty}
  Z_{ij}
  \bigg|_{
    w_{ij}^{\rightarrow} = 0
  }
  &=
  \frac{
    k_i^{\leftarrow} k_j^{\rightarrow}
    \left( 1 - l_i^{\leftrightarrow} l_j^{\leftrightarrow} \right)
    +
    k_i^{\leftrightarrow} k_j^{\leftrightarrow}
    l_i^{\leftrightarrow} l_j^{\leftrightarrow}
  }{
    1 - l_i^{\leftrightarrow} l_j^{\leftrightarrow}
  }
  \frac{
    l_i^{\leftarrow} l_j^{\rightarrow}
  }{
    1 - l_i^{\leftarrow} l_j^{\rightarrow}
  }
  \label{eq:Zij_0N+N}
\end{align}
which can be reduced to a common denominator so that the partition function can be expressed via the following relation
\begin{align}
  Z
  =
  \prod_{i,j: j > i}
  &\bigg\{
  \left( 1 - l_i^{\leftrightarrow} l_j^{\leftrightarrow} \right)
  \Big[
    \left( 1 - l_i^{\leftarrow} l_j^{\rightarrow} \right)
    \left(
      1 - l_i^{\rightarrow} l_j^{\leftarrow}
      +
      k_i^{\rightarrow} k_j^{\leftarrow}
      l_i^{\rightarrow} l_j^{\leftarrow}
    \right)
    +
    \left( 1 - l_i^{\rightarrow} l_j^{\leftarrow} \right)
    k_i^{\leftarrow} k_j^{\rightarrow}
    l_i^{\leftarrow} l_j^{\rightarrow}
  \Big]
  \nonumber\\
  &+
  l_i^{\leftrightarrow} l_j^{\leftrightarrow}
  k_i^{\leftrightarrow} k_j^{\leftrightarrow}
  \left(
    1
    -
    l_i^{\rightarrow} l_j^{\leftarrow}
    l_i^{\leftarrow} l_j^{\rightarrow}
  \right)
  \bigg\}
  \bigg/
  \bigg\{
  \left( 1 - l_i^{\rightarrow} l_j^{\leftarrow} \right)
  \left( 1 - l_i^{\leftarrow} l_j^{\rightarrow} \right)
  \left( 1 - l_i^{\leftrightarrow} l_j^{\leftrightarrow} \right)
  \bigg\}
  \label{eq:final_Z_recm}
\end{align}
where the last summand in the curly braces is the sum of the (expanded) last summands in the numerators of eqs. (\ref{eq:Zij_00N}) - (\ref{eq:Zij_0N+N}) and captures the tendency to reciprocate links, while the first term factorizes the respective first summands and describes the tendency towards topological non-reciprocation.

\medskip

As a consequence, the weighted- and unweighted-reciprocity preserving probability distribution $P(G) := P(\{w_{ij}\} | \{k_i^{\rightarrow}, k_i^{\leftarrow}, k_i^{\leftrightarrow}, l_i^{\rightarrow}, l_i^{\leftarrow}, l_i^{\leftrightarrow}\})$ expressed as $P(G) = e^{-H} / Z$ reads
\begin{align}
  \hspace{-2.0cm}
  P(G)
  =
  \prod_{i,j: j > i}
  \frac{
    \left(
      k_i^{\rightarrow} k_j^{\leftarrow}
    \right)^{
      a_{ij}^{\rightarrow}
    }
    \left(
      k_i^{\leftarrow} k_j^{\rightarrow}
    \right)^{
      a_{ij}^{\leftarrow}
    }
    \left(
      k_i^{\leftrightarrow} k_j^{\leftrightarrow}
    \right)^{
      a_{ij}^{\leftrightarrow}
    }
    \left(
      l_i^{\rightarrow} l_j^{\leftarrow}
    \right)^{
      w_{ij}^{\rightarrow}
    }
    \left(
      l_i^{\leftarrow} l_j^{\rightarrow}
    \right)^{
      w_{ij}^{\leftarrow}
    }
    \left(
      l_i^{\leftrightarrow} l_j^{\leftrightarrow}
    \right)^{
      w_{ij}^{\leftrightarrow}
    }
    \left( 1 - l_i^{\rightarrow} l_j^{\leftarrow} \right)
    \left( 1 - l_i^{\leftarrow} l_j^{\rightarrow} \right)
    \left( 1 - l_i^{\leftrightarrow} l_j^{\leftrightarrow} \right)
  }{
      \left( 1 - l_i^{\leftrightarrow} l_j^{\leftrightarrow} \right)
    \left[
      \left( 1 - l_i^{\leftarrow} l_j^{\rightarrow} \right)
      \left(
        1 - l_i^{\rightarrow} l_j^{\leftarrow}
        +
        k_i^{\rightarrow} k_j^{\leftarrow}
        l_i^{\rightarrow} l_j^{\leftarrow}
      \right)
      +
      \left( 1 - l_i^{\rightarrow} l_j^{\leftarrow} \right)
      k_i^{\leftarrow} k_j^{\rightarrow}
      l_i^{\leftarrow} l_j^{\rightarrow}
    \right]
    +
    l_i^{\leftrightarrow} l_j^{\leftrightarrow}
    k_i^{\leftrightarrow} k_j^{\leftrightarrow}
    \left(
      1
      -
      l_i^{\rightarrow} l_j^{\leftarrow}
      l_i^{\leftarrow} l_j^{\rightarrow}
    \right)
  }
  \label{eq:final_P_recm}
\end{align}
with the standard reciprocity mappings, $a_{ij}^{\rightarrow} = \one (w_{ij}) [1 - \one (w_{ji})]$ for unweighted non-reciprocated outgoing links, $a_{ij}^{\leftarrow} = [1 - \one (w_{ij})] \one (w_{ji})$ for unweighted non-reciprocated incoming links, and $a_{ij}^{\leftrightarrow} = \one (w_{ij}) \one (w_{ji})$ for unweighted reciprocated links, as well as $w_{ij}^{\rightarrow} = w_{ij} - \min \{w_{ij}, w_{ji}\}$ for weighted non-reciprocated outgoing links,  $w_{ij}^{\leftarrow} = w_{ji} - \min \{w_{ij}, w_{ji}\}$ for weighted non-reciprocated incoming links, and $w_{ij}^{\leftrightarrow} = \min \{w_{ij}, w_{ji}\}$ for weighted reciprocated links.
Parameter estimation can then be performed again by maximum likelihood estimation or equivalently by solving the following system of non-linear first-order conditions
\begin{align}
  &\hspace{-0.3cm}
  d_i^{\rightarrow}
  =
  \sum_{j: j \neq i}
  k_i^{\rightarrow} k_j^{\leftarrow}
  \left( 1 - l_i^{\leftrightarrow} l_j^{\leftrightarrow} \right)
  l_i^{\rightarrow} l_j^{\leftarrow}
  \left( 1 - l_i^{\rightarrow} l_j^{\leftarrow} \right)
  \Big/
  X_{ij}
  \\
  &\hspace{-0.3cm}
  d_i^{\leftarrow}
  =
  \sum_{j: j \neq i}
  k_i^{\leftarrow} k_j^{\rightarrow}
  \left( 1 - l_i^{\leftrightarrow} l_j^{\leftrightarrow} \right)
  l_i^{\leftarrow} l_j^{\rightarrow}
  \left( 1 - l_i^{\leftarrow} l_j^{\rightarrow} \right)
  \Big/
  X_{ij}
  \\
  &\hspace{-0.3cm}
  d_i^{\leftrightarrow}
  =
  \sum_{j: j \neq i}
  k_i^{\leftrightarrow} k_j^{\leftrightarrow}
  l_i^{\leftrightarrow} l_j^{\leftrightarrow}
  \left(
    1
    -
    l_i^{\rightarrow} l_j^{\leftarrow}
    l_i^{\leftarrow} l_j^{\rightarrow}
  \right)
  \Big/
    X_{ij}
  \\
  &\hspace{-0.3cm}
  s_i^{\rightarrow}
  =
  \sum_{j: j \neq i}
  \Big[
    k_i^{\rightarrow} k_j^{\leftarrow}
    \left( 1 - l_i^{\leftrightarrow} l_j^{\leftrightarrow} \right)
    +
    k_i^{\leftrightarrow} k_j^{\leftrightarrow}
    l_i^{\leftrightarrow} l_j^{\leftrightarrow}
  \Big]
  l_i^{\rightarrow} l_j^{\leftarrow}
  \left( 1 - l_i^{\rightarrow} l_j^{\leftarrow} \right)
  \Big/
  \Big[
    \left( 1 - l_i^{\rightarrow} l_j^{\leftarrow}\right)
    X_{ij}
  \Big]
  \\
  &\hspace{-0.3cm}
  s_i^{\leftarrow}
  =
  \sum_{j: j \neq i}
  \Big[
    k_i^{\leftarrow} k_j^{\rightarrow}
    \left( 1 - l_i^{\leftrightarrow} l_j^{\leftrightarrow} \right)
    +
    k_i^{\leftrightarrow} k_j^{\leftrightarrow}
    l_i^{\leftrightarrow} l_j^{\leftrightarrow}
  \Big]
  l_i^{\leftarrow} l_j^{\rightarrow}
  \left( 1 - l_i^{\leftarrow} l_j^{\rightarrow} \right)
  \Big/
  \Big[
    \left( 1 - l_i^{\leftarrow} l_j^{\rightarrow}\right)
    X_{ij}
  \Big]
  \\
  &\hspace{-0.3cm}
  s_i^{\leftrightarrow}
  =
  \sum_{j: j \neq i}
  k_i^{\leftrightarrow} k_j^{\leftrightarrow}
  l_i^{\leftrightarrow} l_j^{\leftrightarrow}
  \left(
    1
    -
    l_i^{\rightarrow} l_j^{\leftarrow}
    l_i^{\leftarrow} l_j^{\rightarrow}
  \right)
  \Big/
  \Big[
    \left( 1 - l_i^{\leftrightarrow} l_j^{\leftrightarrow}\right)
    X_{ij}
  \Big]
\end{align}
for all nodes $i \in \{1,...,n\}$ where the symmetric ($X_{ij} = X_{ji}$) term
\begin{align}
  X_{ij}
  &:=
  \left( 1 - l_i^{\leftrightarrow} l_j^{\leftrightarrow} \right)
  \left[
    \left( 1 - l_i^{\leftarrow} l_j^{\rightarrow} \right)
    \left(
      1 - l_i^{\rightarrow} l_j^{\leftarrow}
      +
      k_i^{\rightarrow} k_j^{\leftarrow}
      l_i^{\rightarrow} l_j^{\leftarrow}
    \right)
    +
    \left( 1 - l_i^{\rightarrow} l_j^{\leftarrow} \right)
    k_i^{\leftarrow} k_j^{\rightarrow}
    l_i^{\leftarrow} l_j^{\rightarrow}
  \right]
  \nonumber\\
  &\,\,+
  l_i^{\leftrightarrow} l_j^{\leftrightarrow}
  k_i^{\leftrightarrow} k_j^{\leftrightarrow}
  \left(
    1
    -
    l_i^{\rightarrow} l_j^{\leftarrow}
    l_i^{\leftarrow} l_j^{\rightarrow}
  \right)
  \label{eq:def_Xij}
\end{align}

is the numerator of the factors in the partition function $Z$ (eq. \ref{eq:final_Z_recm}).
In order to derive an exact sampling scheme from the distribution, let us first decompose $X_{ij}$ into the following fractions

\begin{align}
  p_{ij}^{\not\leftrightarrow}
  &=
  \left( 1 - l_i^{\leftrightarrow} l_j^{\leftrightarrow} \right)
  \left( 1 - l_i^{\rightarrow} l_j^{\leftarrow} \right)
  \left( 1 - l_i^{\rightarrow} l_j^{\leftarrow} \right)
  \Big/
  X_{ij}
  \\
  p_{ij}^{\rightarrow}
  &=
  k_i^{\rightarrow} k_j^{\leftarrow}
  \left( 1 - l_i^{\leftrightarrow} l_j^{\leftrightarrow} \right)
  l_i^{\rightarrow} l_j^{\leftarrow}
  \left( 1 - l_i^{\rightarrow} l_j^{\leftarrow} \right)
  \Big/
  X_{ij}
  \\
  p_{ij}^{\leftarrow}
  &=
  k_i^{\leftarrow} k_j^{\rightarrow}
  \left( 1 - l_i^{\leftrightarrow} l_j^{\leftrightarrow} \right)
  l_i^{\leftarrow} l_j^{\rightarrow}
  \left( 1 - l_i^{\leftarrow} l_j^{\rightarrow} \right)
  \Big/
  X_{ij}
  \\
  p_{ij}^{\leftrightarrow}
  &=
  k_i^{\leftrightarrow} k_j^{\leftrightarrow}
  l_i^{\leftrightarrow} l_j^{\leftrightarrow}
  \left(
    1
    -
    l_i^{\rightarrow} l_j^{\leftarrow}
    l_i^{\leftarrow} l_j^{\rightarrow}
  \right)
  \Big/
  X_{ij}
\end{align}
so that $p_{ij}^{\not\leftrightarrow} + p_{ij}^{\rightarrow} + p_{ij}^{\leftarrow} + p_{ij}^{\leftrightarrow} = 1$ which we can immediately verify by expanding eq. (\ref{eq:def_Xij}). Notice that the last three equations match with the first three first-order constraints and the general definitions on (non-)reciprocal degrees. They can thus be interpreted as existence probabilities for the specific type of an unweighted (non-)reciprocated edge.

\medskip

As we also deal with a weighted notion of reciprocity, we need to decompose the last term, $p_{ij}^{\leftrightarrow}$, further into
\begin{align}
  p_{ij}^{\not\leftrightarrow | \leftrightarrow}
  &=
  k_i^{\leftrightarrow} k_j^{\leftrightarrow}
  l_i^{\leftrightarrow} l_j^{\leftrightarrow}
  \left( 1 - l_i^{\rightarrow} l_j^{\leftarrow} \right)
  \left( 1 - l_i^{\leftarrow} l_j^{\rightarrow} \right)
  \Big/
  X_{ij}
  \\
  p_{ij}^{\rightarrow | \leftrightarrow}
  &=
  k_i^{\leftrightarrow} k_j^{\leftrightarrow}
  l_i^{\leftrightarrow} l_j^{\leftrightarrow}
  l_i^{\rightarrow} l_j^{\leftarrow}
  \left( 1 - l_i^{\leftarrow} l_j^{\rightarrow} \right)
  \Big/
  X_{ij}
  \\
  p_{ij}^{\leftarrow | \leftrightarrow}
  &=
  k_i^{\leftrightarrow} k_j^{\leftrightarrow}
  l_i^{\leftrightarrow} l_j^{\leftrightarrow}
  l_i^{\leftarrow} l_j^{\rightarrow}
  \left( 1 - l_i^{\rightarrow} l_j^{\leftarrow} \right)
  \Big/
  X_{ij}
\end{align}
so that the probability of an unweighted reciprocal link $p_{ij}^{\leftrightarrow} = p_{ij}^{\not\leftrightarrow | \leftrightarrow} + p_{ij}^{\rightarrow | \leftrightarrow} + p_{ij}^{\leftarrow | \leftrightarrow}$ can be separated into the probability of an exact weight-reciprocated link $p_{ij}^{\not\leftrightarrow | \leftrightarrow}$, a weight-reciprocated link with stronger out-weight $p_{ij}^{\rightarrow | \leftrightarrow}$ or a weight-reciprocated link with a stronger in-weight $p_{ij}^{\leftarrow | \leftrightarrow}$.

\medskip

If we factorize the maximum entropy distribution of the Reciprocal Enhanced Configuration Model (eq. \ref{eq:final_P_recm}), into dyad probabilities $P(G) = \prod_{i,j: j > i} P_{ij}$, with
\begin{align}
  X_{ij}
  P_{ij}
  &=
  \left(
    k_i^{\rightarrow} k_j^{\leftarrow}
  \right)^{
    \one \left( w_{ij}^{\rightarrow} + w_{ij}^{\leftrightarrow} \right)
    \left[
    1
    -
    \one \left( w_{ij}^{\leftarrow} + w_{ij}^{\leftrightarrow} \right)
    \right]
  }
  \nonumber\\
  &\,\times
  \left(
    k_i^{\leftarrow} k_j^{\rightarrow}
  \right)^{
    \left[
    1
    -
    \one \left( w_{ij}^{\rightarrow} + w_{ij}^{\leftrightarrow} \right)
    \right]
    \one \left( w_{ij}^{\leftarrow} + w_{ij}^{\leftrightarrow} \right)
  }
  \nonumber\\
  &\,\times
  \left(
    k_i^{\leftrightarrow} k_j^{\leftrightarrow}
  \right)^{
    \one \left( w_{ij}^{\rightarrow} + w_{ij}^{\leftrightarrow} \right)
    \one \left( w_{ij}^{\leftarrow} + w_{ij}^{\leftrightarrow} \right)
  }
  \nonumber\\
  &\,\times
  \left(
    l_i^{\rightarrow} l_j^{\leftarrow}
  \right)^{
    w_{ij}^{\rightarrow}
  }
  \left(
    l_i^{\leftarrow} l_j^{\rightarrow}
  \right)^{
    w_{ij}^{\leftarrow}
  }
  \left(
    l_i^{\leftrightarrow} l_j^{\leftrightarrow}
  \right)^{
    w_{ij}^{\leftrightarrow}
  }
  \nonumber\\
  &\,\times
  \left( 1 - l_i^{\rightarrow} l_j^{\leftarrow} \right)
  \left( 1 - l_i^{\leftarrow} l_j^{\rightarrow} \right)
  \left( 1 - l_i^{\leftrightarrow} l_j^{\leftrightarrow} \right)
\end{align}

we can express it in terms of the different types of (un-)weighted (non-)reciprocal links

\begin{align}
  \hspace{-0.75cm}
  P_{ij} =
  \begin{cases}
    p_{ij}^{\not\leftrightarrow}
    &\text{, if}\,\,
    w_{ij}^{\rightarrow} = 0,\,
    w_{ij}^{\leftarrow} = 0,\,
    w_{ij}^{\leftrightarrow} = 0
    \\
    p_{ij}^{\rightarrow}
    \left(
      l_i^{\rightarrow} l_j^{\leftarrow}
    \right)^{
      w_{ij}^{\rightarrow} - 1
    }
    \left(
      1 - l_i^{\rightarrow} l_j^{\leftarrow}
    \right)
    &\text{, if}\,\,
    w_{ij}^{\rightarrow} > 0,\,
    w_{ij}^{\leftarrow} = 0,\,
    w_{ij}^{\leftrightarrow} = 0
    \\
    p_{ij}^{\leftarrow}
    \left(
      l_i^{\leftarrow} l_j^{\rightarrow}
    \right)^{
      w_{ij}^{\leftarrow} - 1
    }
    \left(
      1 - l_i^{\leftarrow} l_j^{\rightarrow}
    \right)
    &\text{, if}\,\,
    w_{ij}^{\rightarrow} = 0,\,
    w_{ij}^{\leftarrow} > 0,\,
    w_{ij}^{\leftrightarrow} = 0
    \\
    p_{ij}^{\not\leftrightarrow | \leftrightarrow}
    \left(
      l_i^{\leftrightarrow} l_j^{\leftrightarrow}
    \right)^{
      w_{ij}^{\leftrightarrow} - 1
    }
    \left(
      1 - l_i^{\leftrightarrow} l_j^{\leftrightarrow}
    \right)
    &\text{, if}\,\,
    w_{ij}^{\rightarrow} = 0,\,
    w_{ij}^{\leftarrow} = 0,\,
    w_{ij}^{\leftrightarrow} > 0
    \\
    p_{ij}^{\rightarrow | \leftrightarrow}
    \left(
      l_i^{\leftrightarrow} l_j^{\leftrightarrow}
    \right)^{
      w_{ij}^{\leftrightarrow} - 1
    }
    \left(
      1 - l_i^{\leftrightarrow} l_j^{\leftrightarrow}
    \right)
    \left(
      l_i^{\rightarrow} l_j^{\leftarrow}
    \right)^{
      w_{ij}^{\rightarrow} - 1
    }
    \left(
      1 - l_i^{\rightarrow} l_j^{\leftarrow}
    \right)
    &\text{, if}\,\,
    w_{ij}^{\rightarrow} > 0,\,
    w_{ij}^{\leftarrow} = 0,\,
    w_{ij}^{\leftrightarrow} > 0
    \\
    p_{ij}^{\leftarrow | \leftrightarrow}
    \left(
      l_i^{\leftrightarrow} l_j^{\leftrightarrow}
    \right)^{
      w_{ij}^{\leftrightarrow} - 1
    }
    \left(
      1 - l_i^{\leftrightarrow} l_j^{\leftrightarrow}
    \right)
    \left(
      l_i^{\leftarrow} l_j^{\rightarrow}
    \right)^{
      w_{ij}^{\leftarrow} - 1
    }
    \left(
      1 - l_i^{\leftarrow} l_j^{\rightarrow}
    \right)
    &\text{, if}\,\,
    w_{ij}^{\rightarrow} = 0,\,
    w_{ij}^{\leftarrow} > 0,\,
    w_{ij}^{\leftrightarrow} > 0
  \end{cases}
  \label{eq:final_Pij_samplingdecomposition}
\end{align}
where we can set
$
p_{ij}^{\not\leftrightarrow}
=
1
- p_{ij}^{\rightarrow}
- p_{ij}^{\leftarrow}
- p_{ij}^{\not\leftrightarrow | \leftrightarrow}
- p_{ij}^{\rightarrow | \leftrightarrow}
- p_{ij}^{\leftarrow | \leftrightarrow}
$
due to one degree of freedom of probabilities as we have shown before.

\medskip

This lends itself to a two-step sampling scheme, in which we draw in a first step a purely topological reciprocal edge $\hat{\omega}_{ij}^{(1)} := \left( \one (w_{ij}^{\rightarrow}), \one (w_{ij}^{\leftarrow}), \one (w_{ij}^{\leftrightarrow}) \right)$,
\begin{align*}
  \hat{\omega}_{ij}^{(1)}
  \sim
  \textit{Categorical}
  \left(
    p_{ij}^{\not\leftrightarrow},
    p_{ij}^{\rightarrow}, p_{ij}^{\leftarrow},
    p_{ij}^{\not\leftrightarrow | \leftrightarrow},
    p_{ij}^{\rightarrow | \leftrightarrow},
    p_{ij}^{\leftarrow | \leftrightarrow}
  \right)
\end{align*}
from a Categorical distribution with support given by the six cases in eq.(\ref{eq:final_Pij_samplingdecomposition}), and in a second step we sample a vector of (non-)reciprocal weights $\hat{\omega}_{ij}^{(2)} := \left( w_{ij}^{\rightarrow}, w_{ij}^{\leftarrow}, w_{ij}^{\leftrightarrow} \right)$,
\begin{align*}
  \hat{\omega}_{ij}^{(2)}
  \sim
  \textit{Geometric}(1 - l_i^{\rightarrow} l_j^{\leftarrow})
  \times
  \textit{Geometric}(1 - l_i^{\leftarrow} l_j^{\rightarrow})
  \times
  \textit{Geometric}(1 - l_i^{\leftrightarrow} l_j^{\leftrightarrow})
\end{align*}
 from a Cartesian product of (shifted) Geometric distributions. Both samples can be combined by elementwise multiplication, $\circ$, to yield a vector
\begin{align}
  \hat{\omega}_{ij}
  :=
  \left(
    \hat{\omega}_{ij}^{\rightarrow},
    \hat{\omega}_{ij}^{\leftarrow},
    \hat{\omega}_{ij}^{\leftrightarrow}
  \right)
  =
  \hat{\omega}_{ij}^{(1)} \circ \hat{\omega}_{ij}^{(2)}
\end{align}
with $0$, $1$ or $2$ non-zero entries, which is able to generate a valid weighted dyadic relationship
\begin{align}
    \left(\hat{w}_{ij}, \hat{w}_{ji}\right) = \left(
    \hat{\omega}_{ij}^{\rightarrow} + \hat{\omega}_{ij}^{\leftrightarrow},
    \hat{\omega}_{ij}^{\leftarrow} + \hat{\omega}_{ij}^{\leftrightarrow}
  \right)
\end{align}
Applying this sampling procedure for all pairs of nodes $j > i$ yields sample dyads $(\hat{w}_{ij}, \hat{w}_{ji})$ that together form a sample $\widehat{W} = \{\left(\hat{w}_{ij}, \hat{w}_{ji}\right)\}_{j > i}$ of the reciprocity-preserving ensemble over weighted networks.

\newpage
\section{Interbank Markets}
We are going to apply the randomized graph ensembles to a time series of interbank networks. In such networks, banks act as nodes and their lending volume is represented by weighted edges. In this section we briefly describe the institutional details of interbank markets and the dataset of the Italian market for interbank deposits (e-MID).
As the \textit{Reciprocal Enhanced Configuration Model} (RECM) which we have developed in the previous section preserves constraints that form strict supersets of those of the other models, we can investigate in more detail the role that reciprocity plays in a market for interbank deposits.

\medskip

Unlike many conventional markets for real goods or financial assets, the products on wholesale credit market are fundamentally heterogeneous.
The underlying contracts are bilaterally negotiated in terms of credit volume, interest rate and maturity and may reflect different levels of counterparty risk.
This in turn makes the default risk of the borrowing bank an essential part of the product.
Despite the inherent risk, the largest part of such bilateral interbank deposit contracts is based on unsecured loans.
Banks, however, can also finance their operations using collateralized loans, which can be acquired on the secured segments of the interbank market or as part of the standard refinancing operations of the central bank.
This has the advantage of reducing credit risk due to seizable collateral, but to the detriment of higher financing costs.

\medskip

Commercial banks usually need to hold a certain amount of liquidity so as to manage their daily operations as well as to fulfill regulatory liquidity requirements. The former consists of idiosyncratic shocks to customer deposits as a result of the myriad of bank transfers which customers of the banks regularly perform throughout the day.
The latter, on the other hand, highlights the role of minimum reserves that commercial banks have to deposit on their central bank accounts as a fraction of their average customer deposits (e.g., currently 1\% for banks in the Euro Area).
Maintaining balance over such a reserve maintenance period is greatly facilitated by a liquid, well-functioning market for interbank deposits.
The existence of reliable credit lines with many other banks therefore insures a commercial bank against a potential shortfall of funds.
Nevertheless, this tendency to establishing a complete network of interbank liabilities among any active bank is hampered by the concept of credit risk. This is because the costs of a haircut in case of a debtor's default may outweigh the benefit of both short-term yield on surplus liquidity and long-term advantages of relationship banking.

\medskip

Liquidity however also depends on two other crucial features that are less relevant in other markets: search costs and informational frictions.
Search costs are particularly relevant in opaque markets, where negotiations may be conducted bilaterally, e.g., over the phone.
An institutional marketplace such as an official exchange can greatly mitigate search costs and thus enable mutually beneficial trades by aggregating and distributing information on different contracts and trading partners.
In contrast to search frictions, informational frictions typically remain even on established exchanges.
Whereas search costs are based on the fundamental uncertainty of the nature of possible links (i.e., credit lines) in the deposit network, informational costs are related to uncertainty on the nature of the nodes (i.e., default risk of the partner bank).
With high search costs, a bank is unaware of which credit lines can be activated or created. But even without any search costs, a liquidity-long bank that consequently knows about the liquidity needs of the other banks may not be able to fully assess the solvency of their potential trading partners.

\medskip

While most banks are part of an electronic automated payment system with accounts at their respective central bank to facilitate automatic settlement, e.g., real time gross settlement (RTGS) systems like TARGET2 in Europe or Fedwire in the US, the negotiation process for funds between two banks is oftentimes less standardized as offers are often made over the counter (OTC) in opaque bilateral agreements or on trading floors.

\subsection{The Italian Market for Interbank Deposits (e-MID)}
A notable exception to the search-cost-ridden OTC markets was the European electronic Market for Interbank Deposits (e-MID).\footnote{
  Alternative regulated market settings in Europe consisted of the trading platform MTS and the MIC (founded in 2009) that depend on the clearing house Cassa Compensazione e Garanzia (CC\&G) which takes the role as central clearing party so that counterparty risk in these markets is reduced by netting transactions while transaction partners remain anonymous \citep{cappelletti2011theimpact}.
}
The Italian-based e-MID was a screen-based market for uncollateralized interbank loans with maturities ranging in discrete steps from overnight to weekly or monthly loans up to a maturity of one year.
Founded in 1990 as an Italian platform, the e-MID transitioned to a centralized market for European banks with the introduction of the Euro in 1999, and in 2006 made up for 17\% of the entire unsecured trading volume in the Euro Area interbank market with most trades denominated in Euro (95\%) on an overnight basis ($>$90\%) as of 2010 \citep{fricke2015coreperipherystructure}.

\medskip

Throughout its opening hours, 9am - 6pm, banks have the option to post public bid or ask quotes, i.e. proposals that state volume, rate and timing as well as the bank's identity - there is an option to disclose identity later but most banks prefer the former option \citep{hatzopoulos2015quantifyingpreferential} - which are then collected similar to a limit order book and displayed in descending order for bid quote rates and ascending order for ask quote rates.
Unlike a conventional limit order book there is no consolidation however.
In fact, other banks may \textit{hit} these proposals by submitting orders. The banks which submit orders on existing proposals are also called \textit{aggressors} in the context of e-MID data.

\medskip

If the quoter was liquidity-short, thus submitting a \textit{bid} proposal, and a liquidity-long aggressor hit that bid-quote with a corresponding sell order, the trade gets executed automatically.
If on the other hand the quoter was liquidity-long with an \textit{ask} proposal and got hit by a liquidity-short aggressor submitting a buy (market) order, the quoter gets in turn the possibility to accept or reject the order.
As rejections can be based on self-defined credit line limits, the lender still has the upper hand in the negotiation process.

\medskip

Nevertheless, independent of bid or ask position, the aggressor also has the possibility to underwrite an adjusted counter-offer based on a proposal she wants to hit, which the quoter in turn may reject or accept \citep{brunetti2011effectsof}.
Proposals, i.e. quotes, are therefore similar to limit orders while hits, i.e. aggressions, are similar to market orders in financial markets.
Bid limit orders, i.e. sell market orders, in fact dominate the number of actual trades on the e-MID market with 81\% of total interbank lending \citep{schwarz2019mindthe}.
Minimum quote size in general is 1.5M Euro, but minimum realized trade size is only 50000 Euro. Besides posting quotes on the book, banks also have the possibility to directly contact a preferred counterparty by sending them a quote request or an issued order that the latter may accept or decline \citep{beaupain2011inferringtrading}.

\subsection{Dataset}
Our dataset consists of tick data on realized trades on the e-MID platform from 2005-01-01 to 2011-12-31.
Each transaction record contains details on date, time (in seconds), maturity, currency, trading volume, interest rates, anonymized quoter and aggressor ID, and the side of the hit order, i.e., a label \textit{sell} if the aggressor hit a bid-quote with her sell-order or a label \textit{buy} if the aggressor hit an ask-quote with her buy-order \citep{iori2015networkedrelationships}.
We will follow the literature \citep{fricke2015coreperipherystructure, raddant2014structurein} in focusing on trades between Italian banks because (i) the amount of foreign lending deteriorates tremendously over our sample period covering the financial crisis and (ii) Italian and non-Italian banks have been found to form fairly separate clusters which justifies an isolated investigation. In the same vein we concentrate exclusively on the overnight segment of the Italian money market as it comprises the majority of trades (>90\%) on the e-MID platform whereas contracts with longer maturities were nearly evaporated from the market once the financial crisis set in.

\medskip

\cite{finger2013networkanalysis} have shown that the network properties of the e-MID crucially depend on the time scale and aggregation period.
The authors argue that longer aggregation periods (than a daily scale at which many properties appear to be randomly generated, thus with too low of a signal-to-noise ratio) are necessary to uncover the underlying relationships in the network and find quarterly aggregations to yield stable network properties.
Since our focus is on credit lines among banks, we follow this advice which leaves us with 28 consecutive quarterly, directed and weighted networks between Italian banks on the unsecured market for overnight deposits, $W(t) := \{w_{ij}(t)\}$, indicating the total lending volume (in Euro) from bank $i$ to $j$ in the respective quarter $t$.
For each quarterly network we will also approximate the corresponding probability distributions with a sample of a thousand networks drawn from the Directed Enhanced Configuration Model (DECM), Reciprocal Weighted Configuration Model (RWCM) and Reciprocal Enhanced Configuration Model (RECM) to calculate synthetic network statistics based on the underlying empirical constraints of the respective model.

\subsection{Crisis Events}
\begin{figure}[H]
  \centering
  \includegraphics[scale=0.5, trim = 0cm 0cm 0cm 0cm, clip]{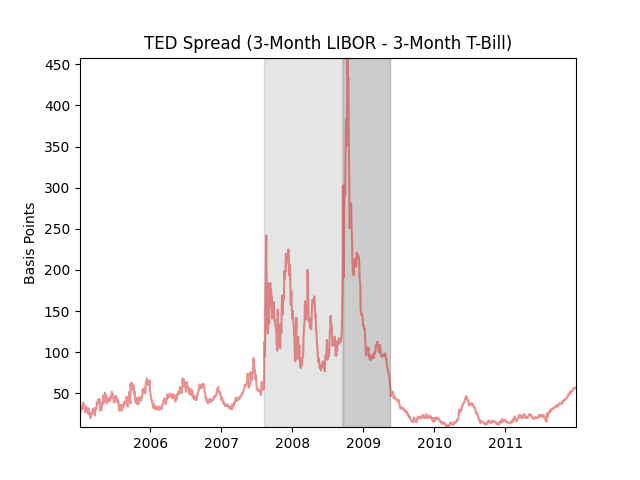}
  \caption{\textit{Counterparty Risk - TED Spread (Source: Federal Reserve Bank of St. Louis).}
  }
  \label{fig:ted_spread}
\end{figure}

As the financial crisis is in the center of our sample period and is certainly going to have a sizeable impact on our interbank networks, we need a working definition to mark out relevant events.
Even though we deal with European data, its inception lies in US money markets.
Figure (\ref{fig:ted_spread}) shows the TED Spread, i.e. the difference between 3-Month LIBOR and a 3-Month Treasury Bill, as a measure of counterparty risk.
LIBOR, the London interbank offered rate, is a benchmark rate at which selected larger banks are able to borrow on the unsecured interbank market, while T-Bills as government debt obligations possess a similar liquidity but are generally considered safe assets.
While the spread should be small in normal times, it should widen up in times of (financial) crisis as counterparty or default risk soars and liquidity providing banks need compensation for taking up that elevated risk in form of higher lending rates.

\medskip

We mark the beginning of the crisis as August 7th, 2007, the date on which BNP Paribas froze redemptions on three of their investment funds as they were not able to provide a stable Net Asset Value \citep{kacperczyk2010whensafe} which swiftly dried up liquidity in securitized markets and lead the TED spread to jump from 57 to 83 and 113 basis points over the next three days.
While interruptions on the US interbank market already surfaced a couple of months earlier, this date marks a quantitative starting point of elevated counterparty risk in the market.

Due to international repercussions, it is even less trivial to find a definitive end date of the financial crisis. We will therefore stick to the previous criterion and define it as the day at which the TED spread is back at 57 basis points, which is May 19th, 2009 (and at the same time coincides with the last quarter of the crisis-induced recession in the US).
As our network observations are also in quarterly frequency we therefore set the crisis episode as 2007Q3 - 2009Q2.
The insolvency of Lehman brothers on September 15th, 2008 obviously marked one of the defining events of the crisis. It lead spreads to jump from 179 to 303bp within the next 2 days making the TED spread (and by proxy counterparty risk) reach a new all-time high. We thus use this event to define the first stage of the crisis as 2007Q3 - 2008Q3 and the second (post-Lehman) stage as 2008Q4 - 2009Q2.\footnote{
  Spreads based on Overnight Index Swaps (OIS) as safe rates instead of government papers display similar dynamics and would hence lead to a similar classification of crisis episodes.
}

\subsection{Stylized Facts}
Our dataset and aggregation procedure is based on \cite{finger2013networkanalysis}, \cite{fricke2015coreperipherystructure}, \cite{fricke2013onassortative} or \cite{raddant2014structurein} who provide excellent overviews of the descriptive features of the data.
We compile in this section the main features of the e-MID data that are also in line with what the literature has found for other unsecured interbank markets.\footnote{This paper concentrates on unsecured money markets as our dataset only comprises uncollateralized loans. Nevertheless, we should emphasize that over the last decade the trend has moved from unsecured to secured money markets. For a more recent discussion on collateralized markets we refer to \cite{coz2024stylizedfacts} who concisely compile new regularities for these markets. In particular, they highlight the role of collateral re-use in the form of persistent ("evergreening") bilateral trading relationships. This is similar to what has been found for unsecured markets but in this case is likely to be the outcome of increased regulatory constraints on short-term liquidity holdings.}

\medskip

The key characteristic of empirical interbank networks at any time aggregation level \citep{finger2013networkanalysis} is their incompleteness as (i) it stands in contrast to earlier theoretical models \citep{allen2000financialcontagion} which show in an explicit microeconomic framework that completely connected networks are less susceptible to contagion in case of shocks to the interbank market and (ii) it provides a justification for the use of graph theory even for unweighted networks.
Whereas daily networks tend to be \textit{sparse} as banks may only contact a subset of their usual trading partners which would grant credit lines to them, aggregation unravels these relationships and makes them measurable.
In case of the e-MID network, densities range on average from approximately 2.5\% for daily networks over 10\% for monthly and 20\% for quarterly networks to 30\% for yearly networks.
The quantitative nature of these values however varies considerably across countries.
For quarterly aggregation which will be our focus from now on, \cite{craig2014interbanktiering} report values between 0.41\% and 0.66\% for the German interbank market and \cite{intveld2014findingthe} calculate approximately 8\% for the Dutch interbank market.\footnote{
Note however that the German interbank data contained information on 2182 banks which is an order of magnitude larger than in the Italian or Dutch datasets.
}

\medskip

Empirical interbank networks exhibit the small-world phenomenon \citep{watts1998collectivedynamics}, i.e.
paths between two banks are short, see e.g. \cite{boss2004networktopology} for Austrian data.
\cite{finger2013networkanalysis} find for the Italian e-MID case that the average length of shortest paths between two banks is 1.912, indicating that the average pair of banks is connected via one intermediary bank.
They show that this result is quantitatively in line both with simple probabilistic models such as the Erd\"os-Renyi random graph with constant linking probability, as well as with elementary generative network models like said small-world model.
Small-world networks are typically characterized by a high tendency to form closed triplets.
In this respect the authors show higher clustering in the undirected version of the empirical network than in such benchmark models.
Once accounting for directionality, they however show that there is less of a tendency to form cyclic triplets and more towards forming transitive triplets than expected under these synthetic models.

\medskip

This heterogeneity also translates into degree distributions. Undirected as well directed degree distributions in interbank networks display extremely positive skewness at any level of aggregation. The level of skewness is generally inconsistent with the family of Poisson distributions that results as the limiting distribution of the constant probability random graph. Earlier literature \citep{boss2004networktopology, demasi2006fitnessmodel} highlighted a possibly scale-free nature of the data by estimating power law coefficients, which was overturned later on by \cite{iori2008anetwork} who found (wide) exponential tails in the data. Indeed, \cite{bech2010thetopology} also find evidence for the negative binominal distribution in the market for US federal funds, a fact that has been confirmed as well for the e-MID by \cite{fricke2015onthe} on the daily level, whereas quarterly data seems to be best characterized by a family of exponential distributions.\footnote{
  \cite{lux2020onthe} shows that the tendency of degree distributions to be falsely attributed to power-laws instead of negative binomial models seems to stretch also to other financial markets like bank-based funding networks of corporate firms.
}
\cite{fricke2015coreperipherystructure} highlight in this regard that in- and out-degree distributions display a remarkably low correlation, given that banks on interbank networks have usually been found to act both as a creditor and debtor over longer aggregation periods.

\medskip

Another related, strikingly non-random, characteristic of interbank networks is disassortative link formation: Banks with fewer trading partners preferably connect to banks with many trading partners, see for daily networks e.g. \cite{iori2008anetwork} in case of the e-MID or \cite{bech2010thetopology} for the US Federal Funds Market. \cite{fricke2013onassortative} show disassortative mixing also for quarterly e-MID networks. They furthermore show that, in order to match this empirical fact, scale-free networks generated by preferential attachment \citep{barabasi1999emergenceof} would need an implausibly small scaling coefficient of the degree distribution, i.e. a sufficiently fat tail, that is at odds with empirically estimated coefficients.

\medskip

The failure of capturing the interbank market with elementary network models, lead \cite{craig2014interbanktiering} to apply the Core-Periphery model \citep{borgatti2000modelsof} from sociology, consisting of a densely connected subgroup and another loosely connected subgroup that however may entertain indirect connections via the former. They find that German quarterly interbank networks in fact are well characterized by a small set of large core banks, intermediating trades among the majority of sparsely connected small, peripheral banks. This structure has quickly become an established stylized fact of interbank markets, see e.g. \cite{fricke2015coreperipherystructure} for Italian networks or \cite{intveld2014findingthe} for Dutch markets.
\cite{ho1985amicro} provide an early theoretical model that rationalizes such a dichotomy by assuming that larger banks have more diversification possibilities which makes them less price sensitive in the interbank (funding) market and, as a consequence, become net borrowers while small banks remain net lenders.
Indeed, the distinction into large core banks being net purchasers and small peripheral banks being net suppliers of funds has been empirically confirmed over a wide variety of marketplaces, see \cite{cocco2009lendingrelationships} for the Portuguese interbank market, \cite{raddant2014structurein} for the Italian market or \cite{allen1989banksize} for the US case.
Nevertheless, as \cite{craig2014interbanktiering} note, the majority of banks is usually active on both sides of the market over longer time spans.

\medskip

\begin{figure}[H]
  \centering
  \includegraphics[scale=0.24, trim = 2cm 0.5cm 2cm 0.5cm, clip]{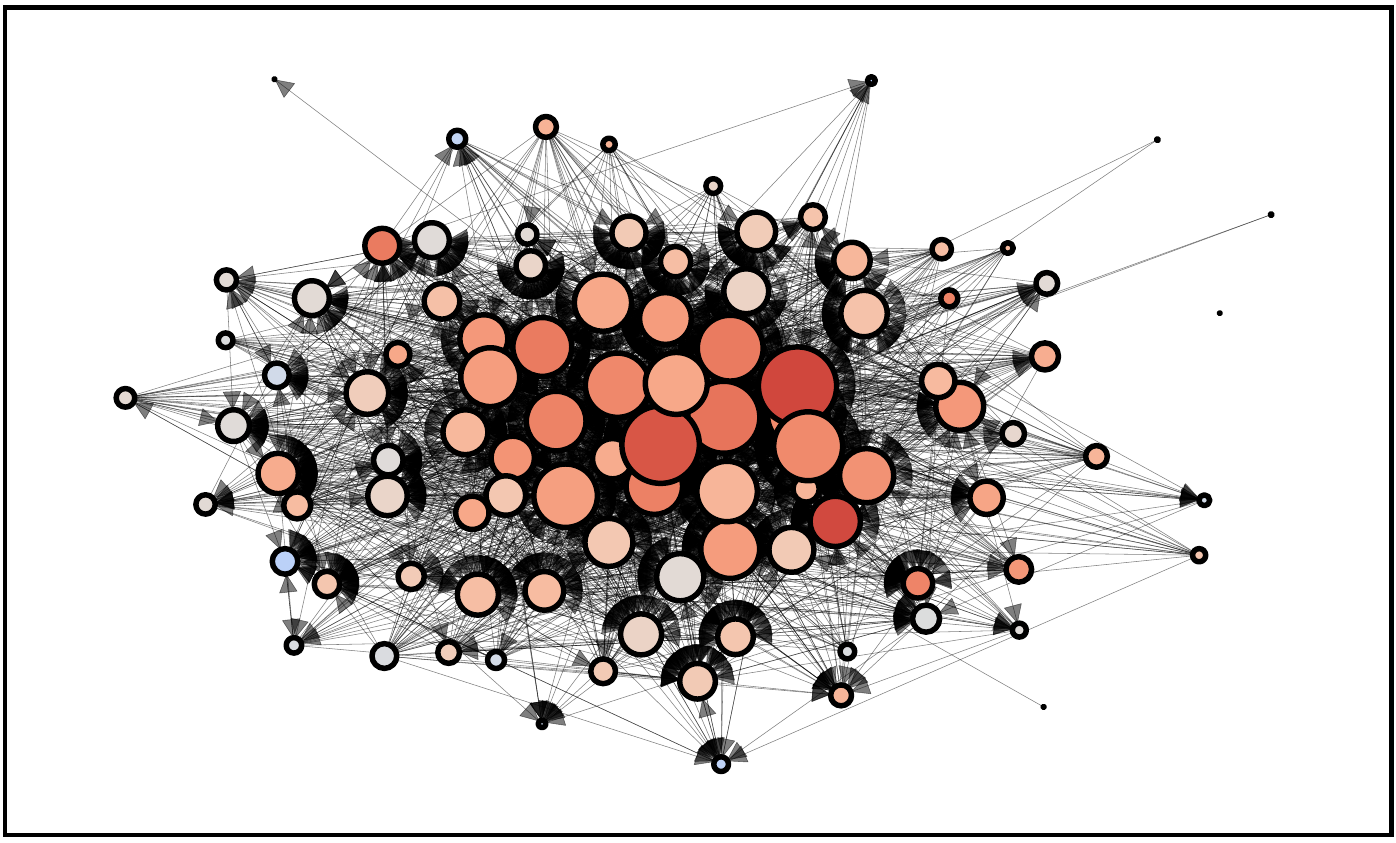}
  \includegraphics[scale=0.24, trim = 2cm 0.5cm 2cm 0.5cm, clip]{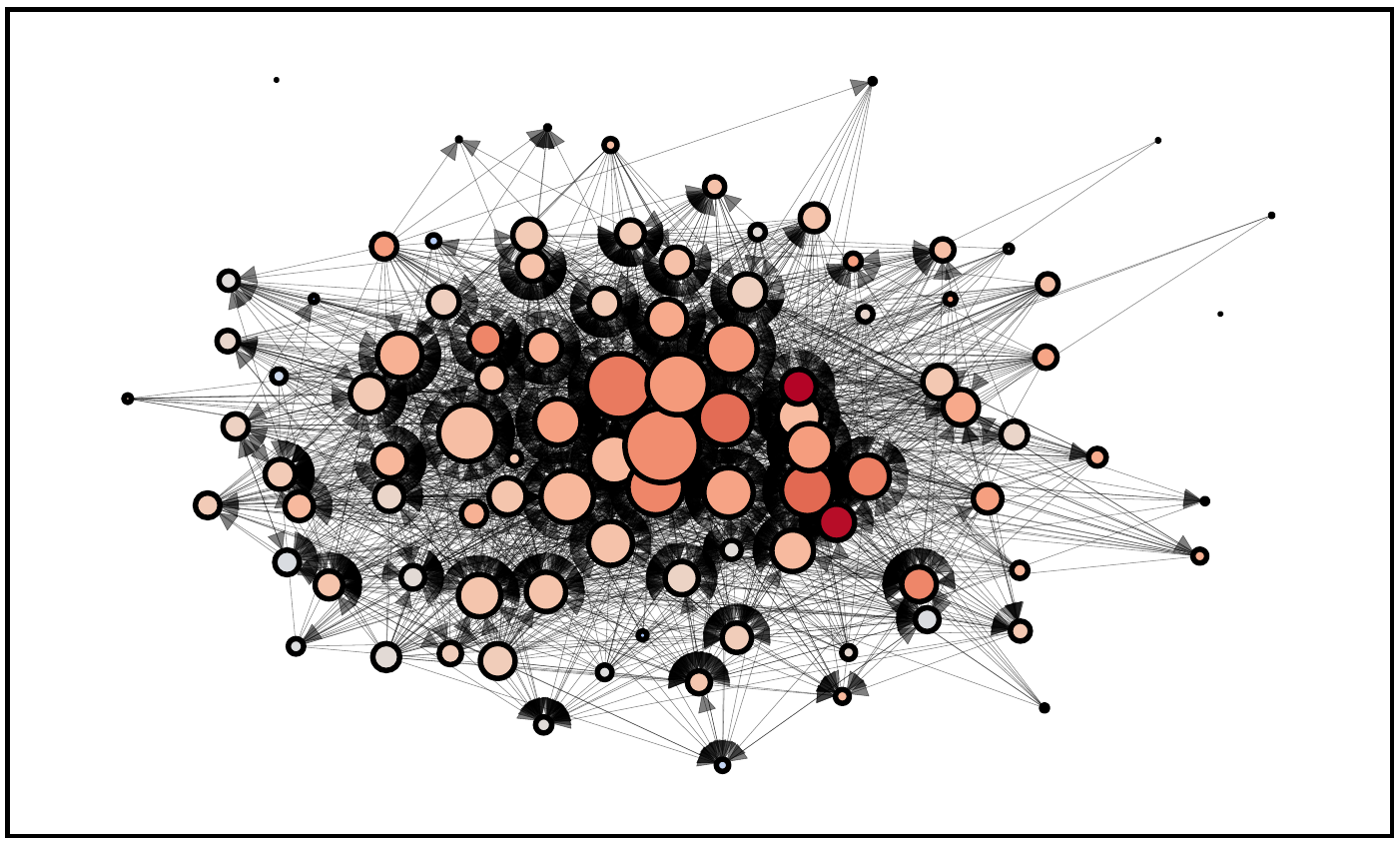}
  \includegraphics[scale=0.24, trim = 2cm 0.5cm 2cm 0.5cm, clip]{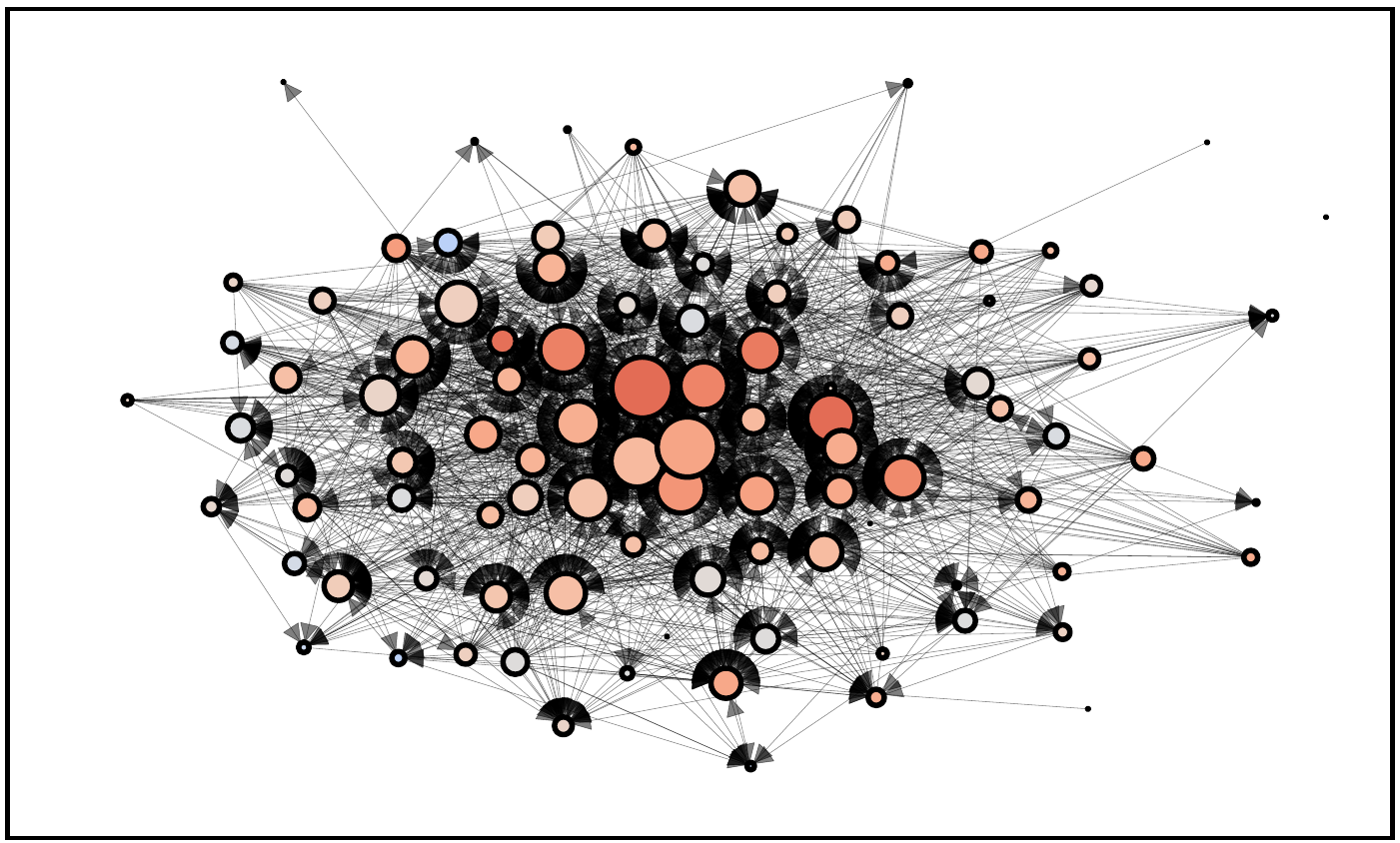}
  \caption{\textit{Interbank Network (2008Q3 - 2008Q4 - 2009Q1).} Node positions are the same across time, node size reflects degree, i.e. total number of trading partners and node color reflects strength, i.e. total volume (blue = low, red = high).
  }
  \label{fig:nets_q3}
\end{figure}

Figure (\ref{fig:nets_q3}) illustrates some of these stylized facts as well as transitional effects due to a large external shock to counterparty risk by the Lehman Brothers default. In particular, it displays the network representation of all Italian e-MID overnight deposit transactions in the third quarter of 2008, at the end of which the actual default materialized, together with the two subsequent quarters.

We immediately recognize the incompleteness of the networks as well as the large heterogeneity of banks. In particular, the core periphery structure stands out. Banks apparently do not have a global constant connection probability, but the larger banks are tightly connected among themselves and possibly intermediating trades from the loosely connected periphery. We however also note that the idealized structure of a completely connected core and an empty intra-periphery network seems also violated as the periphery still entertains a sizeable amount of bilateral transactions, yet most connections seem to be directed at core banks. The general connection patterns however support the idea of heavily skewed degree distributions and disassortative mixing.

\medskip

In the fallout of the Lehman default, we also notice in the figure the direct repercussions on the Italian interbank markets.
The networks become noticeably sparser over the two subsequent quarters.
Banks tend to reduce their connections so as to limit the surge in counterparty risk.
Although the underlying shock to credit risk is external to the Italian market, direct exposure to US money market products and possible devaluations or defaults as well as indirect exposure via intra-Italian credit lines to banks that might be themselves invested in the US market give rise to possible credit rationing in the absence of known risk and pricing measures.
This reduction seems to happen especially for core banks, whose degrees decrease to such an extent that the distinction between core and periphery is hardly visible anymore in the first quarter of 2009.

\medskip

We know from the literature that core banks tend to be mostly net borrowers.
The dry-up of market liquidity should therefore make them particularly susceptible to funding problems.
In the fourth quarter of 2008, we see this fact reflected by two core banks whose trading volume has increased tremendously, i.e.,
whose color has turned to be the most red-intensive of all the nodes in the figure, while at the same time their number of trading partners has noticeably decreased, i.e., their node size has decreased.
In the subsequent quarter both banks are not active anymore in the overnight segment of the e-MID market.

\medskip

Unfortunately our dataset does not allow us to discriminate between different reasons for exiting the market.
While it is possible that leaving banks eventually defaulted, it is equally possible that they refrained from the transparent nature of the e-MID platform in order not to signal enduring liquidity problems and instead moved towards more intransparent contractual settings like pure OTC markets.\footnote{
  The December report on financial stability of the \cite{bancaditalia2010financialstability} states in fact that the share of e-MID to OTC trading in the short-term segment ($\leq 2$ days) over the crisis periods has halved from roughly 60\% to 30\% while the share of central counterparty trading on the collateralized MTS market has risen from 40\% to 70\% (and up to 90\%), indicating the need for anonymity by liquidity-scarce banks.
}
\cite{angelini2011theinterbank} however find in a richer, confidential dataset on the e-MID that in times of crisis the average borrower still exhibits similar values of default risk characteristics (i.e. ratings or profitability measures) compared to normal times, casting doubt on theories of adverse selection alone as main driver in this process.
On the other hand they also find that, as the crisis was progressing, some banks with elevated risk ratings reduced their operations on the e-MID platform while also displaying a significant unwillingness to publish large credit exposure values in their public financial reporting books at the end of reporting periods.

\medskip

Distressed banks may also have turned to the classical lender of last resort, i.e. the European Central Bank and its discount window.
Liquidity from these standing facilities however is based on collateralized loans.
While eligible collateral may face large haircuts, liquidity-short banks may also want to stick to uncollateralized loans from the interbank market to limit their financing costs.
In the following sections we take an agnostic view on the nature of entry and exit dynamics and focus on the state and transitions of the network characteristics.
We will in particular focus on the role of reciprocity in the market and how it interacts with the aforementioned stylized facts.

\newpage
\section{Empirical Results}
We recalled in the previous section the stylized facts of interbank markets with a particular focus on the literature on the Italian interbank market.
What stood out was the large heterogeneity across bank nodes and the resulting inability of the Erd\"os-Renyi model, in which each bank has the same probability to connect to any other bank (so that the number of links in the network is preserved on average), to generate most of these stylized facts.
While the generative mechanisms of the different scale-free models have been shown to explain skewed degree distributions, interbank networks do not seem to be scale-free and so these models in their present form are likewise unable to account for these elementary features.
Based on these observations, we conclude that network models of the interbank markets likely need to be heterogeneous if they want to account for these fundamental characteristics.

\medskip

In the tradition of the Erd\"os-Renyi random graphs, we have recalled and developed in the theoretical sections of this paper randomized graph ensembles with a closed-form solution that are heterogeneous on the node-level.
All of these models, \textit{DECM}, \textit{RWCM} and \textit{RECM}, have in common that they preserve the strength distribution in the ensemble average while our two preferred models, \textit{DECM} and \textit{RECM}, also preserve the degree distribution which appears to be a main driver of the stylized facts of interbank markets according to the literature.
Two of the three models, \textit{RWCM} and \textit{RECM}, are also able to preserve the distribution of weighted reciprocities across nodes, while only our newly developed model, \textit{RECM}, is also able to simultaneously match the unweighted reciprocity distribution.
Our main comparison will thus consist of contrasting the two models, \textit{DECM} and \textit{RECM}, to distill out the effect of (un+)weighted reciprocity as that is the only difference between the models with the latter model being a strict superset of the former in terms of constraints we imposed on the probability distribution of interbank networks.

\subsection{0-Paths: Total Nodes}
\begin{figure}[H]
  \centering
  \includegraphics[scale=0.4]{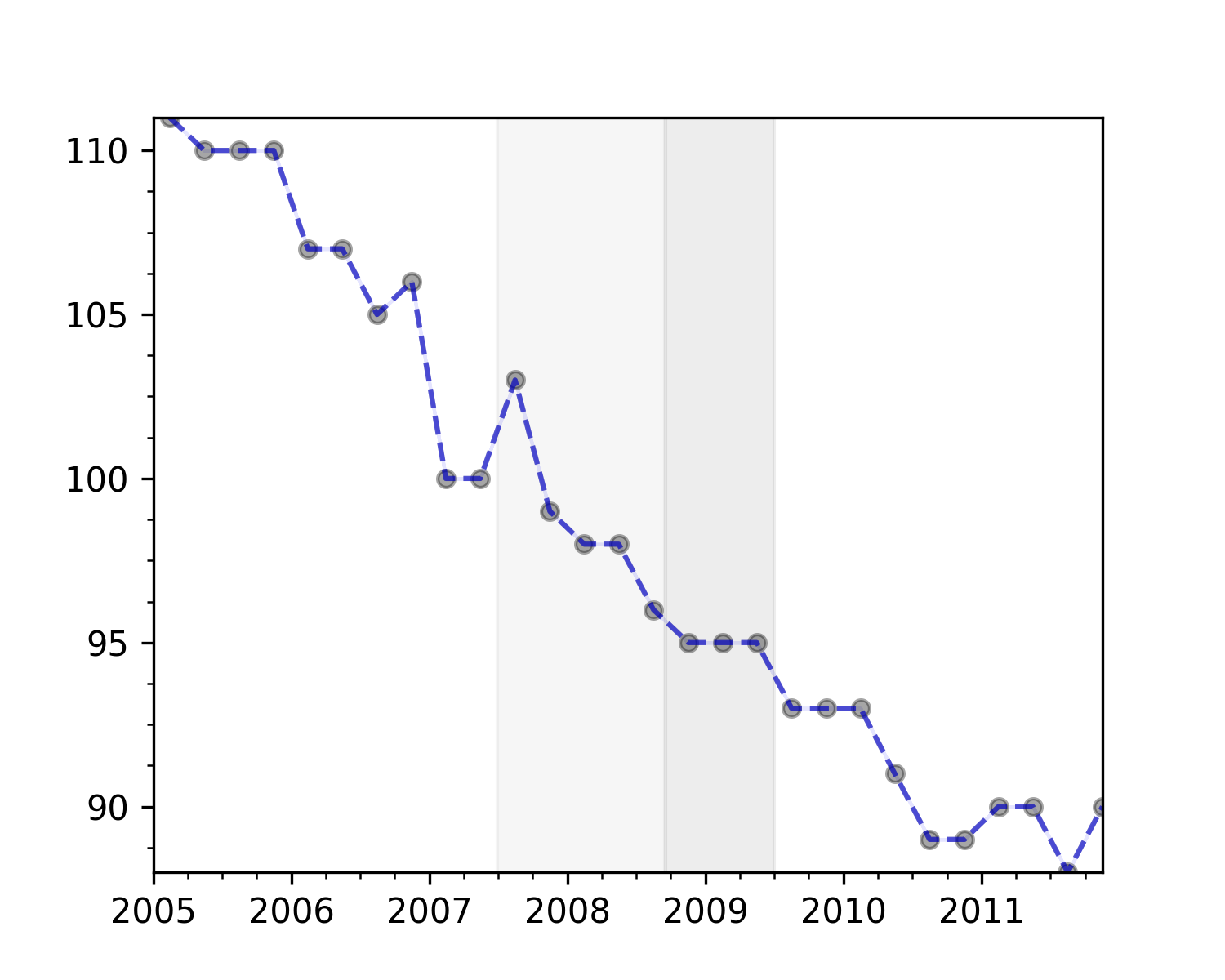}
  \includegraphics[scale=0.4]{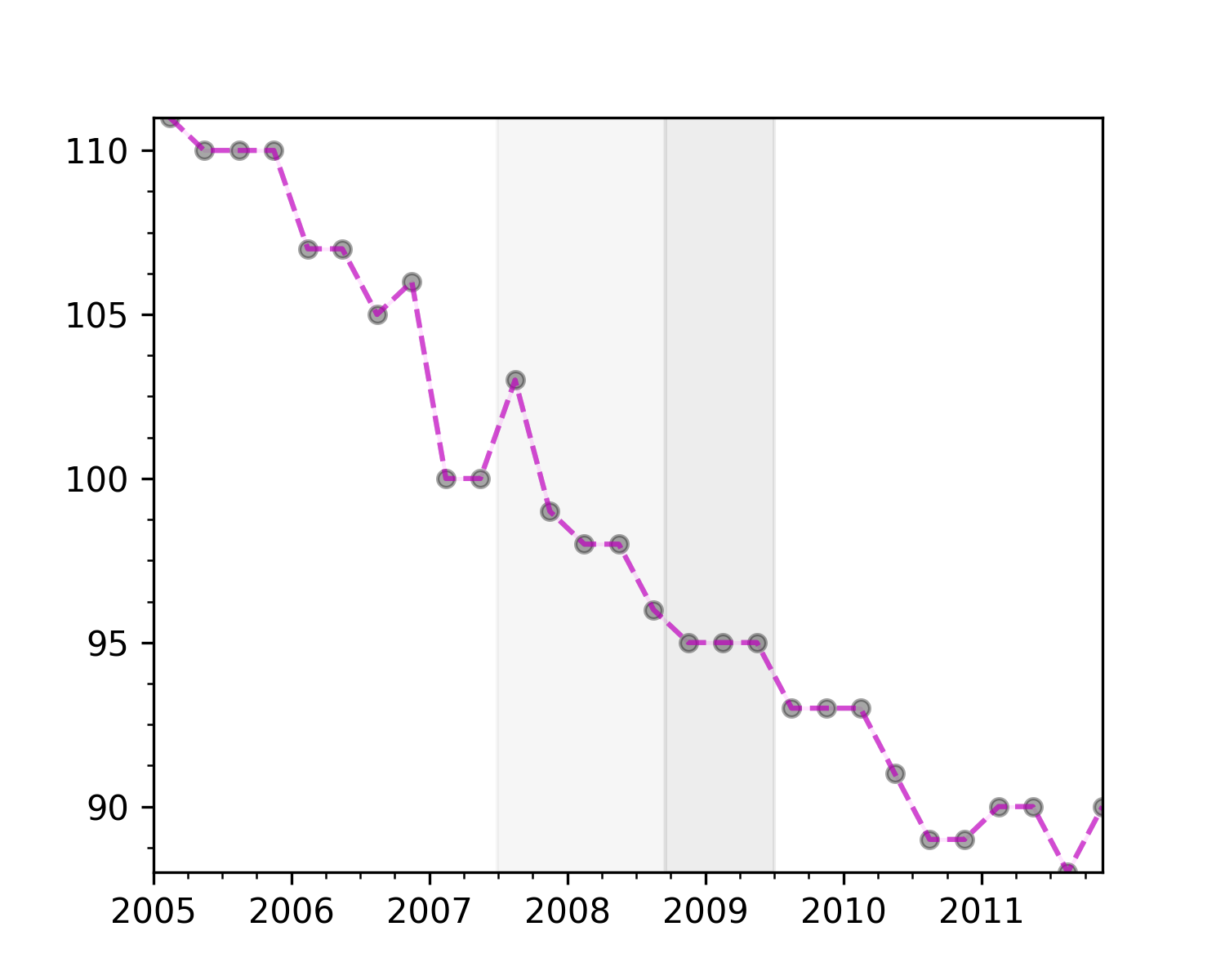}
  \includegraphics[scale=0.4]{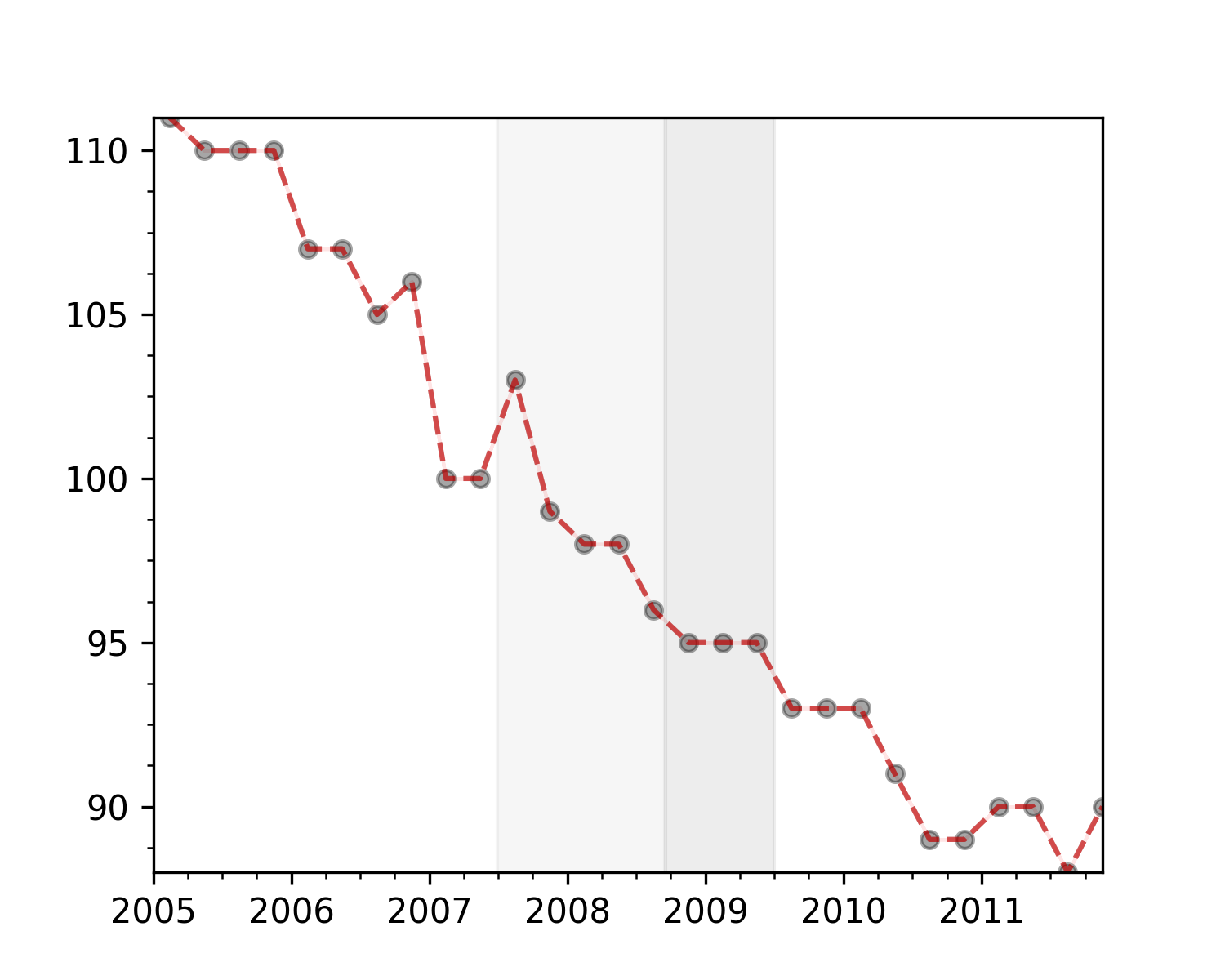}
  \caption{\textit{0-Paths.} Total Nodes. Empirical time series (black dots) are shown alongside ensemble means and 95\% intervals of DECM (blue) that preserves only un+weighted degrees, RWCM (pink) that preserves only weighted reciprocity, and RECM (red) that preserves un+weighted reciprocity on the node-level.
  }
  \label{fig:total_nodes}
\end{figure}

As we have seen in the previous network illustrations the number of banks is non-constant. While there exist individual banks entering the market, the global dynamics of active banks in Figure (\ref{fig:total_nodes}) display a strong downward trend.

Starting with 111 banks in the first quarter of 2005 and ending with 90 banks in the fourth quarter of 2011, the decline in the number of banks on the overnight segment of the e-MID platform is almost monotonic.
In particular the crisis episodes display almost the same rate of decline as pre- and post-crisis periods.
For the sake of comparability with later graphics we show the three different probabilistic ensembles in separate panels.
As each of the models ---  degree- and strength-preserving \textit{DECM} in the left panel, weighted reciprocity-preserving \textit{RWCM} in the middle panel, and un+weighted reciprocity-preserving \textit{RECM} in the right panel --- takes as given the number of nodes $n(t)$ at any given quarter $t$ and distributes probability mass over any conceivable non-negative integer-weighted $n$-node network without self-loops according to the imposed empirical constraints, the ensemble is effectively a point mass in $n$, as it preserves the number of nodes by construction exactly.\footnote{
  As the number of parameters scales linearly in $n$ for all models, albeit with a different scaling factor, we had to estimate, e.g. in 2007-Q1 for $n=100$, quarterly models with $400$ parameters (\textit{DECM}), with $300$ parameters (\textit{RWCM}), and with $600$ parameters (\textit{RECM}).
  All estimations have been implemented in the C programming language using automatic differentiation with the help of Fortran solvers by \cite{hsl2002acollection} using a supremum norm with accuracy $10e^{-4}$ for each constraint.
}

\medskip

Most network statistics will directly or indirectly depend on the number of nodes in the network.
We will thus need to be careful in interpreting non-normalized network statistics in the upcoming sections.
Our focus however lies on investigating network effects that are the result of structural, non-random processes.
Even if our purely statistical approach does not allow us to investigate the theoretical mechanisms of such a process, we aim at providing future research with a categorization of which network effects are likely to be a mere byproduct of elementary features and which features necessitate separate explanatory mechanisms.
For our present dataset this implies that we will investigate effect \textit{levels} in order to assess economic significance, but generally concentrate on \textit{deviations} from the different probabilistic models to assess statistical significance.
Since we re-estimate the models each quarter we thus control exactly for the declining trend in bank activity.

\subsection{1-Paths: Trading Links \& Volume}

\begin{figure}[H]
  \centering
  \includegraphics[scale=0.4]{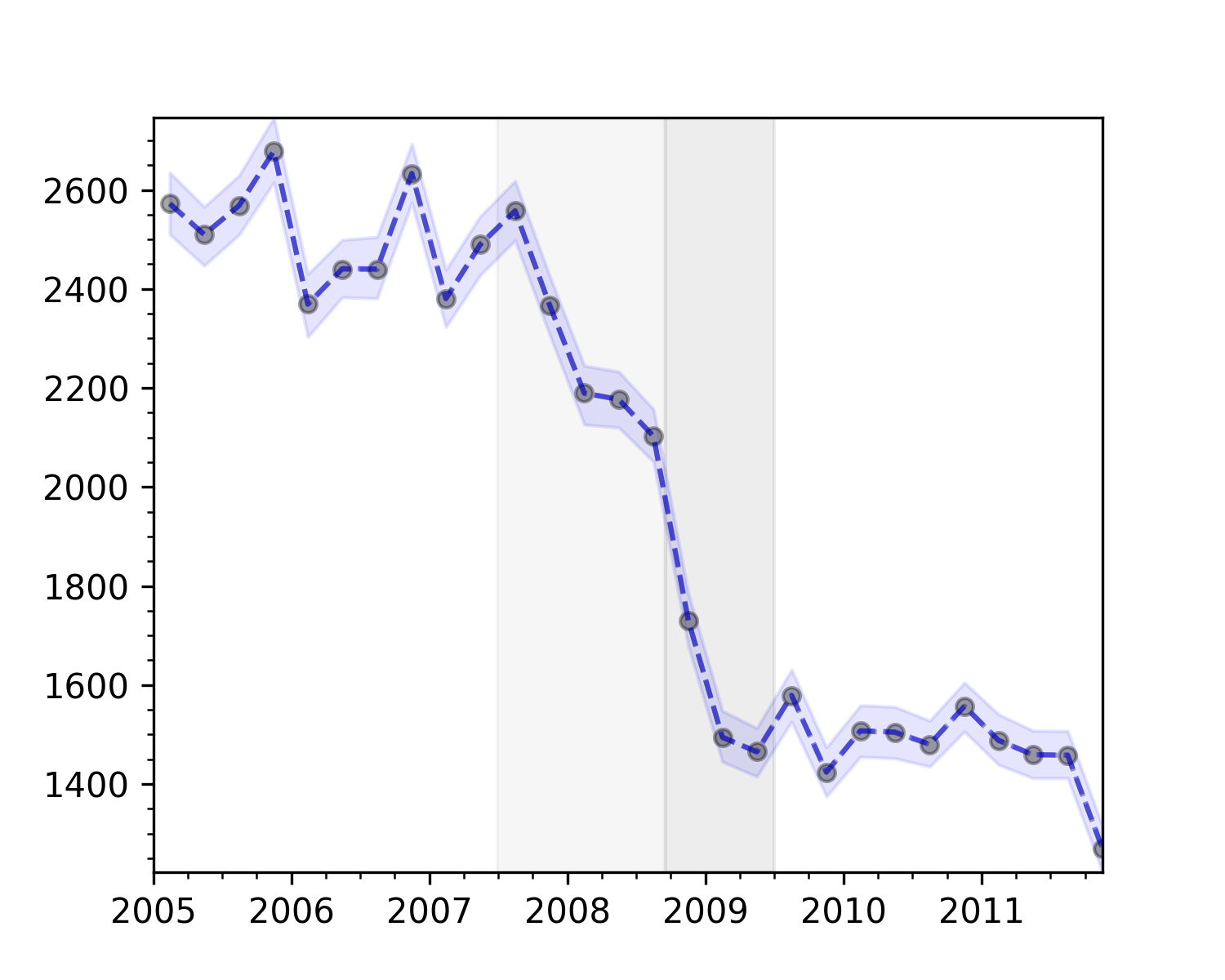}
  \includegraphics[scale=0.4]{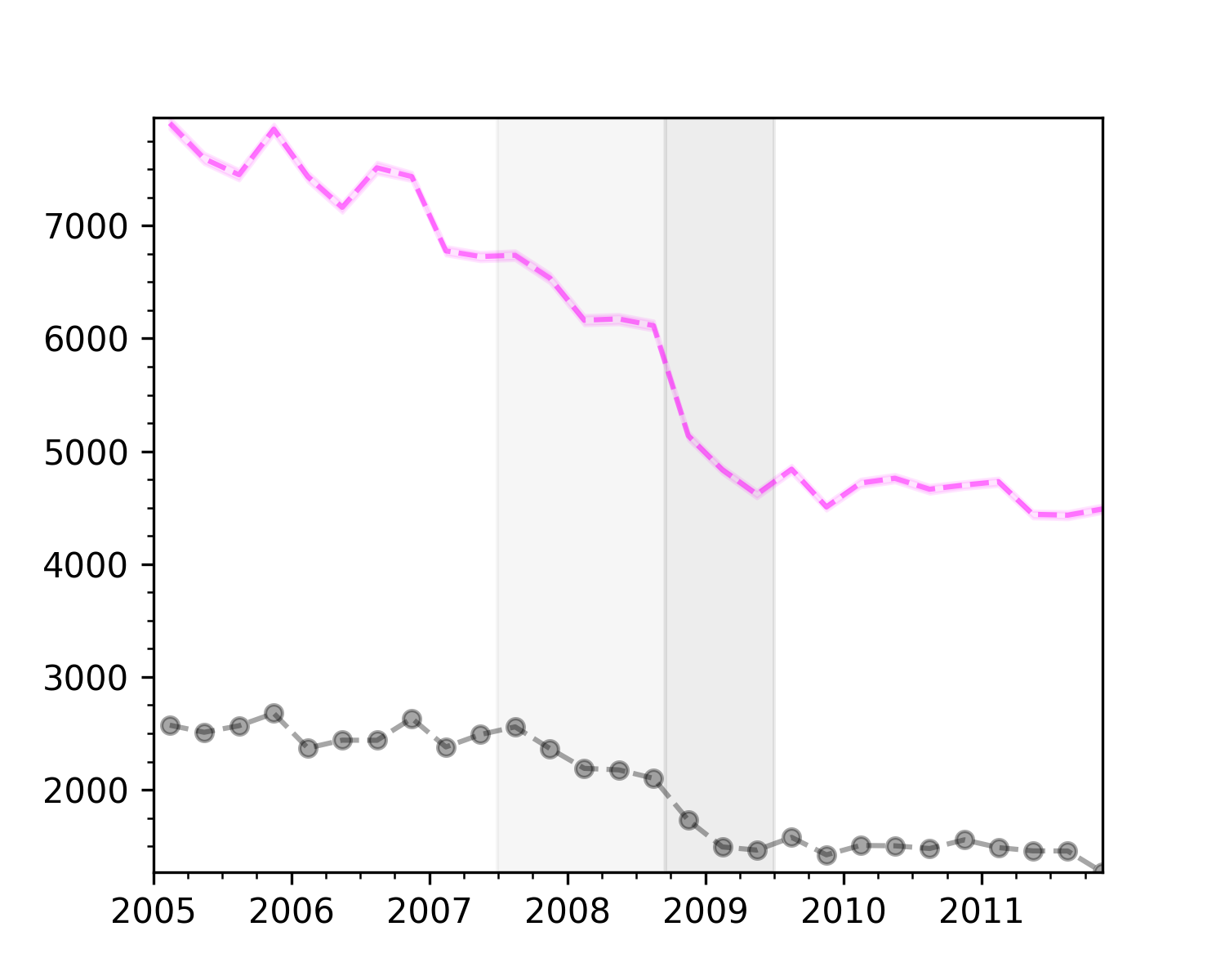}
  \includegraphics[scale=0.4]{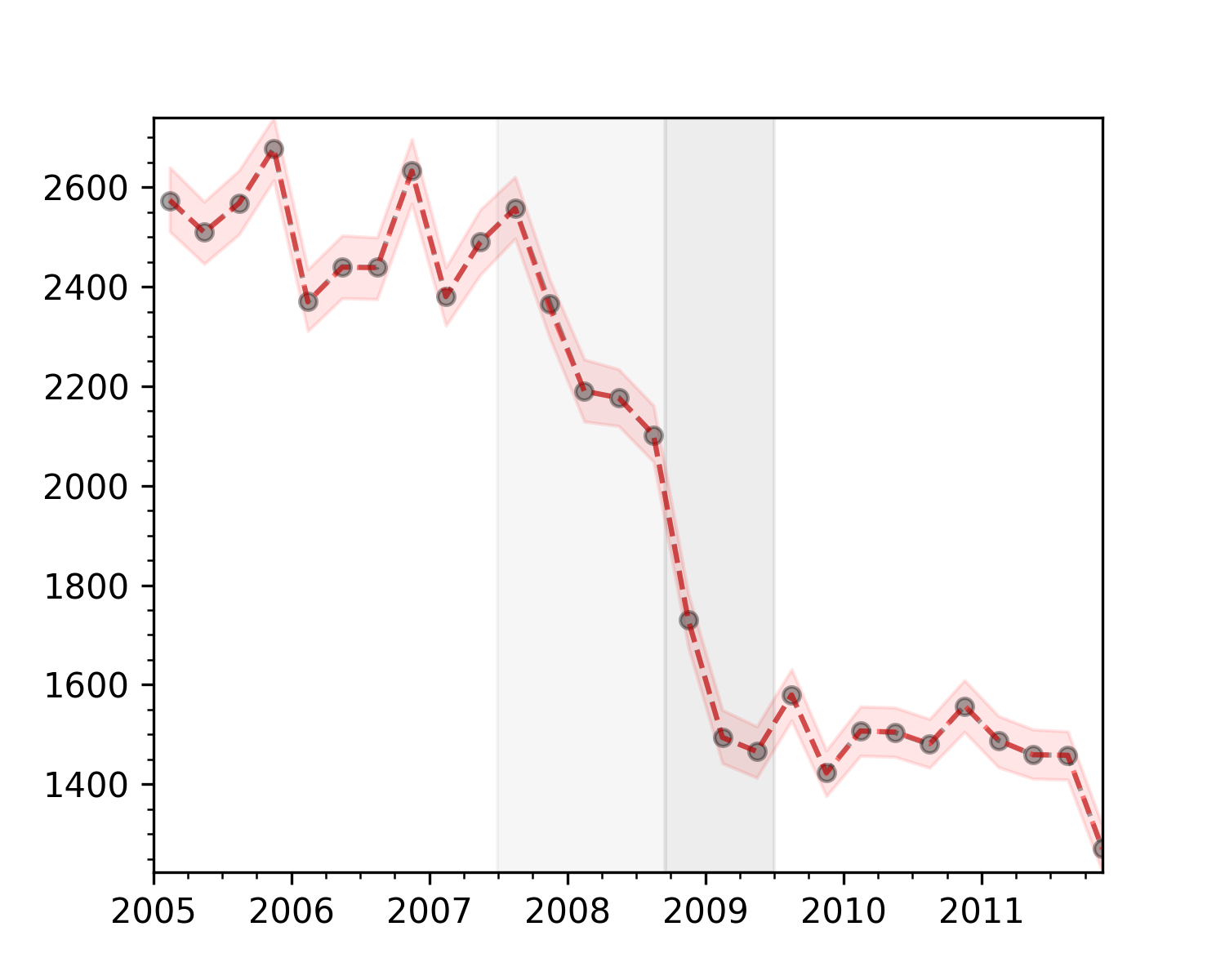}
  \\
  \includegraphics[scale=0.4]{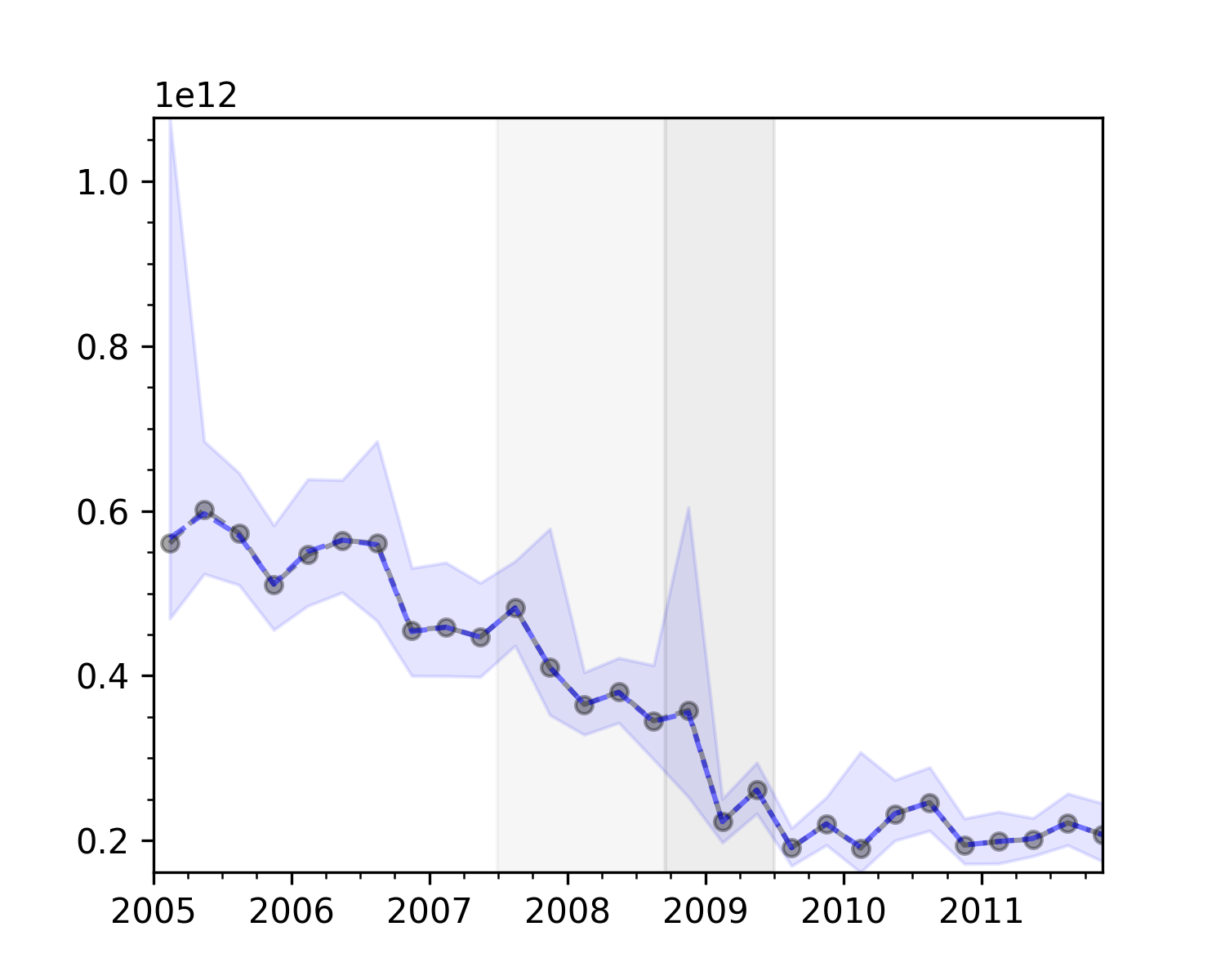}
  \includegraphics[scale=0.4]{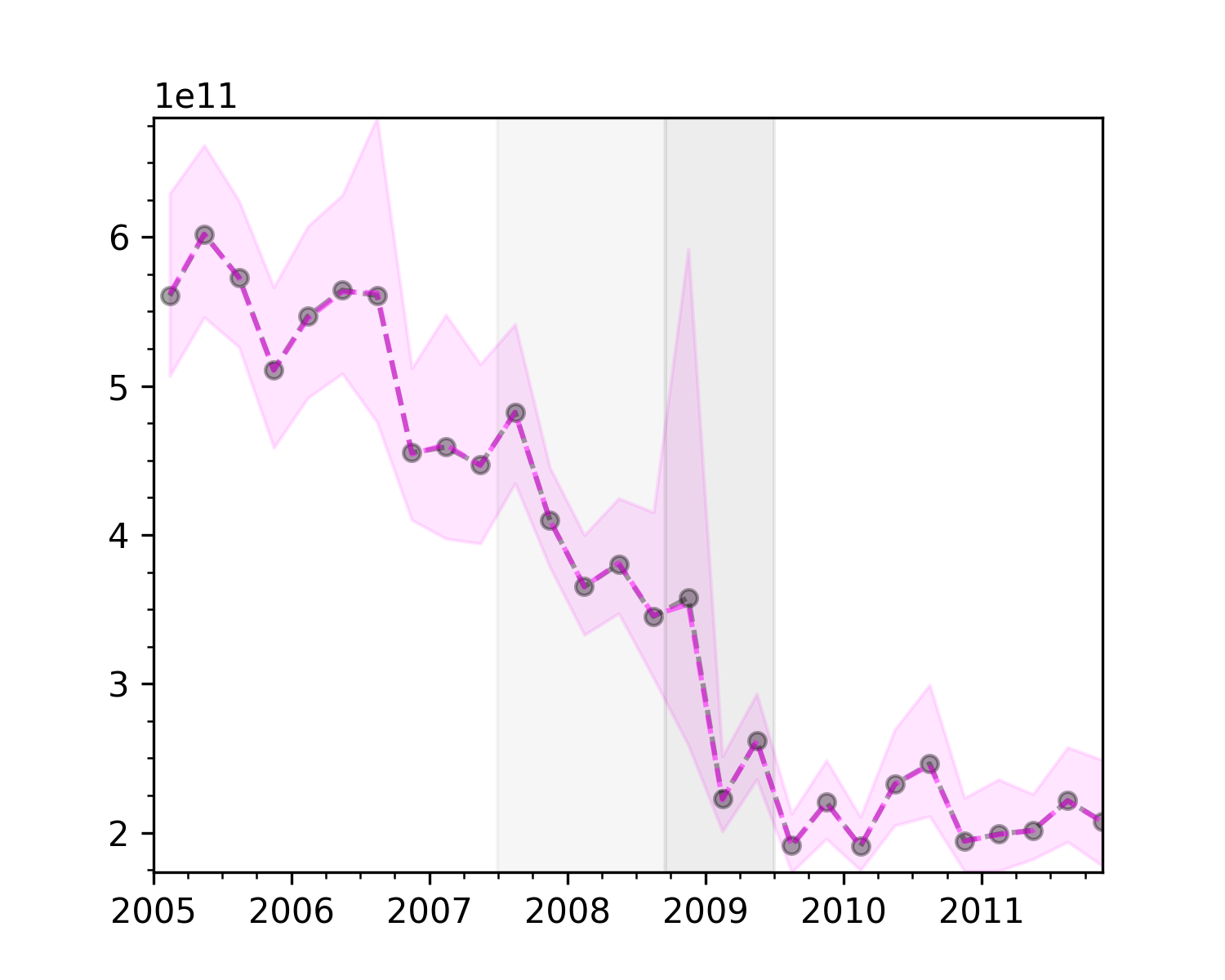}
  \includegraphics[scale=0.4]{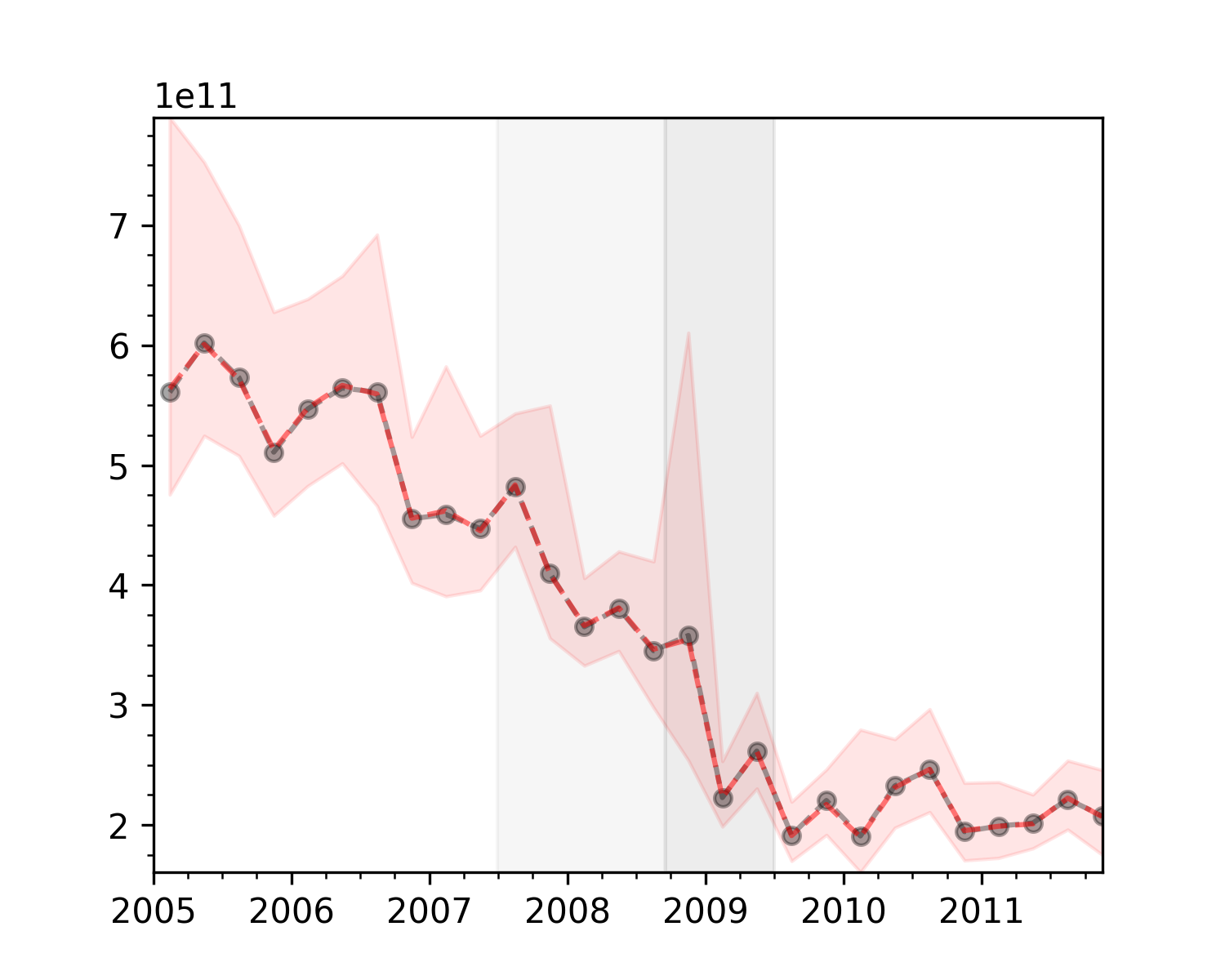}
  \caption{\textit{1-Paths.} Upper panel: Total Links. Lower panel: Total Volume. Empirical time series (black dots) are shown alongside ensemble means and 95\% intervals of DECM (blue) that preserves only un+weighted degrees, RWCM (pink) that preserves only weighted reciprocity, and RECM (red) that preserves un+weighted reciprocity on the node-level.
  }
  \label{fig:ts_total_links_and_volume}
\end{figure}

While the number of nodes is the same in each network configuration of a given quarter, any edge-based statistic follows a non-degenerate distribution.
Figure (\ref{fig:ts_total_links_and_volume}) displays global elementary statistics that involve a path of length one, i.e. a directed edge, between two banks.
The top panel takes an unweighted perspective and depicts the total number of links $(d)$ while the bottom panel takes a weighted perspective, showing the total amount of trading volume $(s)$ in the network. The left panel refers again to the \textit{DECM}, the middle panel to the \textit{RWCM} and the right panel to the \textit{RECM} ensemble, displaying the ensemble averages as well as 2.5\% and 97.5\% percentiles of the respective distributions.

\medskip

Unlike the constant decline in the number of active banks, the number of trading links among these banks remains relatively constant in periods prior to the crisis ($\mu=2508,\,\sigma=99$) and afterwards ($\mu=1473,\,\sigma=80$).\footnote{
    The symbols $\mu$ and $\sigma$ refer to the empirical time averages and standard deviations of the aggregate variable across the respective subperiods of the sample period based on the crisis definition from the previous sections.
}
This implies that banks had a tendency to increase connectivity in those phases.
The crisis period however marked a significant drop in links (41\%) possibly due to elevated counterparty risk and thus greatly surpassed the decline in the number of active banks.
Although the decline in the number of links already started in the first phase of the crisis, the main decline in trading links seems to materialize in the aftermath of the Lehman default with a cumulative decline of 18\% in the following quarter and 29\% in the subsequent quarter.
A look at the network density therefore paints a similar picture around that date. \cite{fricke2015coreperipherystructure} in fact find a formal structural breakpoint using CUSUM and Chow Tests for the density variable.

Turning to trading volume, we see in the bottom panel of Figure (\ref{fig:ts_total_links_and_volume}) that global volume in the network deteriorates over time, but seems to stabilize when the crisis begins to fade out.
Prior to the third quarter of 2007 trading volume amounts on average to $\mu=\text{€}528 \text{bn.}$ ($\sigma=53\text{bn.}$) and decreases by 64\% until the third quarter of 2009.
Interestingly the drop in trading volume due to the default of Lehman Brothers is less pronounced than the drop in trading links.
Not only is the volume effect smaller, it also happens later, with a lag of one quarter after the drop in links.
While trading connections were loosened immediately after the end of the third quarter 2008, trading volume even slightly increased.
Creditors apparently withdrew many credit lines, while serving those debtors whose credit lines were maintained with more liquidity than ever before.
The loss of trading volume (-35\%) in 2009-Q1 relative to 2008-Q3 however reaches similar values like the drop in trading links.
Normalizing total trading volume by trading links yields a similar picture:
The average trading volume per link amounts to $\text{€} 211 \text{M}$ ($\sigma=23\text{M}$) in a pre-crisis quarter, and $\text{€} 144 \text{M}$ ($\sigma=16\text{M}$) in a post-crisis quarter.
This however is not a reflection of a crisis-induced shock to credit risk but follows a more general downward trend in pairwise credit lines within and around the periods of crisis.

\medskip

As we noted in the model section, we preserve trading links and/or volume also for each individual node.
Linearly aggregated global network statistics like total volume in Figure (\ref{fig:ts_total_links_and_volume}) are therefore implicitly preserved as well.
While this holds true for trading volume in all three models, the \textit{RWCM} is unable to match the degree distribution in any quarter and thus also fails to match the total number of trading relationships in the figure.
Although this model is constructed with the empirical distribution of weighted degrees (i.e. strengths) and even weighted reciprocity, this information is not enough to generate a realistic degree distribution.
It imposes too much probability mass on non-existing credit lines and thus generates networks that are far too dense compared to realistic interbank markets.
On the theoretical side this also confirms the \textit{irreducibility conjecture} of \cite{mastrandrea2014enhancedreconstruction}, pointing to the necessity of topological information when modeling incomplete networks. The other two models take these statistics either directly (DECM) or indirectly (RECM) as constraint variables and thus are trivially able to match those in the respective canonical ensembles.

\subsection{2-Paths: Reciprocal Trading Relationships}
\begin{figure}[H]
  \centering
  \includegraphics[scale=0.4, trim = 0cm 0cm 0cm 0cm, clip]{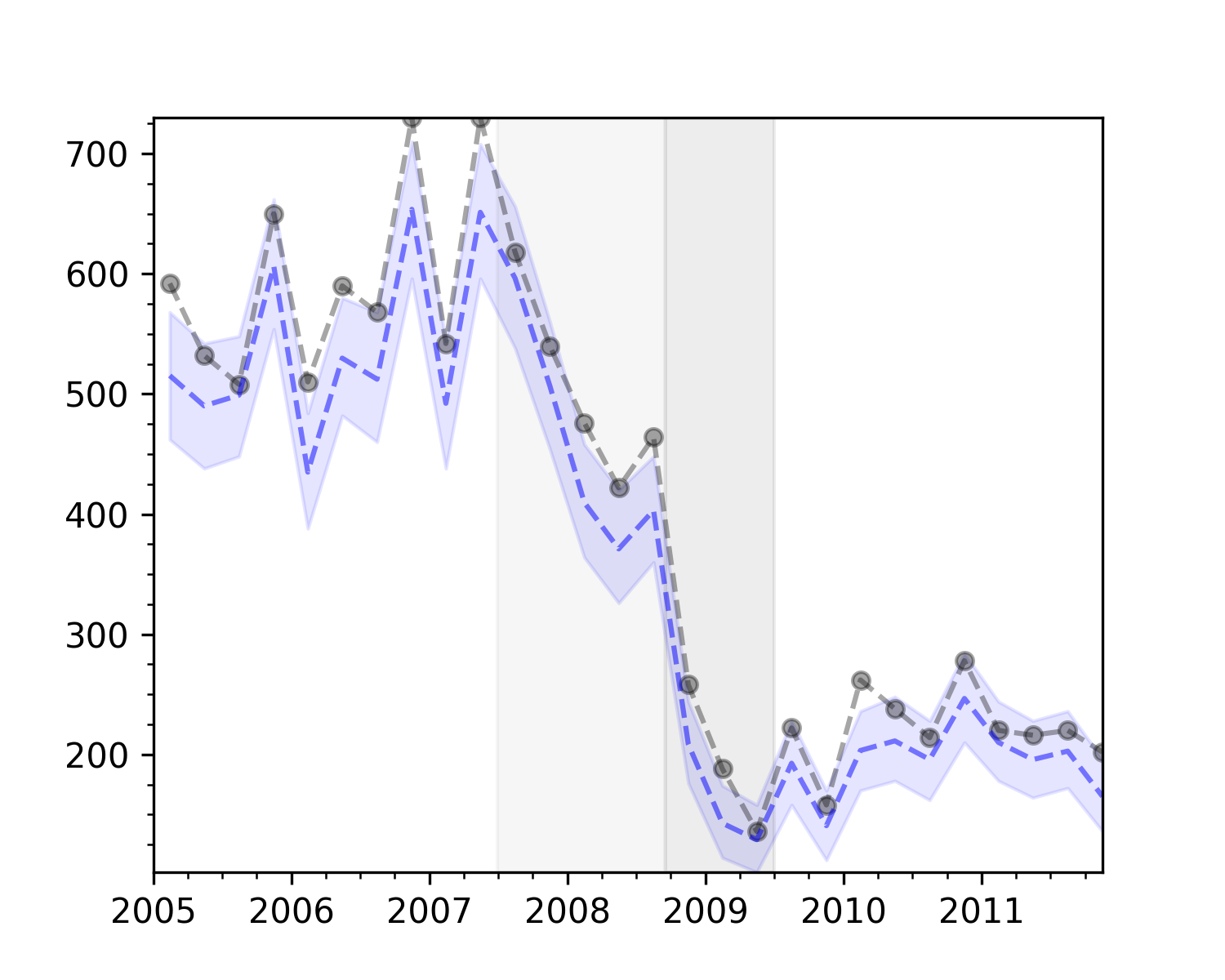}
  \includegraphics[scale=0.4, trim = 0cm 0cm 0cm 0cm, clip]{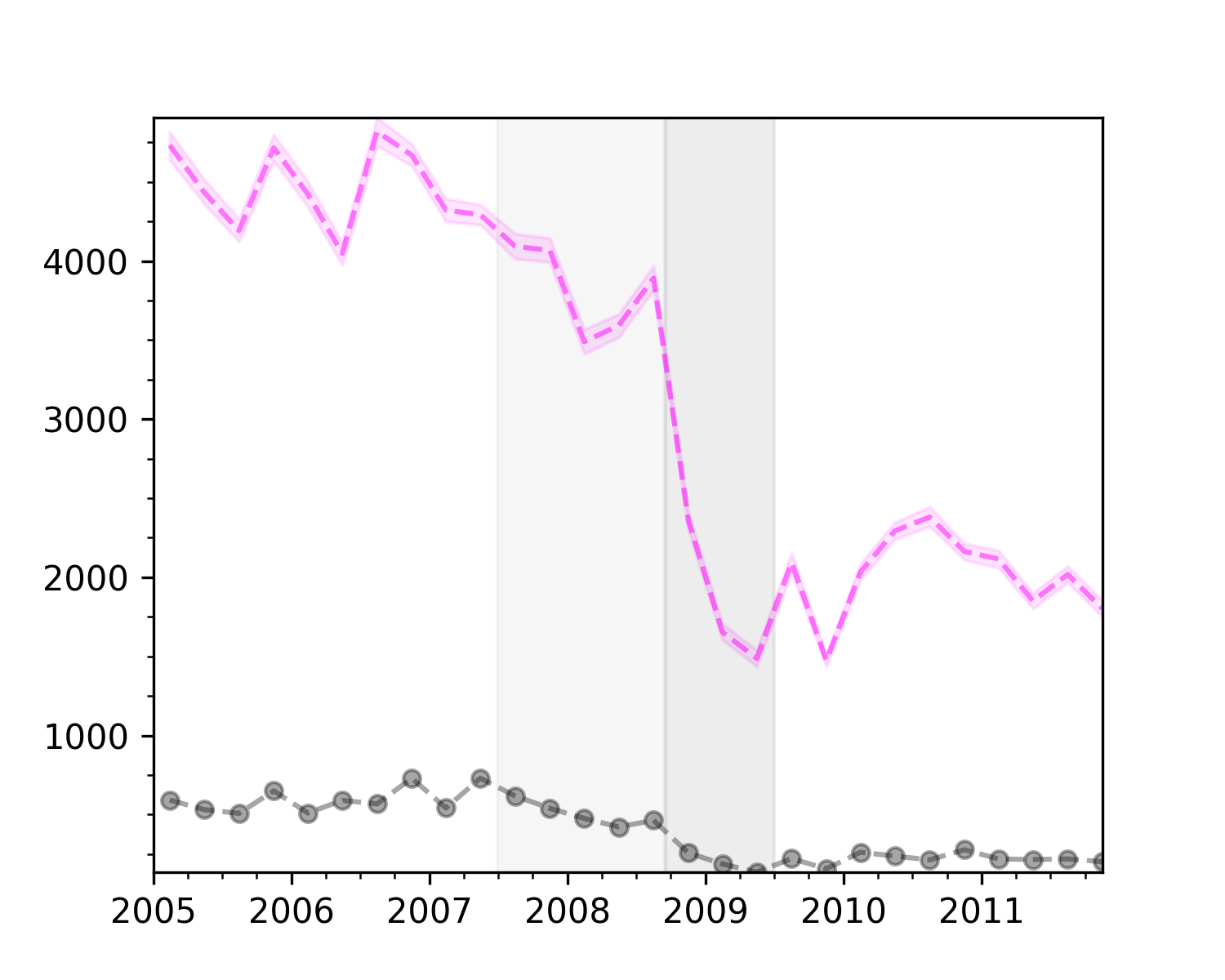}
  \includegraphics[scale=0.4, trim = 0cm 0cm 0cm 0cm, clip]{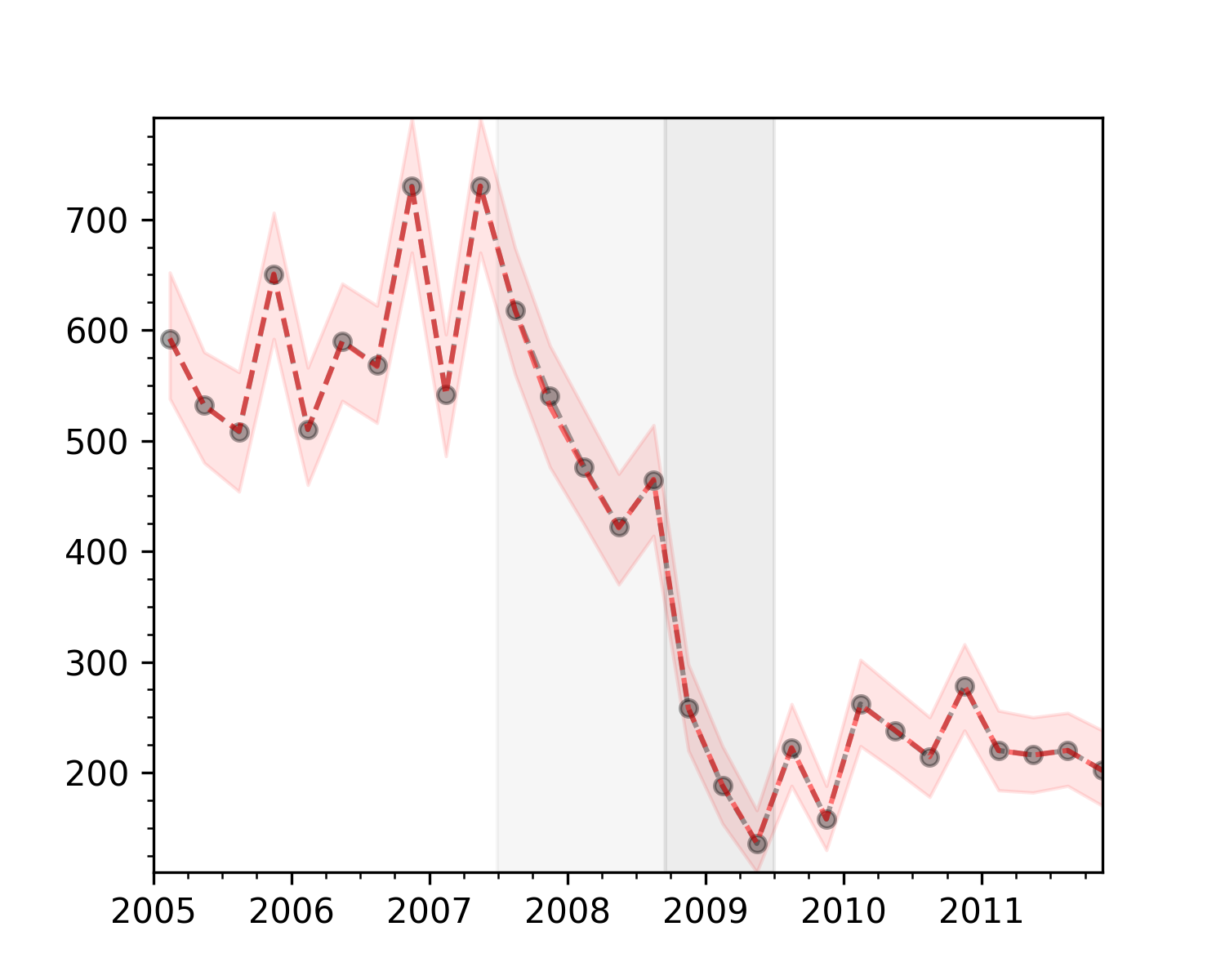}
  \\
  \includegraphics[scale=0.4, trim = 0cm 0cm 0cm 0cm, clip]{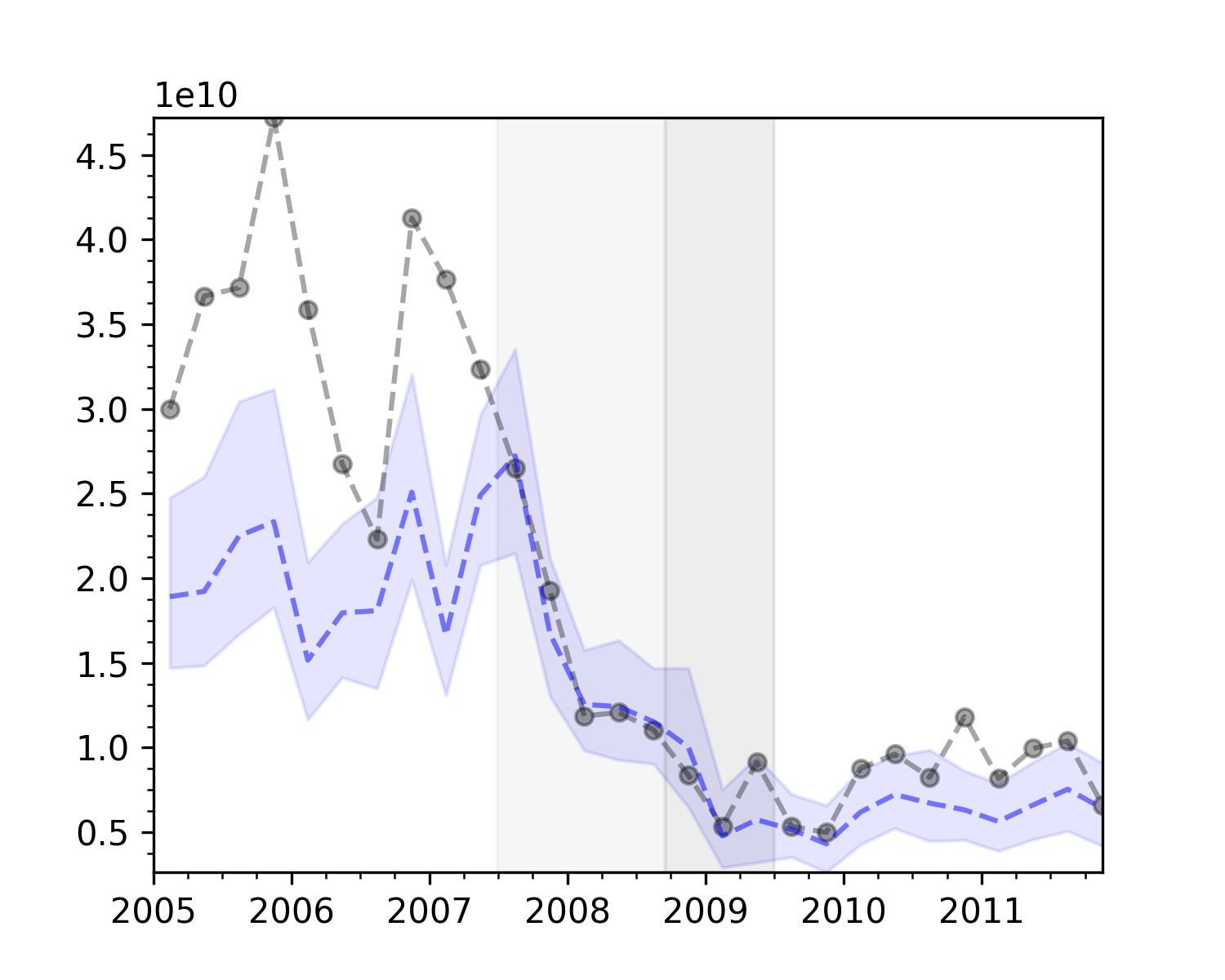}
  \includegraphics[scale=0.4, trim = 0cm 0cm 0cm 0cm, clip]{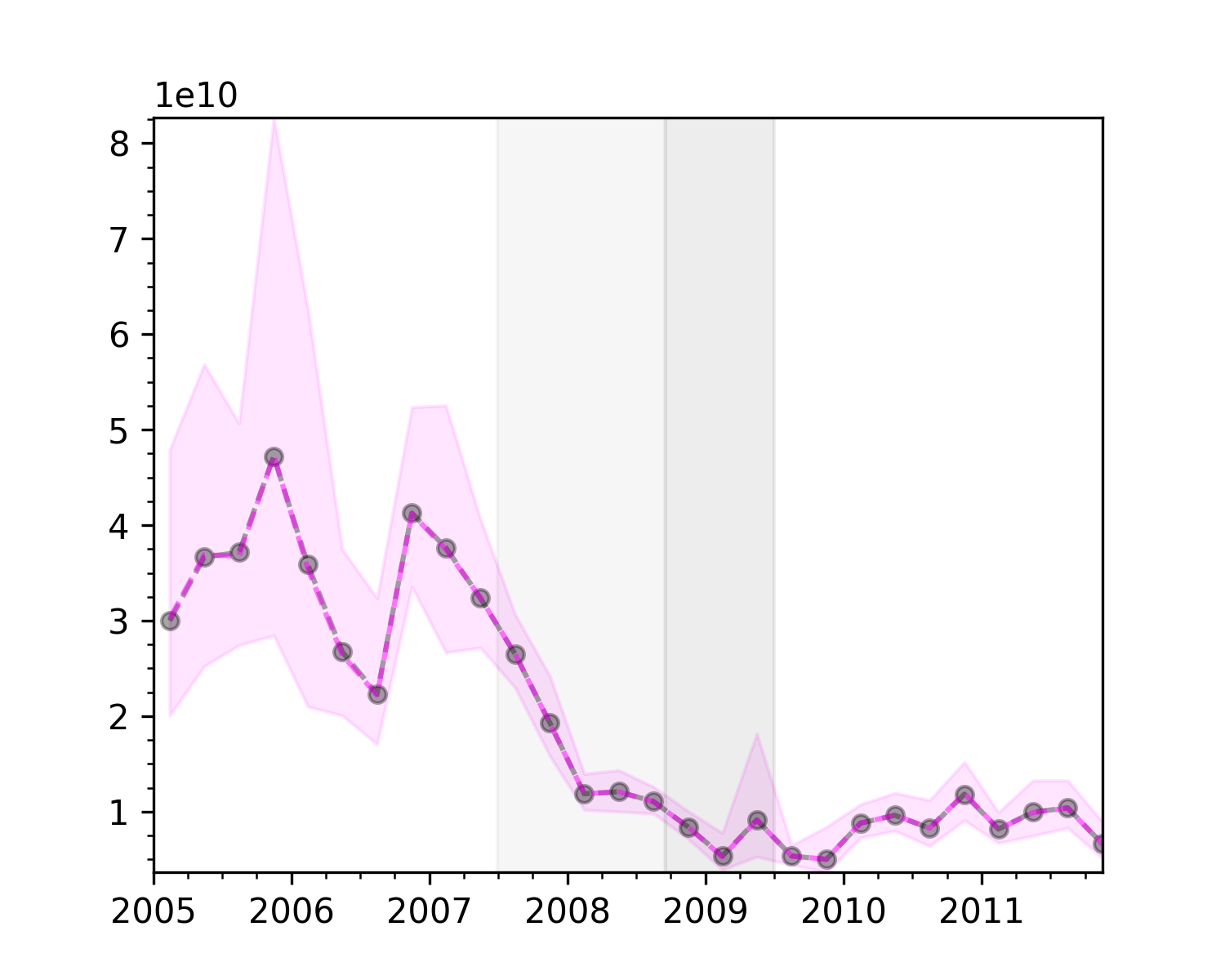}
  \includegraphics[scale=0.4, trim = 0cm 0cm 0cm 0cm, clip]{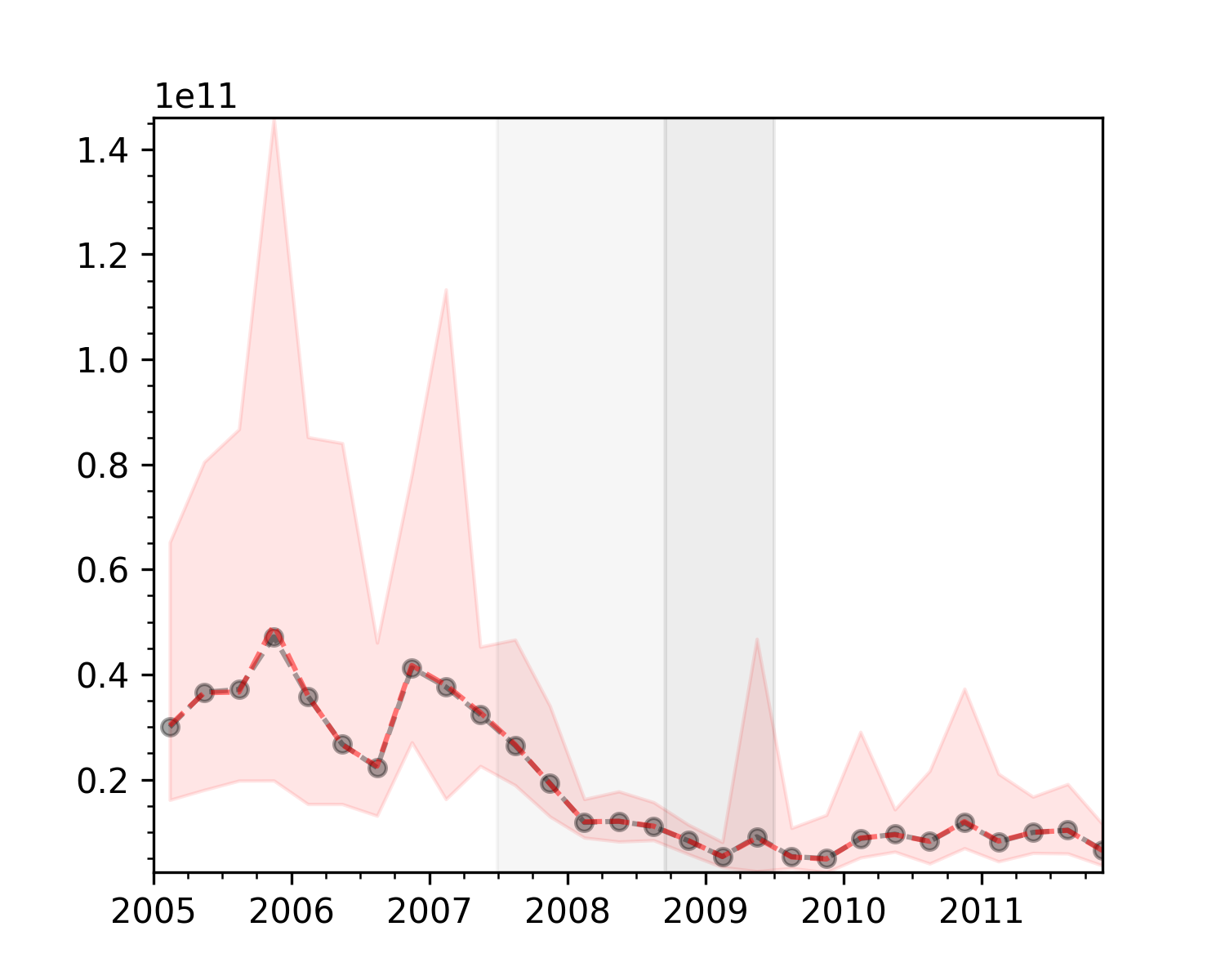}
  \caption{\textit{2-Paths.} Upper panel: Unweighted Reciprocity. Lower panel: Weighted Reciprocity. Empirical time series (black dots) are shown alongside ensemble means and 95\% intervals of DECM (blue) that preserves only un+weighted degrees, RWCM (pink) that preserves only weighted reciprocity, and RECM (red) that preserves un+weighted reciprocity on the node-level.
  }
  \label{fig:ts_unweighted_and_weighted_reciprocity}
\end{figure}

Moving from paths of length one to (closed) paths of length two, we can observe some structural similarities in unweighted and weighted aggregate statistics, cf. Figure (\ref{fig:ts_unweighted_and_weighted_reciprocity}).
Both series, total reciprocal trading links $(d^\leftrightarrow)$ as well total reciprocated trading volume $(s^\leftrightarrow)$, experience a rapid decline in the quarterly networks throughout crisis periods.
As before, the unweighted statistic displays a more immediate reaction to the default of Lehman Brothers with a 44\% drop in reciprocal connections in the Italian interbank network, from 464 reciprocal links in the third quarter down to 258 reciprocal links in the fourth quarter of 2008.
This decline is much larger than the general 18\% loss of trading links in that quarter, which might indicate a fragmentation of the market that goes beyond a mere decline in willingness to trade.

\medskip

Outside of these periods of elevated counterparty risk, the trend in forming reciprocal links seems in fact slightly positive.
As these values are by construction influenced by the number of trading links in the networks, we may also resort to the reciprocity ratio ($D^\leftrightarrow$) defined as the number of reciprocal relative to all links. This variable nevertheless displays similar dynamics as the non-normalized values and even magnifies the observed tendencies of a positive trend, with 0.23 in 2005-Q1 up to 0.29 in 2007-Q2. This growth is interrupted by a large crisis-induced decline (-62\%) from 2007-Q3 to 2009-Q2, leading to lower but steadily increasing post-crisis values, from 0.09 in the last quarter of the crisis up to 0.16 in the last quarter of the sample, which is in line with the results of \cite{finger2013networkanalysis}.

\medskip

The bottom panel of Figure (\ref{fig:ts_unweighted_and_weighted_reciprocity}) displays global reciprocity in trading volume, i.e. the value of bilateral credit lines aggregated across pairs of banks.
Weighted reciprocity fluctuated around a mean value of $\text{€} 34.7 \text{bn}$ $(\sigma = \text{€} 6.8 \text{bn})$ in the pre-crisis periods.
Although it experienced a large decline in the crisis quarters, most of the decline can be attributed to the first stage of the crisis, with a reduction by two thirds from approximately €3bn in the second quarter of 2007 to €1bn in the first quarter of 2008.
The smaller drop in the second period of the crisis is quickly offset afterwards and turned into a slightly increasing trend for the post-crisis periods.

Calculating average reciprocal trading volume $(S^\leftrightarrow)$, i.e. dividing total reciprocal trading volume by the number of reciprocal trading links, again reveals similar dynamics.
In the pre-crisis periods, the average reciprocal pair of banks traded €60.1M ($\sigma = \text{€} 12.2 \text{M}$) in both directions.
This value hits its sample minimum (€24.1M) in the quarter of the Lehman default, 2008-Q3, as a consequence of a still relative high number of reciprocal trading relationships combined with the incipient decline in reciprocal trading volume.
Afterwards, in post crisis episodes, weighted reciprocity ratios reveal an upwards trend along a time average of €37.4M ($\sigma = \text{€} 6.8 \text{M}$).

\medskip

As in the case of the number of trading links in the networks, we again observe the failure of the RWCM to reconstruct topological information, this time in the form of the number of reciprocal trading relationships.
For the same reason as before, i.e. it imposes too much positive probability mass on non-existing credit lines, so that the ensemble considers dense networks to be more likely which in turn also increases the likelihood of (too) many reciprocal links in the network.
Not even the additional preservation of weighted reciprocal trading volume for each bank ameliorates this recurring failure due to the aforementioned irreducibility conjecture.

\medskip

Whereas the DECM was trivially able to generate empirical values of links and trading volume in the previous section, it is no longer obvious a priori whether it is also able to match the observed levels of reciprocity on the interbank market.
This in turn allows the DECM to filter out the pertinent heterogeneous size effects (i.e. each bank's number of trading partners and volume, but not their exact relationship structure) that are prevalent in the non-homogeneous nature of interbank markets, cf. the diverse literature presented in the previous section on interbank markets.
In fact, the DECM distributions on reciprocity in the left panel of Figure (\ref{fig:ts_unweighted_and_weighted_reciprocity}) display some remarkable properties.
The empirically observed number of reciprocal relationships is relatively well predicted by a probabilistic model that is based on degree and strength distributions alone.
Nevertheless, we notice that in most quarters unweighted reciprocity is either close or just above the 95\% interval of the DECM-induced distribution in unweighted reciprocity.
Taken together, the number of reciprocal trading links in the network seems to be governed to a large extent by the in-/out degree correlation.
However the slight overrepresentation of the empirical vs. the theoretical reciprocity values may hint at some non-local (i.e. external or higher-order network-based) influences at work.
Even though the empirical time series undergoes large changes between in- and out-of-crisis episodes, the deviations from the model predictions are nearly constant over time.
This is interesting as it points towards the time-variant values of node degrees and strengths accounting for most of the variation in the number of reciprocal trading relationships across time.

\medskip

If we turn to weighted reciprocity through the lens of the DECM, we immediately observe large, persistent and time-varying deviations of the empirical values from those of the directed enhanced configuration model.
The pre-crisis level of reciprocal trading volume stands out in particular, as the model-implied values not only underestimate the empirical values for almost the entire subperiod, but also because they display a huge quantitative deviation by approximately a factor of two for most of those quarters.
While this volume significantly differs from what can be expected under the DECM which itself is solely based on local degrees and strengths, these deviations completely vanish once market-implied counterparty risk becomes elevated with the beginning of the financial crisis.
Not only do the deviations in reciprocal trading volume disappear with the onset of the crisis, the empirical values even seem to almost perfectly match the mean of the DECM-implied distribution for almost the entire crisis episode.
With the end of the crisis, these deviations enlarge again and remain at or just above the 97.5\% percentile of the distribution, possibly indicating a slight tendency back to the pre-crisis structure but with a quantitatively lower value.
The networks therefore seem to have contracted in the crisis periods to configurations that are the direct result of degree and strength distributions.
In other words, banks seem to have had a preference towards reciprocating credit line values in out-of-crisis times while the level of reciprocated volume in times of crisis are the mere result of degree and strength distribution.

\medskip

As these are statistical models, we unfortunately cannot investigate the causal nature of these findings.
Let us however briefly highlight two possible behavioral explanations.
The first approach is related to trust-based explanations, see e.g. \cite{allen2022trustand} who show that low levels of trust leads to lower interbank borrowing.
Theoretical mechanisms are usually based on the idea that banks face informational asymmetries which can be mitigated via repeated and/or reciprocal interaction.
As a consequence this could lead to a higher probability of lending to an already established (counter-directed) trading partner as opposed to either (i) searching the market for new trading partners on whom there is not much information known in advance or (ii) turning to large intermediaries who may exhibit less information asymmetries due to high liquidity and implicit too-big-too-fail safety nets but which may extract higher rents on rates in return.
If this mechanism mitigates informational frictions in normal times but is too fragile to hold up under extreme events such as a global financial crisis, banks may have turned indifferent between continued reciprocal lending or turning to another trading partner.
The latter motive could be explained by a turn towards larger and therefore possibly safer intermediaries or the simple need for diversifying their lending portfolio as opposed to concentrating their credit risk in few counterparties that were solely backed by fragile reciprocity-induced trust.

\medskip

The second explanation is based on the idea of reciprocity as two-sidedness which we have seen in the dataset as well as in the general literature: Large banks are more likely to be net borrowers while also engaging in intermediation trades between two smaller banks that are themselves otherwise more likely to be net lenders.
Based on the theoretical mechanism of \cite{ho1985amicro} the distinction between large and small banks lies in their different demands for liquidity due a differential exposure to risky investments as a consequence of heterogeneous risk aversion.
While in pre-crisis times, liquidity shocks were likely to be more idiosyncratic, these shocks likely became more correlated in times of crisis, so that resulting liquidity and solvency risks turned overnight loans from safe to unsafe assets.
As a result the main tendencies of large banks with their need for liquidity and small banks with their surplus of liquidity could reduce the tendency in taking the opposite side of a trade.
Small banks have even less incentive to borrow if they tend to loan out only a fraction of their liquidity surplus in times of crisis, while big banks are in even higher need for liquidity if their previous source of funding (i.e. small banks) have restricted lending. This in turn would make it less likely that big banks also take the other side of the trade and provide liquidity to the market again.

\subsection{Core \& Peripheral Banks}
Interbank networks in general, and the e-MID in particular, display huge heterogeneity in trading relationships.
The Core-Periphery model of \cite{borgatti2000modelsof} in turn has been shown to successfully replicate many of these heterogeneous dimensions such as assortativity values and skewed degree distributions, see in particular \cite{fricke2015coreperipherystructure} for more detailed explanations with an application to the Italian interbank market.

\medskip

In order to investigate possibly heterogeneous effects of reciprocity, we will thus decompose our set of banks $\N = \C \sqcup \P$ into a disjoint set of core banks $\C$ and periphery banks $\P$, and analyze reciprocal trading patterns within and across these subgroups.
The topological matrix representation of the network, $A = \{a_{ij}\} := \{a_{ij}\}^{i \in \N}_{j \in \N}$, can then be block-partitioned into a matrix of within-core links, $A^{cc} = \{a_{ij}^{cc}\} := \{a_{ij}\}^{i \in \C}_{j \in \C}$, a matrix of within-periphery links, $A^{pp} = \{a_{ij}^{cc}\} := \{a_{ij}\}^{i \in \P}_{j \in \P}$, a matrix of core-to-periphery links, $A^{cp} = \{a_{ij}^{cp}\} := \{a_{ij}\}^{i \in \C}_{j \in \P}$, and a matrix of periphery-to-core links, $A^{pc} = \{a_{ij}^{pc}\} := \{a_{ij}\}^{i \in \P}_{j \in \C}$.

\medskip

In an idealized Core-Periphery structure of a simple network (i.e. without self-loops or multi-edges), the core would be completely connected, i.e. $A^{cc} = \bm{1}_{|\C| \times |\C|} - I_{|\C|}$ where the first term is a matrix of ones and the second term is an identity matrix of the same shape.
The periphery in turn would be completely unconnected so that the submatrix $A^{pp} = \bm{0}_{|\P| \times |\P|}$ results in a zero-matrix.
Theoretical restrictions on off-diagonal blocks however are not as obvious as on within-blocks.
In the spirit of the pertinent literature, we therefore do not impose restrictions on the ideal matrix structures of $A^{cp}$ and $A^{pc}$.\footnote{
    We note however that \cite{borgatti2000modelsof} suggest as possible restrictions on $A^{cp}$ and $A^{pc}$ one of the following: full connectivity, no connectivity, or some intermediate value.
    With respect to the latter idea, \cite{craig2014interbanktiering} develop row- and column regularity conditions on these off-diagonal blocks so as to better capture the idea of core nodes as intermediaries in the context of German interbank networks.
    Since \cite{fricke2015coreperipherystructure} however have shown for a similar dataset to ours that such a model is statistically indistinguishable from a model without off-diagonal restrictions, we feel comfortable in concentrating on the baseline Core-Periphery model that imposes only within-block restrictions.
}

\medskip

Noting that the ideal core block in a network without self-loops has $|\C|^2 - |\C|$ connections, we can then formulate an error score
\begin{align}
  e(\C) =
  \frac{
    \left(  |\C|^2 - |\C| - \sum_{i,j \in \C} a_{ij} \right) + \sum_{i,j \not\in \C} a_{ij}
  }{
    \sum_{ij} a_{ij}
  }
  \label{eq:def_error}
\end{align}
which takes as argument a subset of possible core banks $\C$ out of a given set of banks $\N$, and enumerates all deviations from an ideal Core-Periphery structure, i.e., any theoretical within-core link that is absent or any within-periphery link that is present in the empirical network.
The normalization factor, i.e. the number of total links in the network $\sum_{ij} a_{ij}$, is independent of $\C$ and thus only scales the solution with respect to a hypothetical maximum error score in a periphery-only network. Hence, it does not alter the optimal solution $\C^* = \text{argmin}_{\C \in \mathbb{C}} e(\C)$ to this NP-hard partitioning problem \citep{ballester2010delinquentnetworks} of searching the space of nontrivial ordered bipartitions, or equivalently the space of possible core node subsets $\mathbb{C} = \{ \C : \emptyset \neq \C \subset \N \}$.

\medskip

As the number of bipartitions scales exponentially in the number of nodes $n$ with $O(2^n)$, we are unable to exhaustively search the solution space of approximately $10^{30}$ possible core-periphery structures.
For this reason the literature has resorted to simulated annealing or evolutionary algorithms \citep{fricke2015coreperipherystructure}. Although this class of algorithms typically performs well for large-scale optimization purposes, global solutions are not guaranteed.
\cite{brusco2011anexact} instead develops a branch-and-bound algorithm to the bipartitioning problem that guarantees an optimal solution.
The downside, however, is the lack of scaling to larger networks.\footnote{
  The authors in fact recommend their method only for networks with up to $n \leq 60$ nodes.
}
Fortunately, based on the previous approach, \cite{lip2011afast} solve the underlying enumeration problem in the algorithm, and develop a sorting method that easily scales to large-scale networks.\footnote{
  We refer to \cite{yanchenko2022coreperipherystructure} for a more detailed overview including strengths and weaknesses of the different core periphery algorithms.
}
In the following, we will rely on the latter approach as it reduces time complexity from exponential to polynomial scaling, $O(n^2)$, by first calculating individual degrees and in a second step sorting these according to a cutoff criterion.
We also checked our results using the original greedy algorithm by \cite{boyd2006computingcoreperiphery} based on a \cite{kernighan1970anefficient} algorithm with our error score (eq. \ref{eq:def_error}) as objective function and find similar results.

\medskip

\begin{figure}[H]
  \centering
  \includegraphics[scale=0.4]{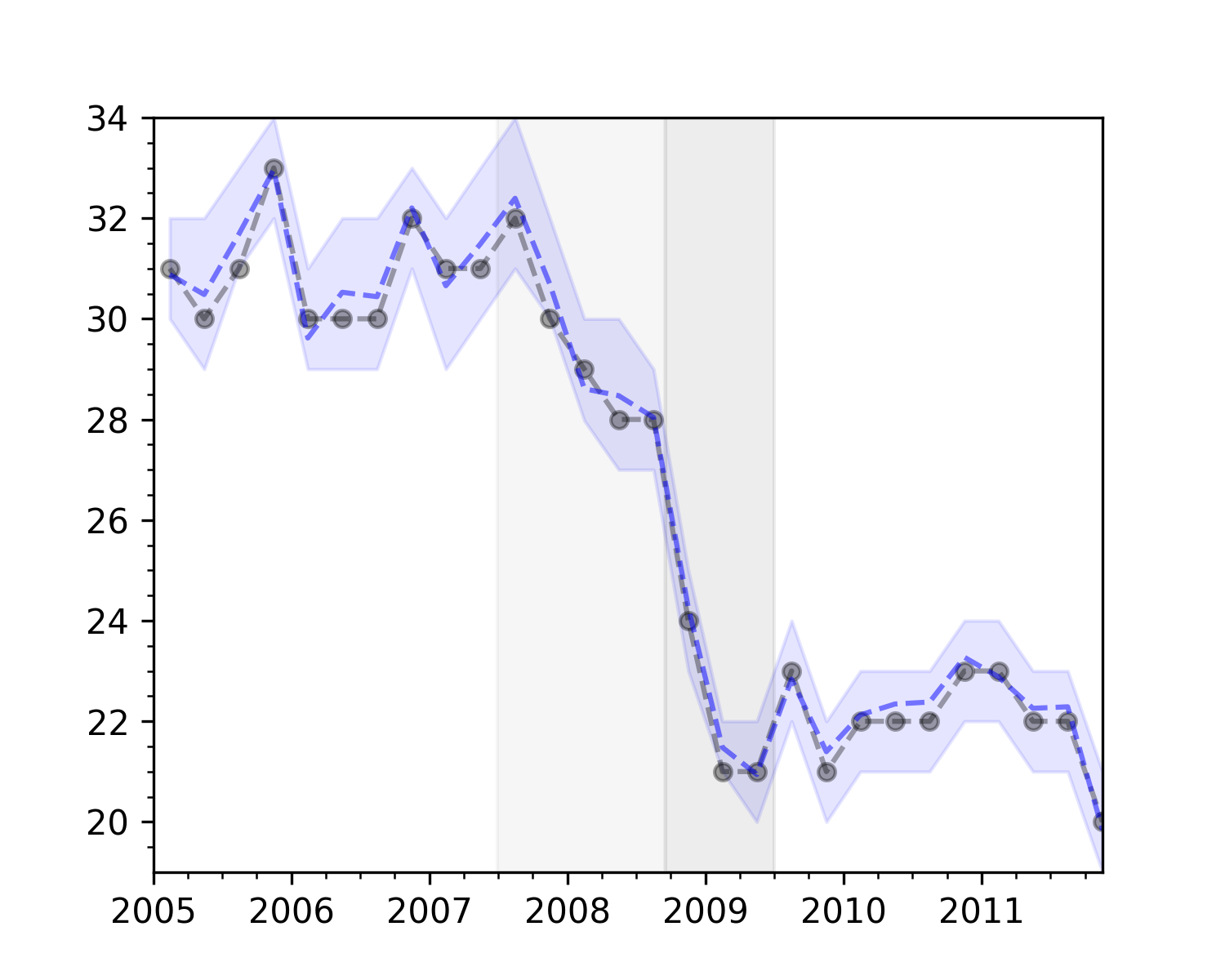}
  \includegraphics[scale=0.4]{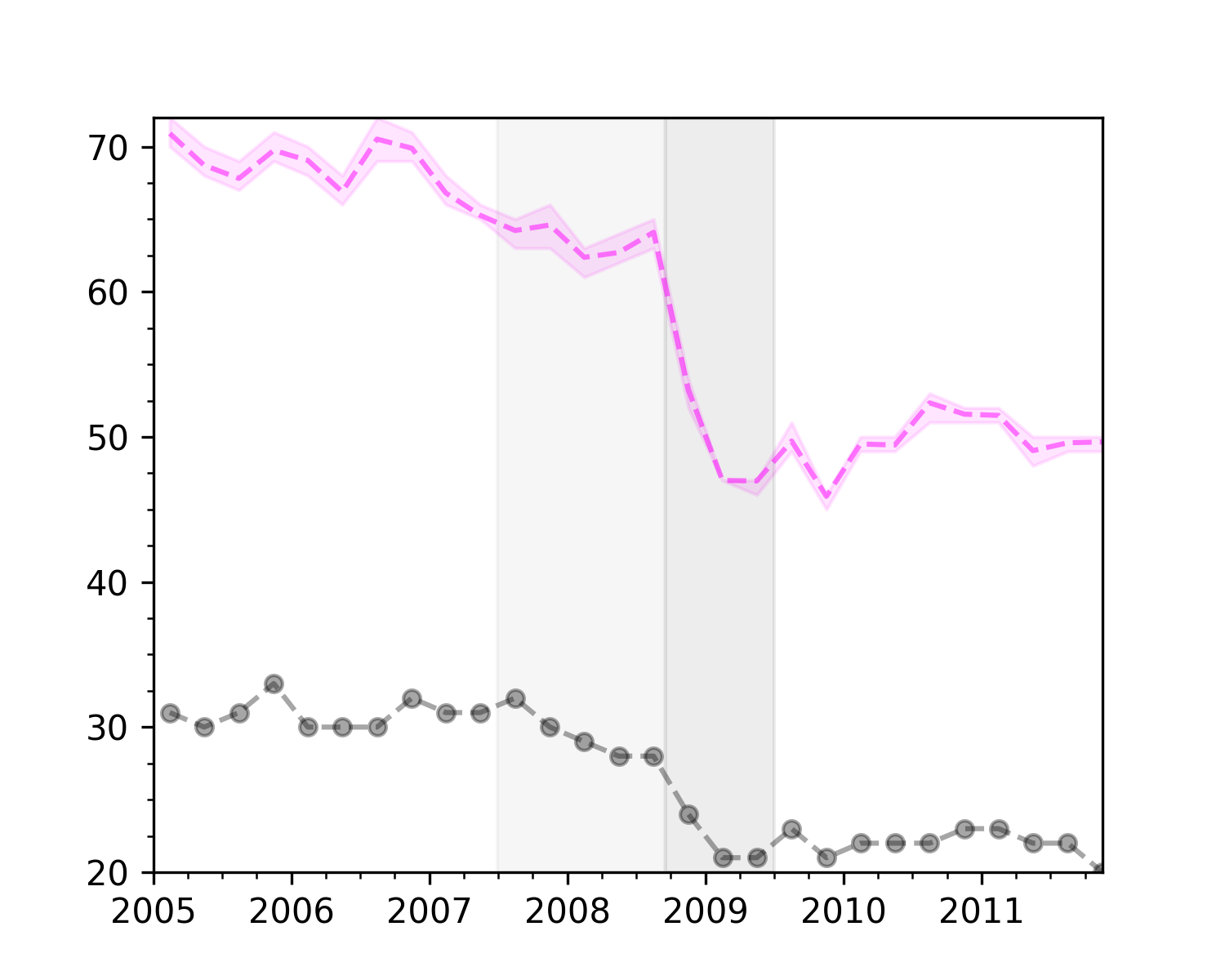}
  \includegraphics[scale=0.4]{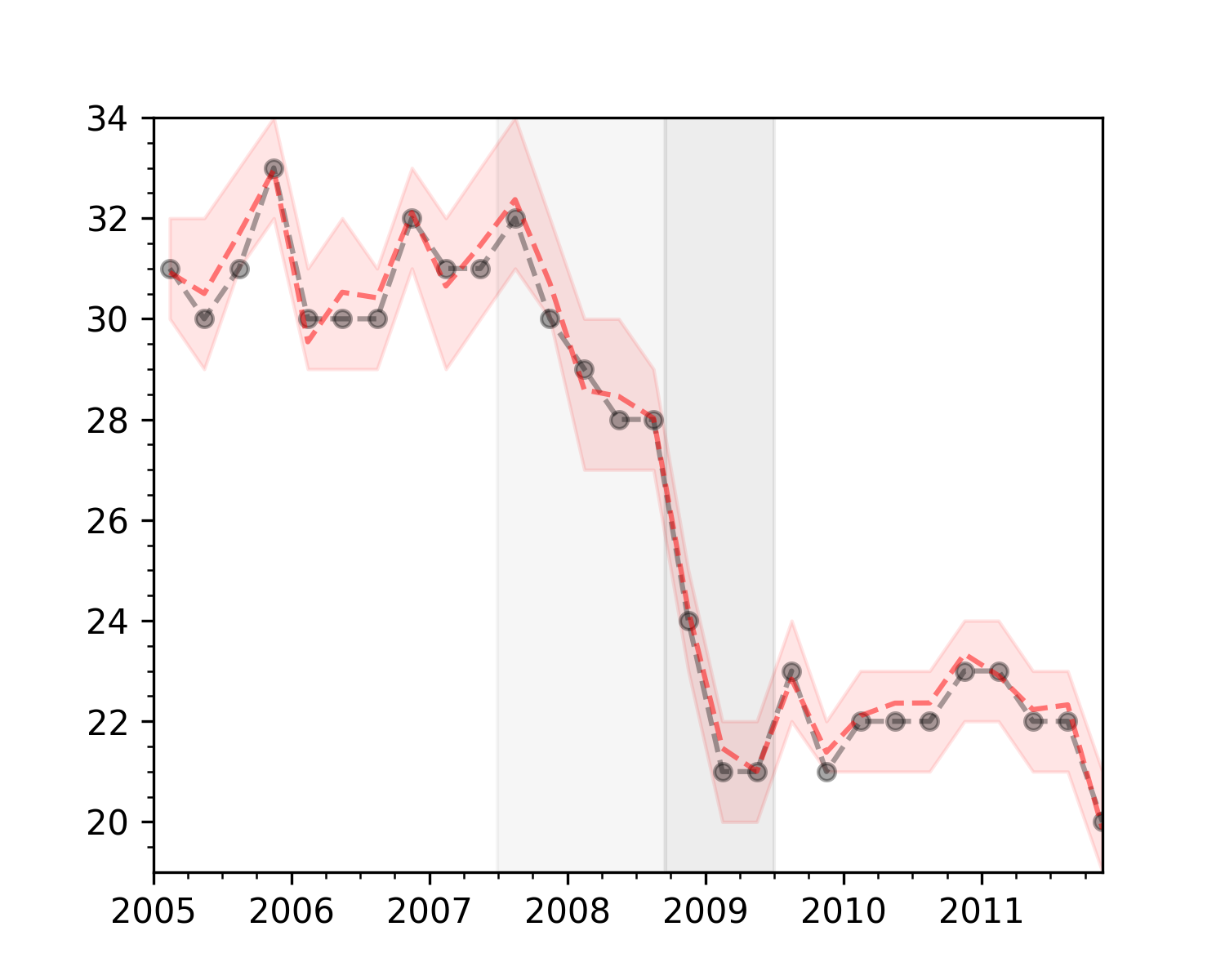}
  \\
  \includegraphics[scale=0.4]{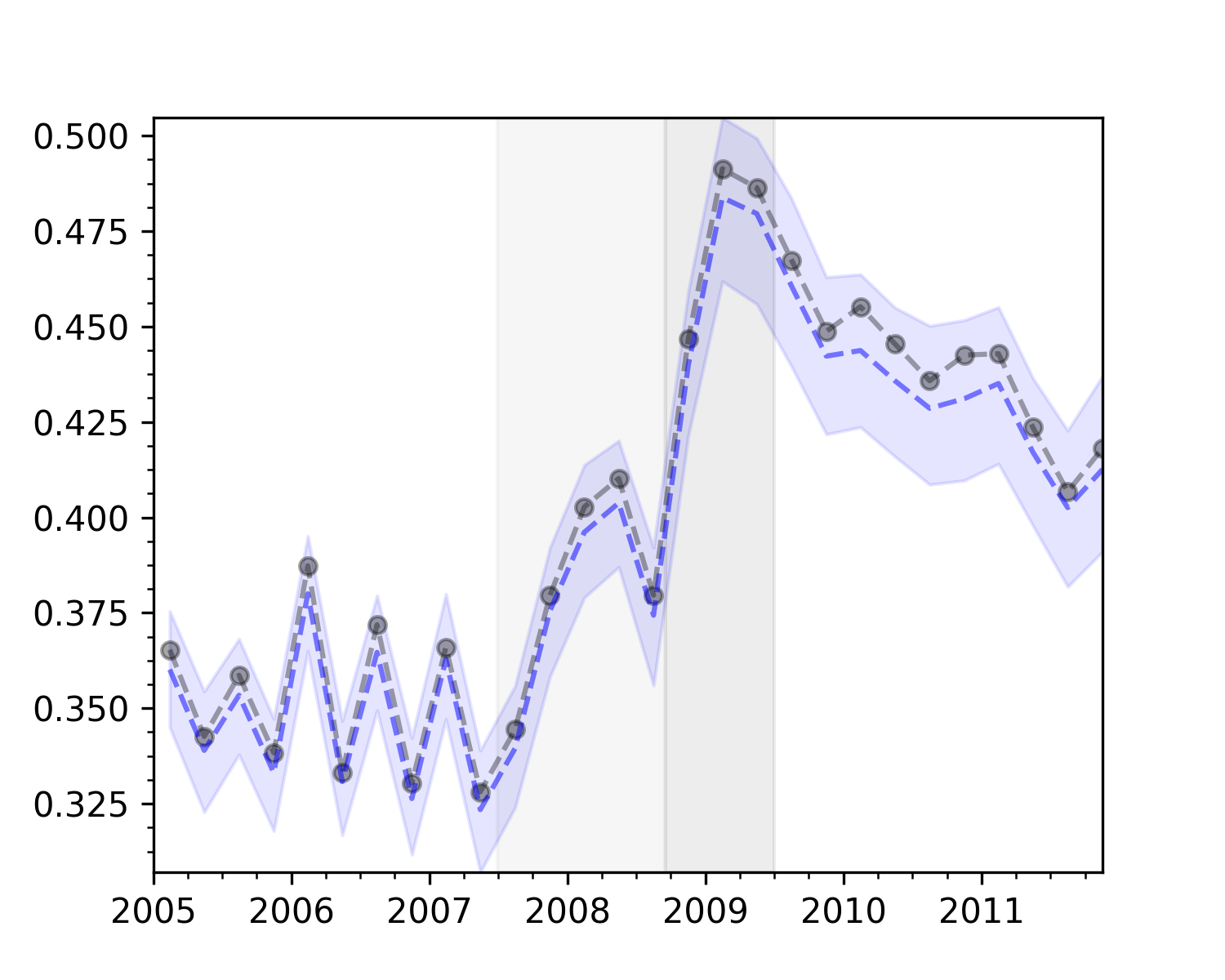}
  \includegraphics[scale=0.4]{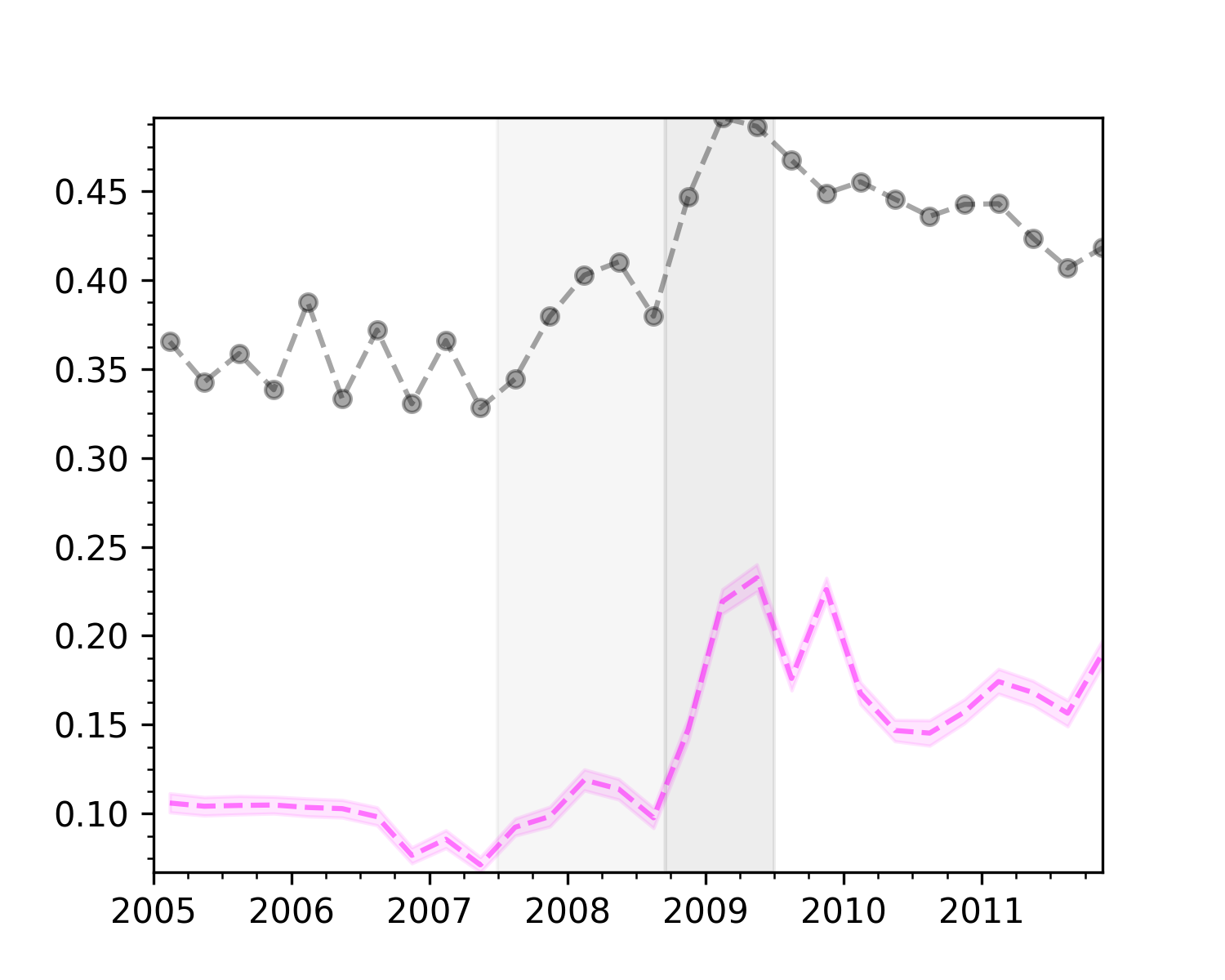}
  \includegraphics[scale=0.4]{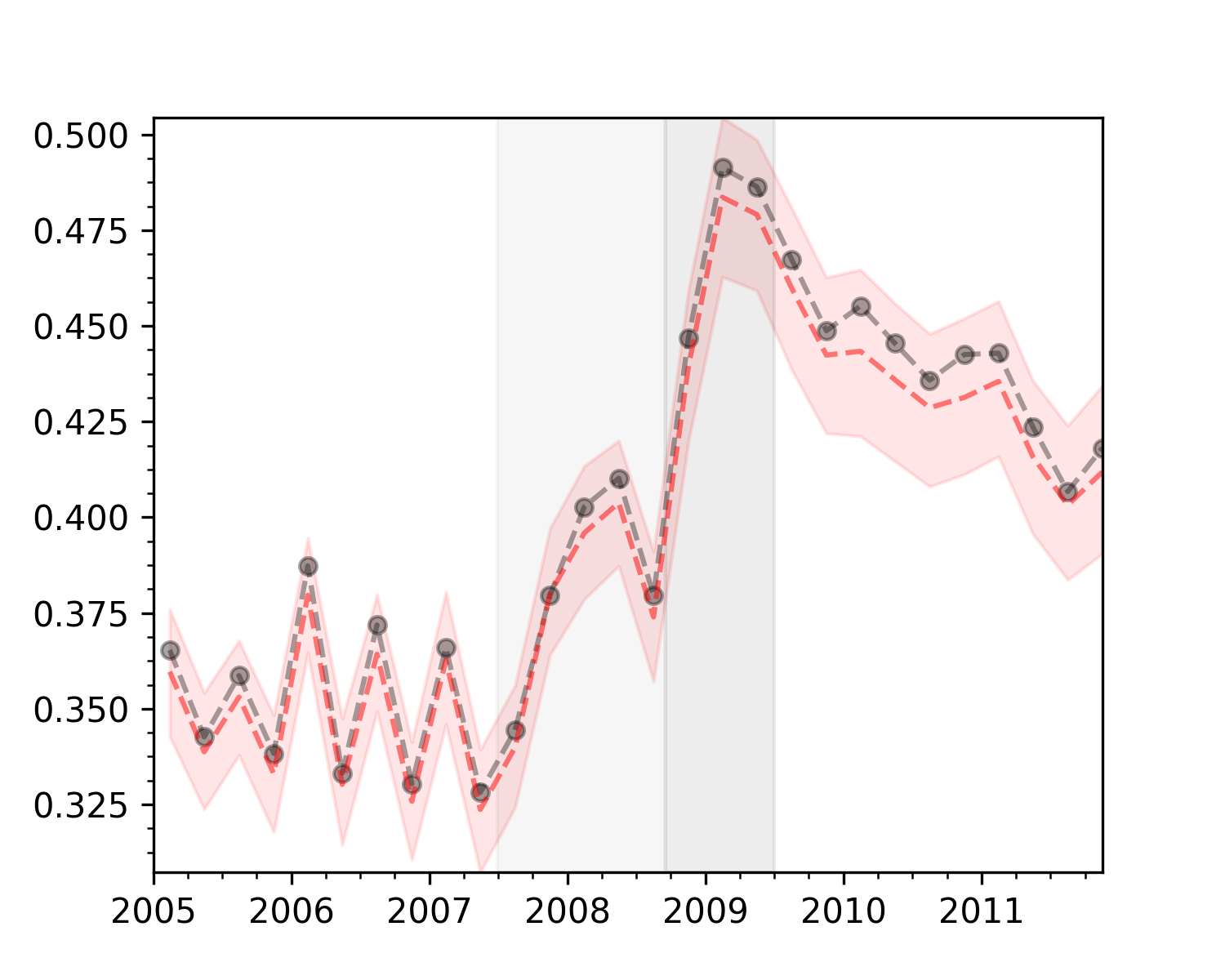}
  \caption{\textit{Core-Periphery Decomposition.} Upper panel: Core size. Lower panel: Error score. Empirical time series (black dots) are shown alongside ensemble means and 95\% intervals of DECM (blue) that preserves only un+weighted degrees, RWCM (pink) that preserves only weighted reciprocity, and RECM (red) that preserves un+weighted reciprocity on the node-level.
  }
  \label{fig:ts_meso_coresize_coreerrors}
\end{figure}

The main results of the core periphery optimization for the three ensembles in terms of core size and error value distributions compared to empirical values can be found in Figure (\ref{fig:ts_meso_coresize_coreerrors}).
Both empirical time series display a major transition in crisis episodes.
While in pre-crisis times the number of core banks seemed to fluctuate around a stable value of 31 core banks with an error score around 0.35 these values deteriorate to 21 core banks with an error score of 0.49 in the last quarter of the crisis.
Afterwards, the number of core banks seems to settle around 22 banks with a slowly improving fit of the core periphery model.

As the core periphery bipartition approach depends fundamentally on the number of links - maximal in the core, minimal in the periphery - we have an explanation why the dynamics of core size resembles those of total links in the network.
In particular we observe that the general loss of connectivity happened to a large extent due to core banks, either remaining in the core with less connectivity and thus more core errors, or becoming peripheral but still maintaining a sizeable amount of connections and thus more periphery errors.
\cite{fricke2015coreperipherystructure} have already pointed out that the magnitude of the error values is not small but still significantly lower than in unrestricted random benchmarks.
The Italian electronic market for interbank deposits can therefore be well characterized by a core periphery architecture, but seems nonetheless also be driven by other linking motivations besides the pure intermediation explanation of the ideal core periphery benchmark.

\medskip

Shifting focus towards the performance of the three configuration models in matching these core sizes and errors, we observe another failure of the RWCM that is unable to match both properties even though it preserves total trading volume and weighted reciprocity of each node.
It is particularly the tendency of the RWCM to generate dense networks which leads it to classify too many banks as core nodes.
As such, the within-core error score becomes very small and pushes the total error strongly below the empirically observed values.
It is therefore not "better" but in fact unable to reconstruct the empirical topology and (impartial) core periphery structure.
The other two models, DECM and RECM, in turn, are in excellent concordance with these two key statistics of the decomposition procedure.
Both distributions in terms of means and confidence intervals in fact look remarkably similar.
This fact however seems a natural consequence of the type of constraints both models have in common.
DECM as well as RECM both preserve the total number of trading partners of each bank.
While both also match each bank's total trading volume (and in the latter case also their un+weighted reciprocal counterparts), it is the degree sequence of the network that is the single ingredient in the ordering approach for core periphery detection by \cite{lip2011afast}.
Since both ensembles are able to match in- and out-degree of each node, matching an empirical core periphery structure is a direct consequence.
With the DECM being a subclass of the RECM, we will call in particular this former model also by the name \textit{Core-Periphery-preserving Configuration Model}.\footnote{
  The RECM of course also preserves the core periphery structure but with more constraints than needed for this goal.
  Strictly speaking, the DECM also imposes too many constraints, i.e. the total in- and out-strengths of each node, but in the class of weighted heterogeneous configuration models it seems a plausible minimal model.
}

\medskip

Table (\ref{tab:linkvol_averages}) illustrates the different connectivity and weight patterns within and across core and periphery partitions.
On the left side, $\overline{A}$ contains the average quarterly within- and across-partition number of links with block density (as the number of realized links divided by the maximum number of possible links in the respective block) in parentheses.
The right side, $\overline{W}$, in turn shows the average quarterly within- and across-partition total trading volume with average trading volume per link in parentheses.\footnote{
  Note that average trading volume per link has been calculated by time-averaging the quarterly statistics, so that a division of the (time-averaged) numbers in $\overline{W}$ by those in $\overline{A}$ yields slightly different but comparable values.
}

\begin{table}[H]
  \[
    \overline{A} =
    \left(
      \begin{array}{c | c}
        387.0 (54.3\%) &295.4 (15.2\%)\\
        \hline
        840.5 (43.4\%) &473.0 (\,9.3\%)
      \end{array}
    \right)
    \hspace{0.2cm}
    \hfill\vline\hfill
    \hspace{0.2cm}
    \overline{W} =
    \left(
      \begin{array}{c | c}
        104.3bn. (259.0M) &35.4bn. (117.2M)\\
        \hline
        178.5bn. (204.0M) &46.6bn. (97.8M)\,\,
      \end{array}
    \right)
  \]
  \caption{\textit{Quarterly Empirical Time Averages}. Left Panel: Total within-block links (density). Right Panel: Total within-block volume (volume-per-link). The top left blocks are core-to-core, the top right blocks are core-to-periphery, the bottom left blocks are periphery-to-core and the bottom right blocks are periphery-to-periphery partitions. }
  \label{tab:linkvol_averages}
\end{table}

Core-periphery algorithms attempt to maximize within-core links and minimize within-periphery links by construction.
This fact is reflected in the density numbers in parenthesis on the left-hand side of the table: The block density in the core is more than five times larger than in the periphery.
At the same time, only every second of all possible links in the core-core block is realized on average, highlighting deviations from an ideal, fully connected core.
Even the total number of links in the periphery is on average slightly larger than in the core.
This seemingly puzzling feature however can be explained by an imperfect fit of the core periphery model in combination with the relatively small number of core banks compared to peripheral banks in the e-MID networks.
Nevertheless, the idea of large core banks vs. smaller peripheral banks remains:
Even though the core displays fewer within links than the periphery, it trades more than 2.5 times as many funds per core-to-core relationship, leading to a higher within-core total trading volume of $\textup{\euro}104.3bn.$ compared to $\textup{\euro}46.6bn.$ that peripheral banks trade among themselves per quarter.

\medskip

What stands out, furthermore, is the overwhelming importance of the core as liquidity absorber and the periphery as liquidity provider, as predicted theoretically by \cite{ho1985amicro}. Quantitatively, the periphery-to-core lending block entertains nearly three times as many links (as well as density values) as the oppositely directed core-to-periphery equivalent.
This discrepancy also translates into the weighted perspective with quarterly periphery-to-core trading volume of $\textup{\euro}178.5bn.$ as opposed to $\textup{\euro}35.4bn.$ in the other direction, a tendency which is also reflected in the normalized quantities (volume-per-link) of the respective blocks.
These findings support the findings in the literature that interbank networks are heterogeneous and asymmetric.

\medskip

Before moving from description to inference, we note that our statistical models would in principle allow for a replication of both of these features. The size heterogeneity driven by the core periphery structure should in fact be well captured by DECM and RECM as both models preserve the degree distribution of the network, which has been shown to determine the classification into core and peripheral banks.\footnote{
    While both models preserve total links and volume aggregated over the \textit{entire neighborhood} for each node, this does not guarantee that they also preserve these statistics aggregated over the \textit{sub-neighborhoods} consisting of only core- or periphery-neighbors respectively.
}
As the results from the previous sections have clearly shown the inability of the RWCM to generate sparse networks and a realistic core periphery structure, we will abstain from presenting results of this particular model in the following.
Instead, we focus on the other two models to check when local information on size, proxied by degrees, and volume (DECM) is enough to reconstruct empirical higher-order network statistics, and, should these features turn out to be insufficient, whether adding reciprocity structure (RECM) provides the missing link in explaining those values.

\medskip

In order to render results comparable in terms of significance across models as well as block partitions of different size, we will use the models as filters for the empirical time-series and treat them as z-scores, i.e. for a given network statistic $x$ we have $z^{(model)} = (x^{(empirical)} - \bar{x}^{(model)}) / \sigma_x^{(model)}$ with $\bar{x}^{(model)}$ and $\sigma_x^{(model)}$ as ensemble (sample) mean and standard deviation of $model \in \{DECM, RECM\}$.
We also provide bounds of 1.96 normal standard deviations as a measure of significance in the corresponding figures.
While such z-tests are only exact for normally distributed variables, they are common in the literature in the absence of exact theoretical tests to assess possible structural, non-random generation mechanisms beyond the constrained network variables.
We furthermore note that in our case significant z-scores happen to coincide nearly always with empirical values that fall outside the 95\% sampling intervals of the theoretical ensembles.

\medskip

\begin{figure}[H]
  \centering
  \subfloat{\includegraphics[scale=0.3]{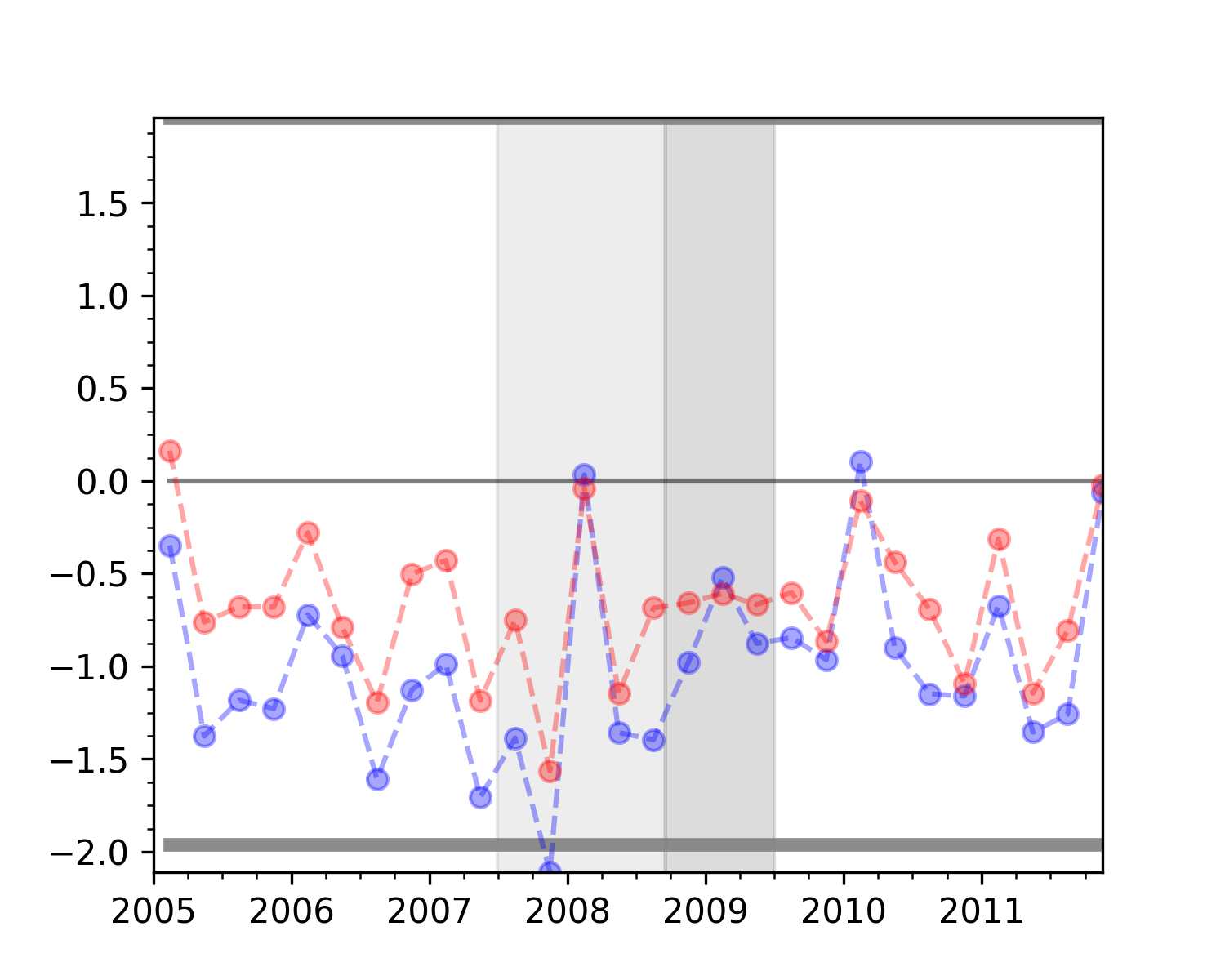}}
  \subfloat{\includegraphics[scale=0.3]{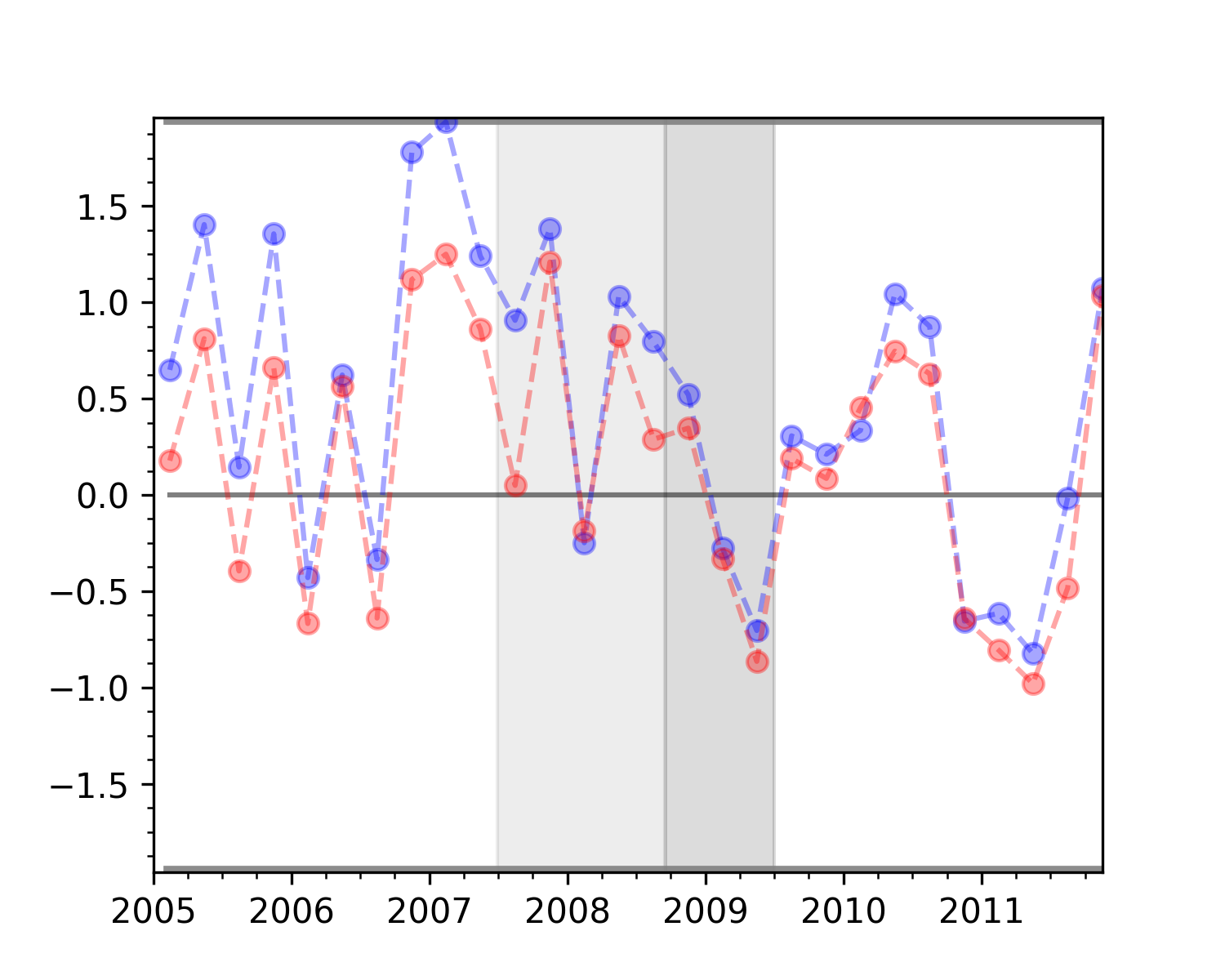}}
  \hfill\vline\hfill
  \subfloat{\includegraphics[scale=0.3]{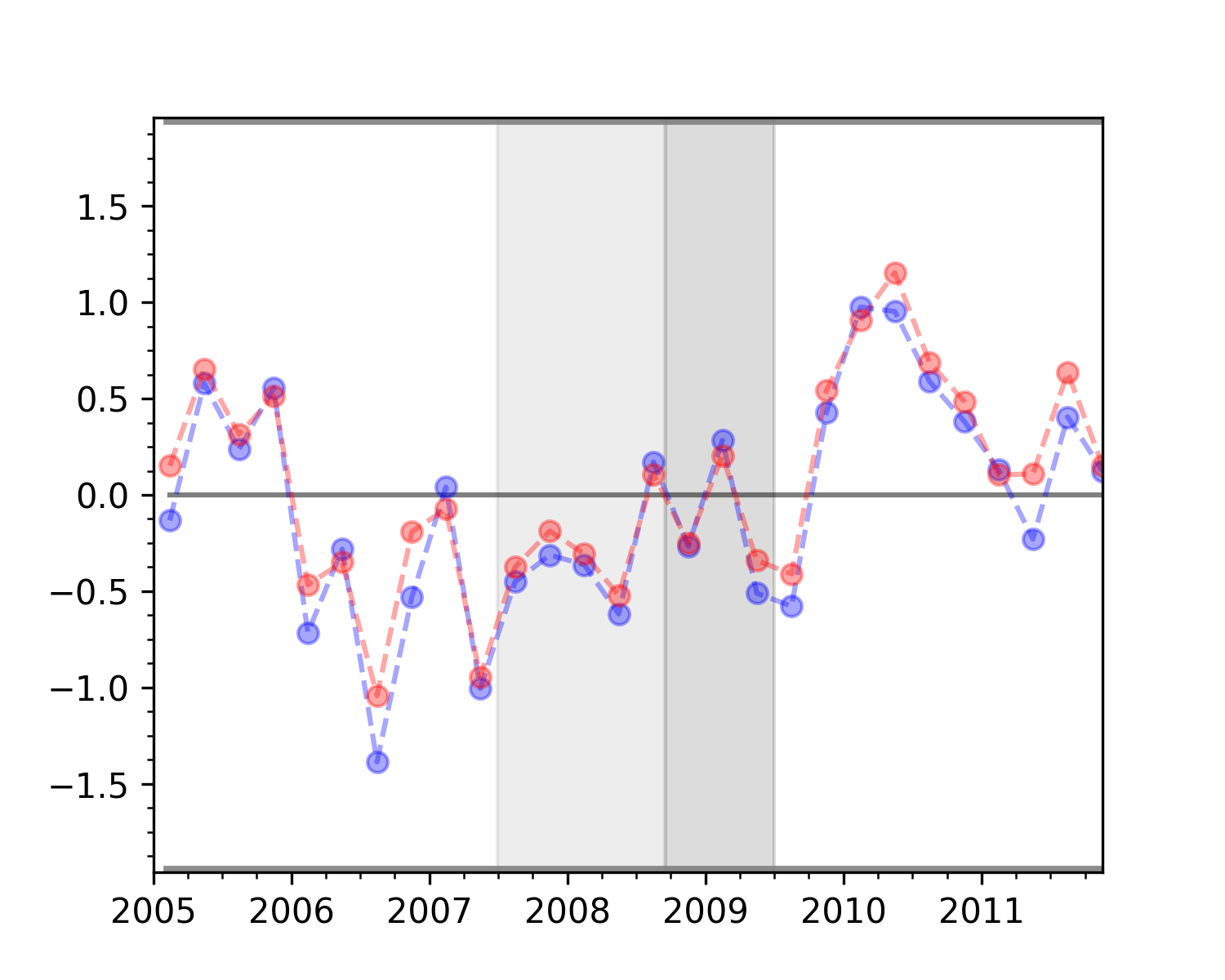}}
  \subfloat{\includegraphics[scale=0.3]{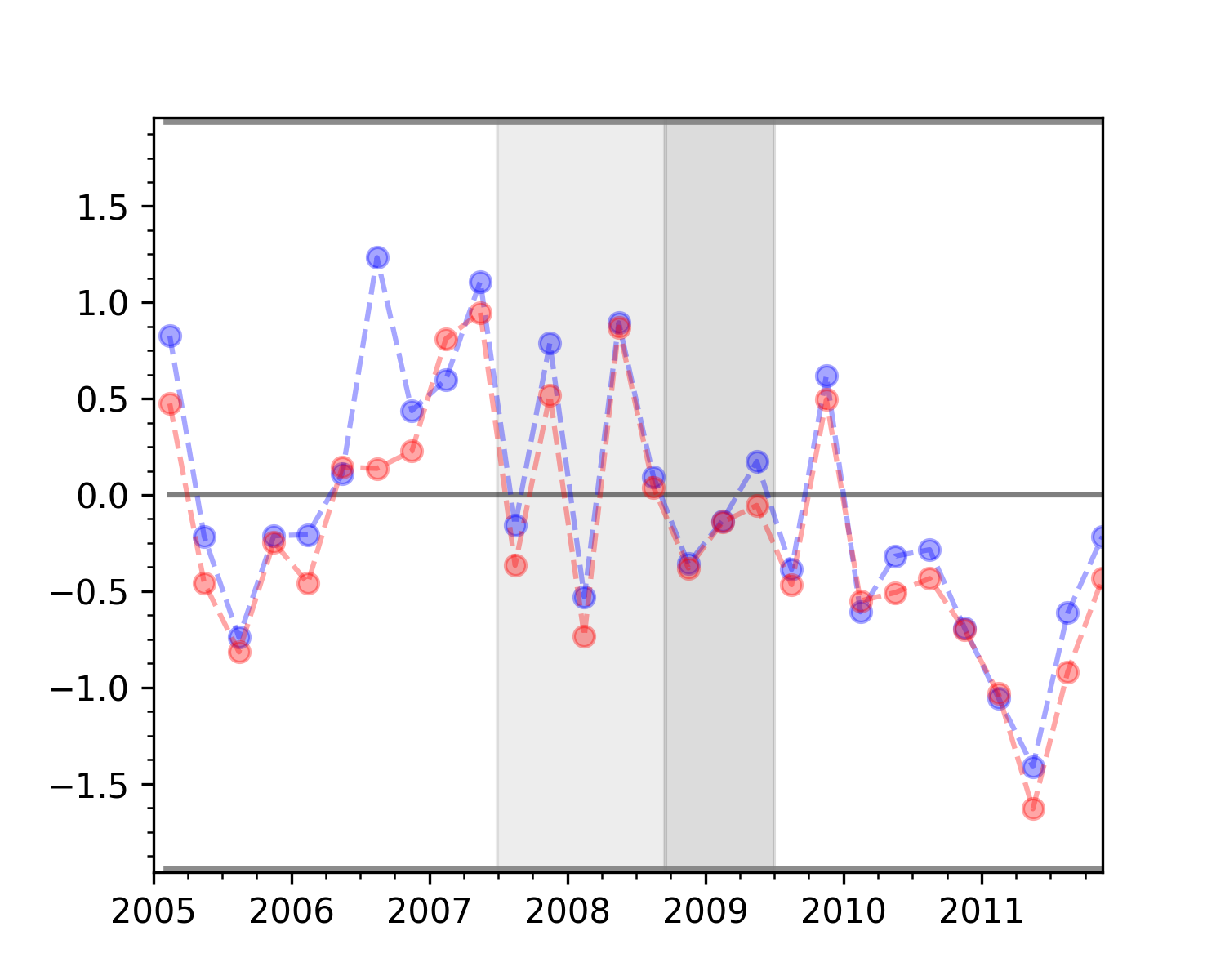}}
  \vspace{-0.4cm}
  \\
  \subfloat{\includegraphics[scale=0.3]{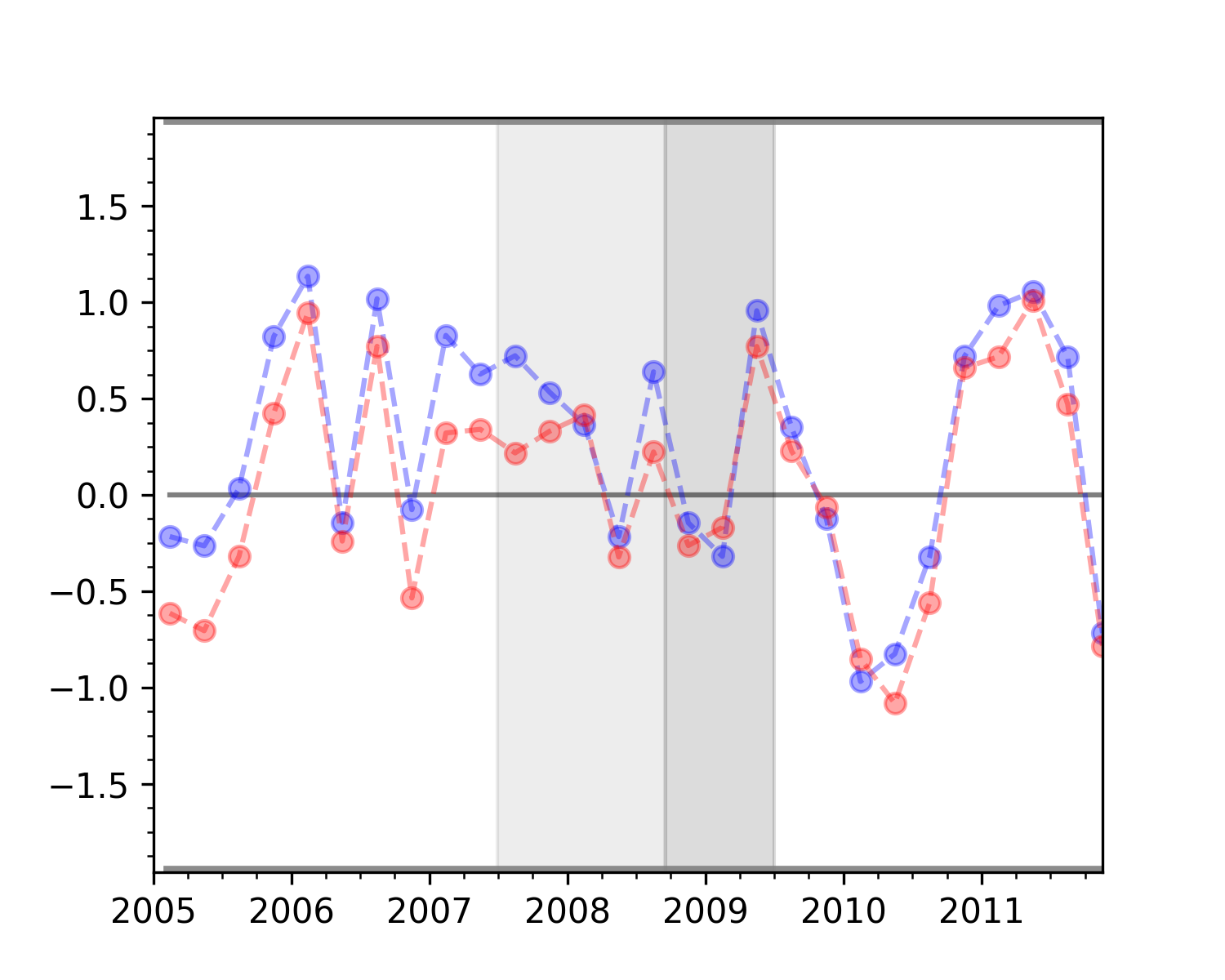}}
  \subfloat{\includegraphics[scale=0.3]{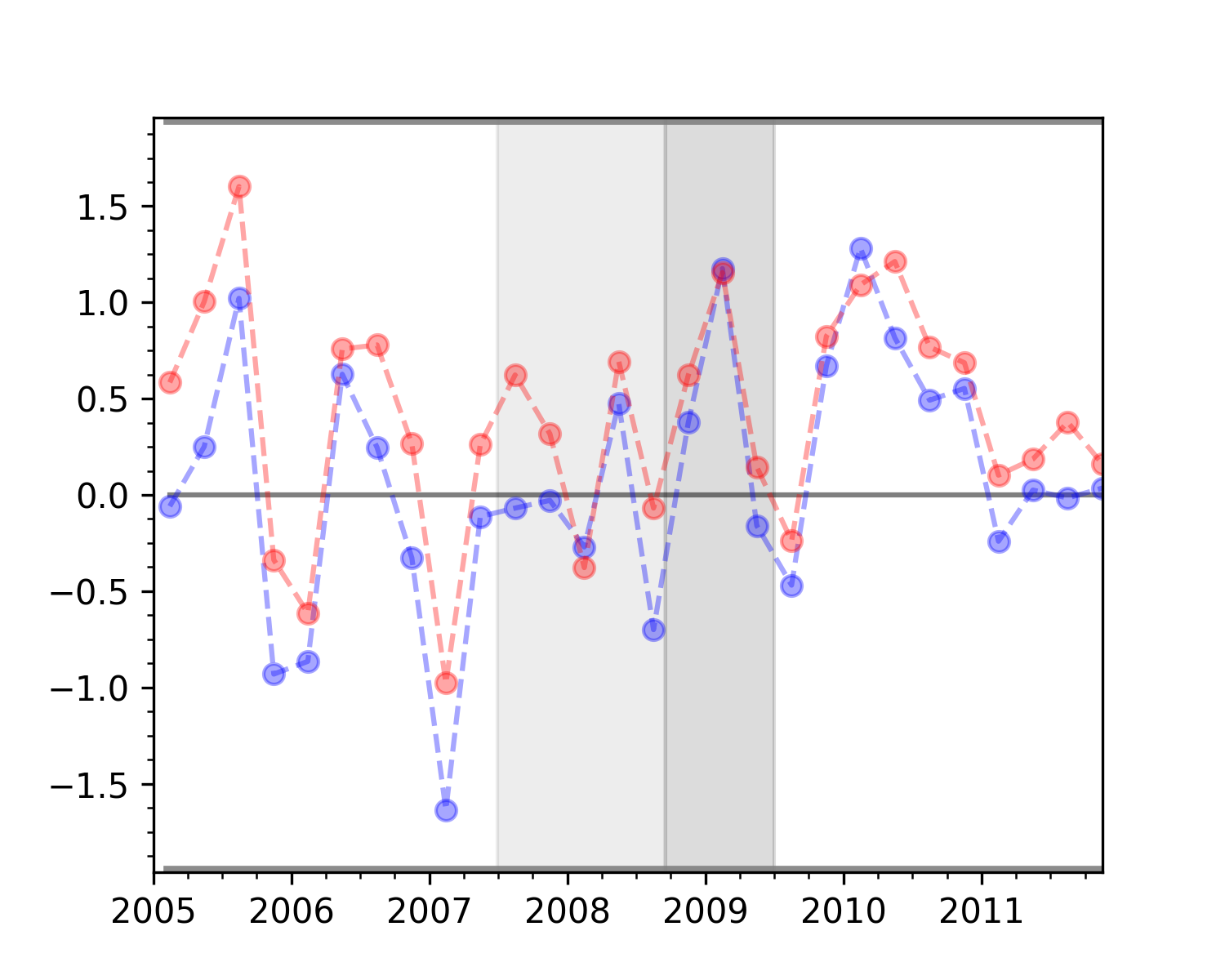}}
  \hfill\vline\hfill
  \subfloat{\includegraphics[scale=0.3]{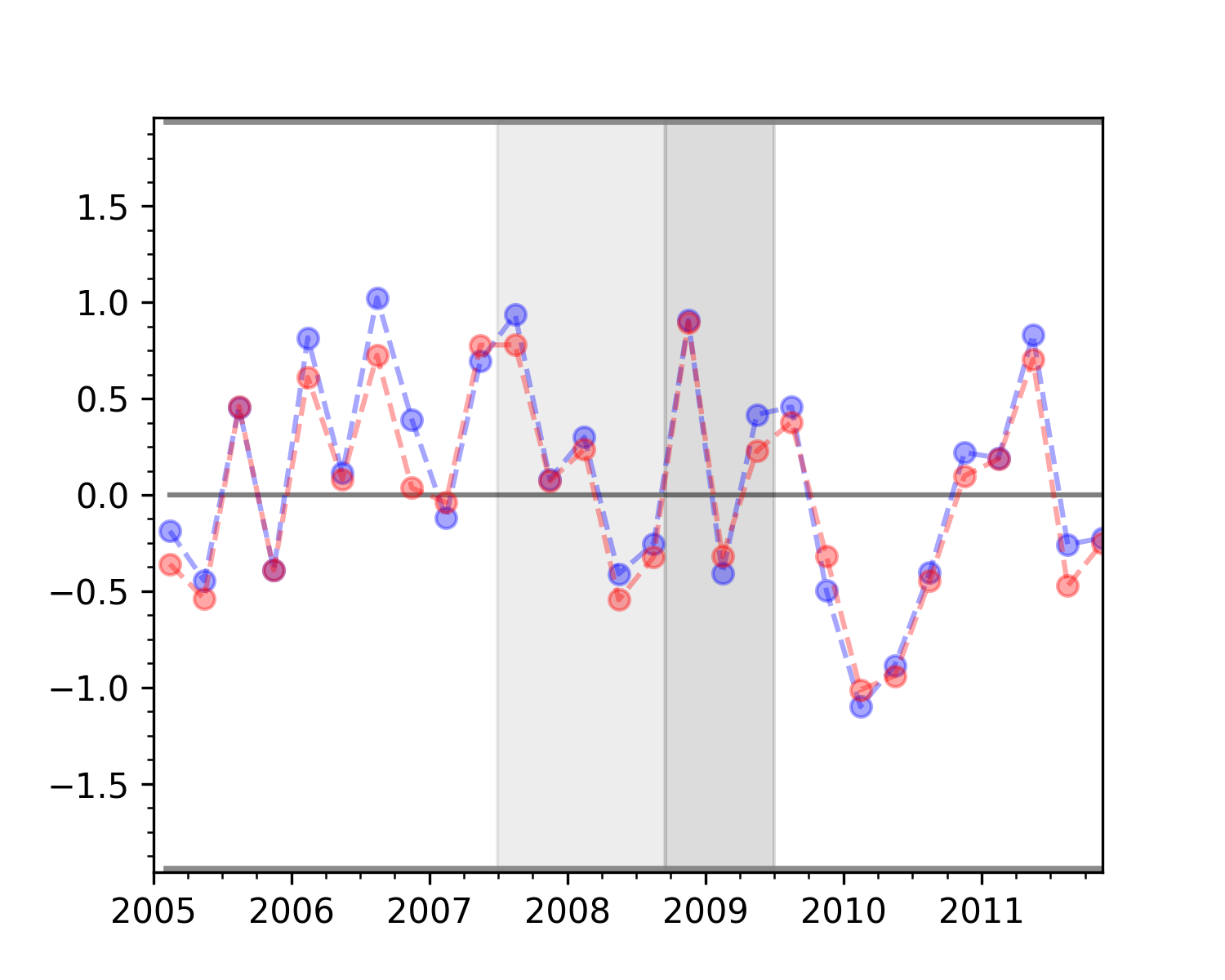}}
  \subfloat{\includegraphics[scale=0.3]{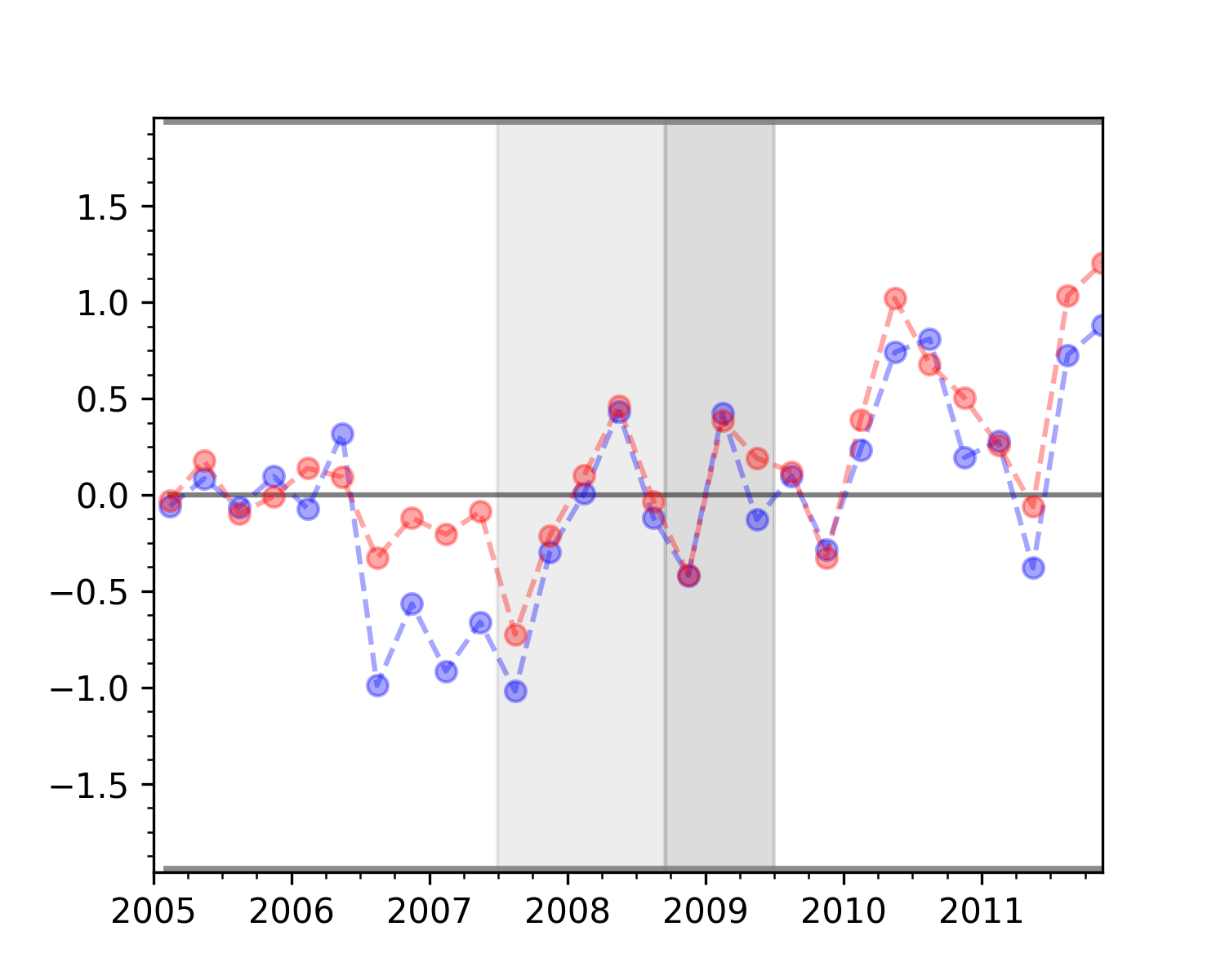}}
  \caption{\textit{Core-Periphery Decomposition (z-scores) - Total Links (left) and Total Volume (right).} Upper left panel: Core-Core. Upper right panel: Core-Periphery. Lower left panel: Periphery-Core. Lower right panel: Periphery-Periphery. DECM-filtered empirical series are shown in blue, RECM-filtered empirical series in red.
  }
  \label{fig:ts_meso_total_links}
\end{figure}

Figure (\ref{fig:ts_meso_total_links}) shows the z-scores of both models for each block, based on the total number of trading links (on the left side) and total trading volume (on the right side) of the respective block.
As we hypothesized, the unweighted statistics are well-matched by both types of models pushing the z-scores to the zero line with random fluctuations.
Similarly, we observe that both models successfully replicate the empirical block-aggregate trading volume dynamics as well.
Although the underlying empirical series exhibit time-variant behavior, such as trends and breaks in and out of crisis episodes, both models capture these effects so that these dynamic patterns do not translate into the filtered series.
In fact, both of these z-scores strongly co-move in each block over time.
Hence, the global degree and strength distributions, which form the basis of both DECM and RECM, appear to be very informative of the more local block-based sub-neighborhoods of exclusively core or periphery trading partners as well.

Turning to reciprocity, we note that in the ideal core periphery structure the within-core block should display full reciprocity as this sub-network is fully connected while an unconnected within-periphery block should display zero reciprocity.
As shown earlier, however, our empirical data matches the basic tendencies of this ideal structure but with noticeable imperfections in the form of elevated error scores.
If these errors were sufficiently small the Core-Periphery-preserving DECM should therefore be able to replicate these patterns.
On the other hand, the DECM has also significantly underestimated reciprocity on the global scale for the e-MID data.
The decomposition into core and periphery now allows us to investigate further which types of banks drive this preference for reciprocity.\footnote{
    In analogy to the relationship of total links and volume on the meso-scale in the DECM, we emphasize that the RECM preserves the reciprocity distributions by definition on the global scale, which however does not guarantee that it is also able to match the distributions of reciprocity values calculated across core and periphery sub-neighborhoods.
}

\medskip

\begin{figure}[H]
  \centering
  \subfloat{\includegraphics[scale=0.3]{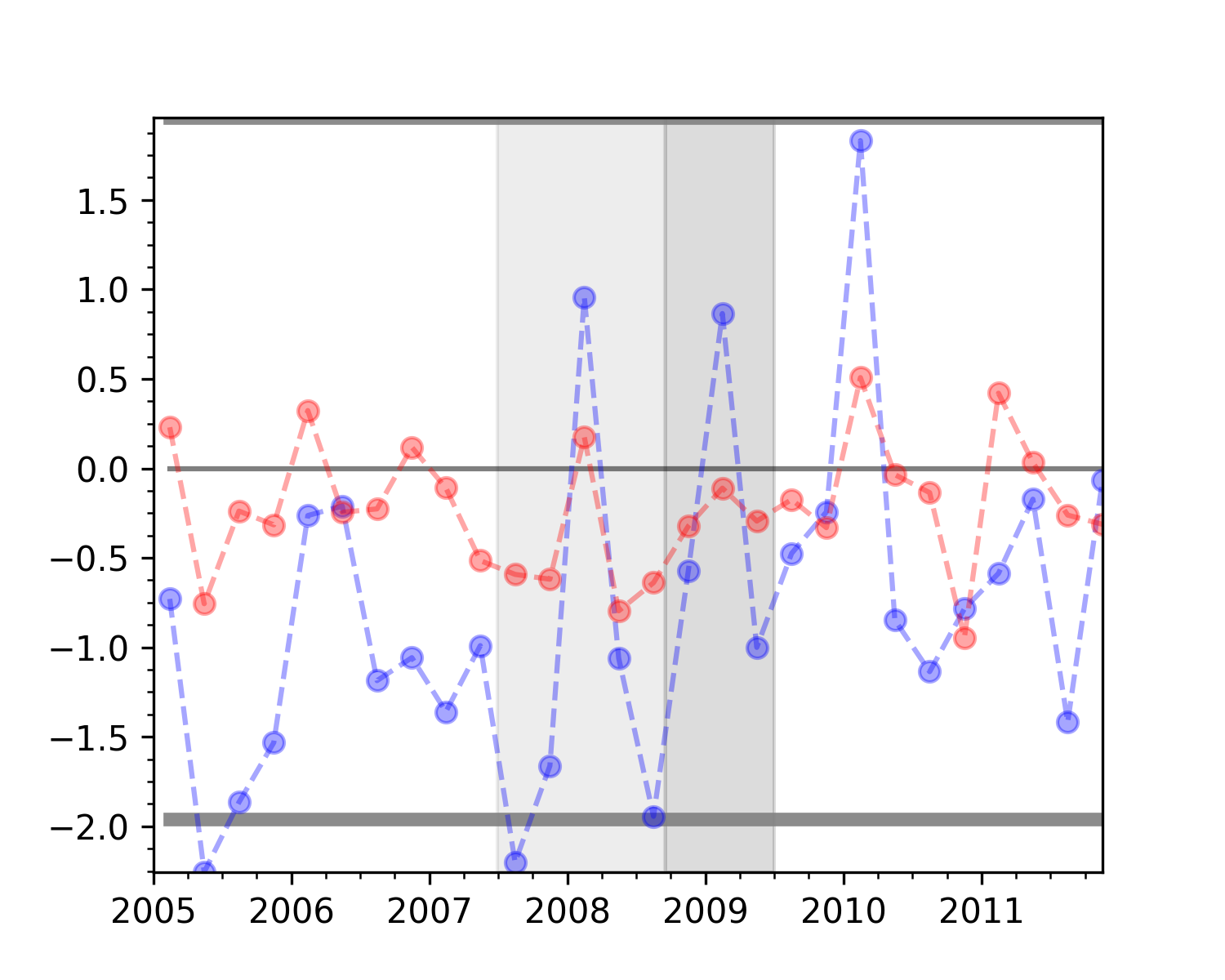}}
  \subfloat{\includegraphics[scale=0.3]{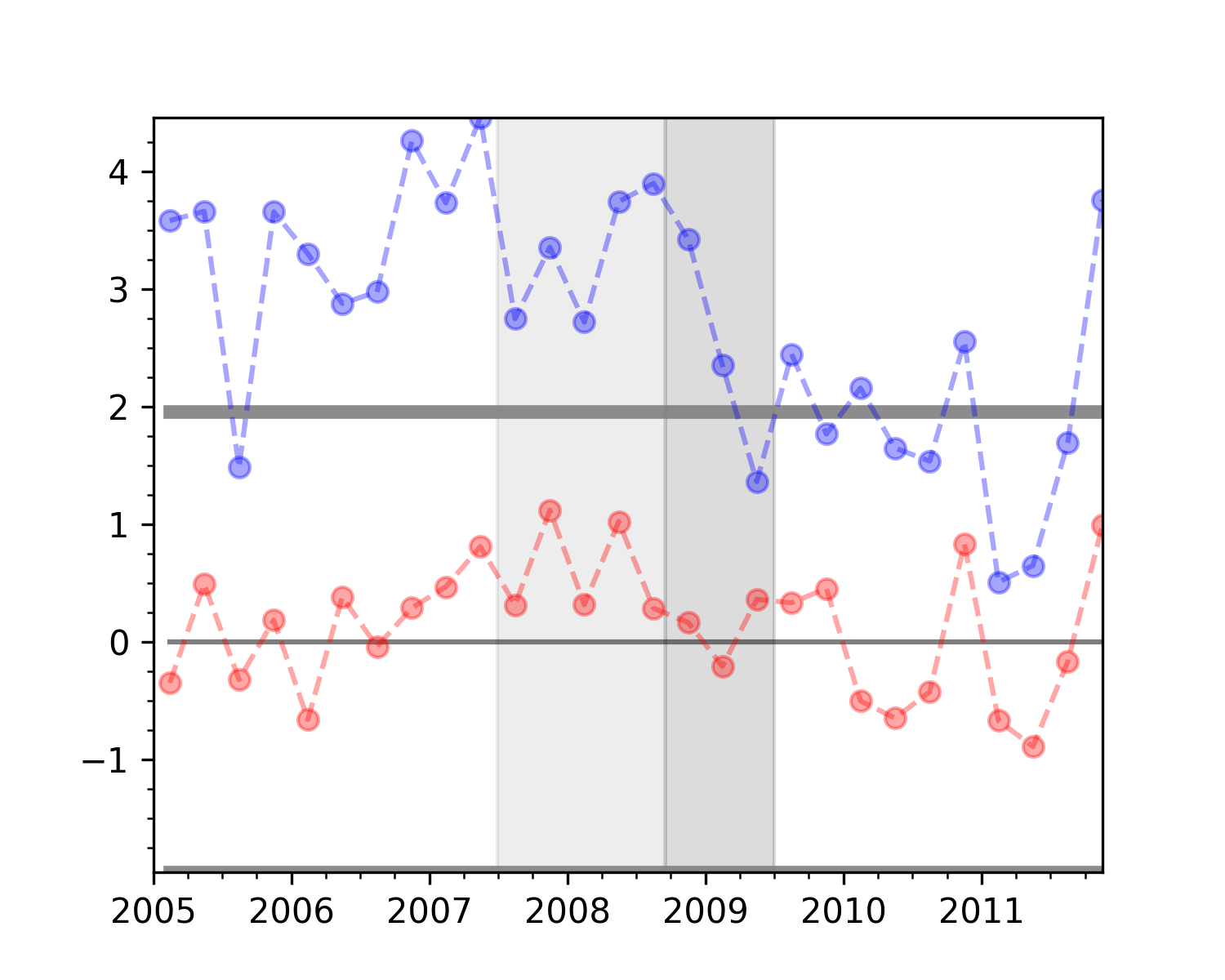}}
  \hfill\vline\hfill
  \subfloat{\includegraphics[scale=0.3]{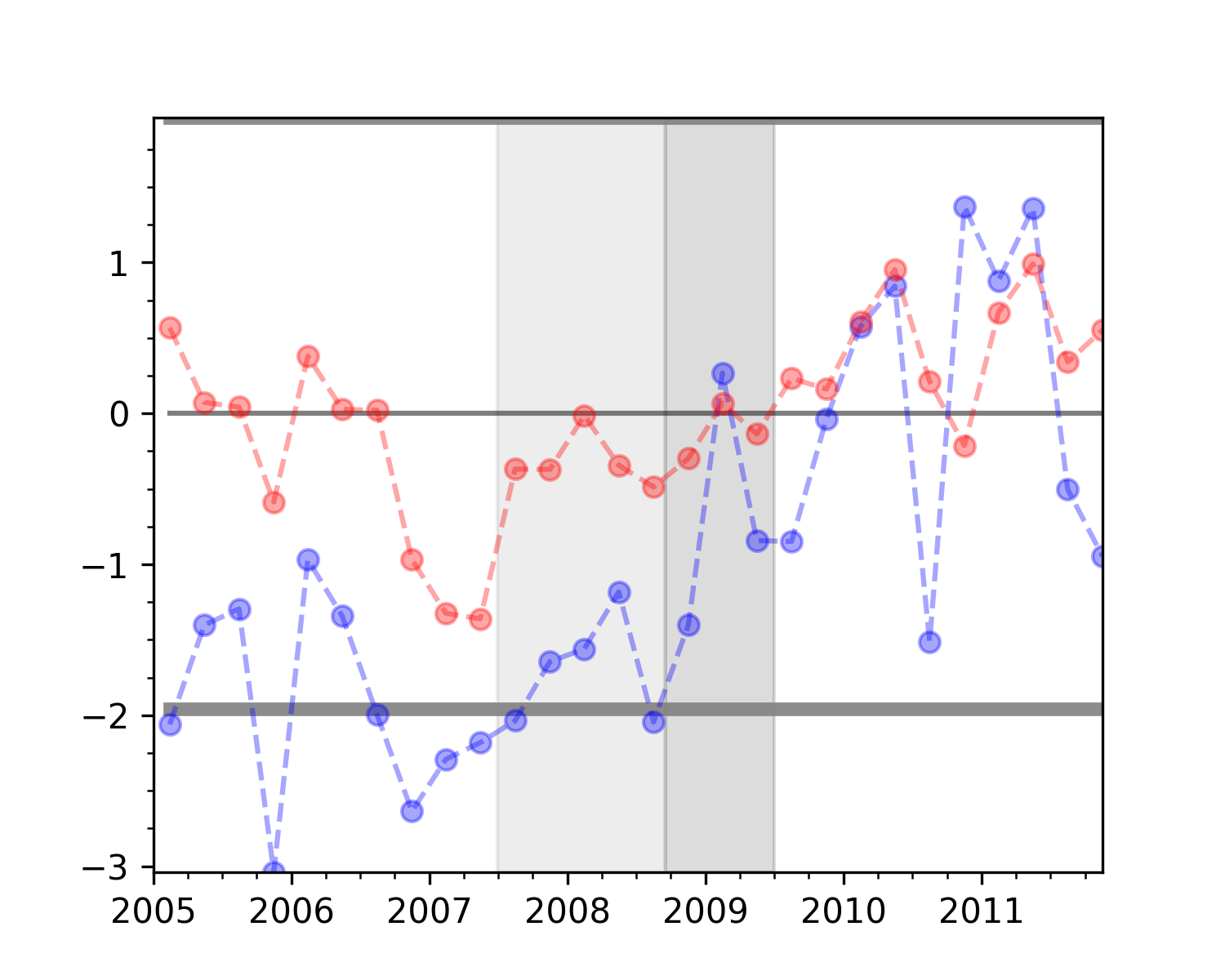}}
  \subfloat{\includegraphics[scale=0.3]{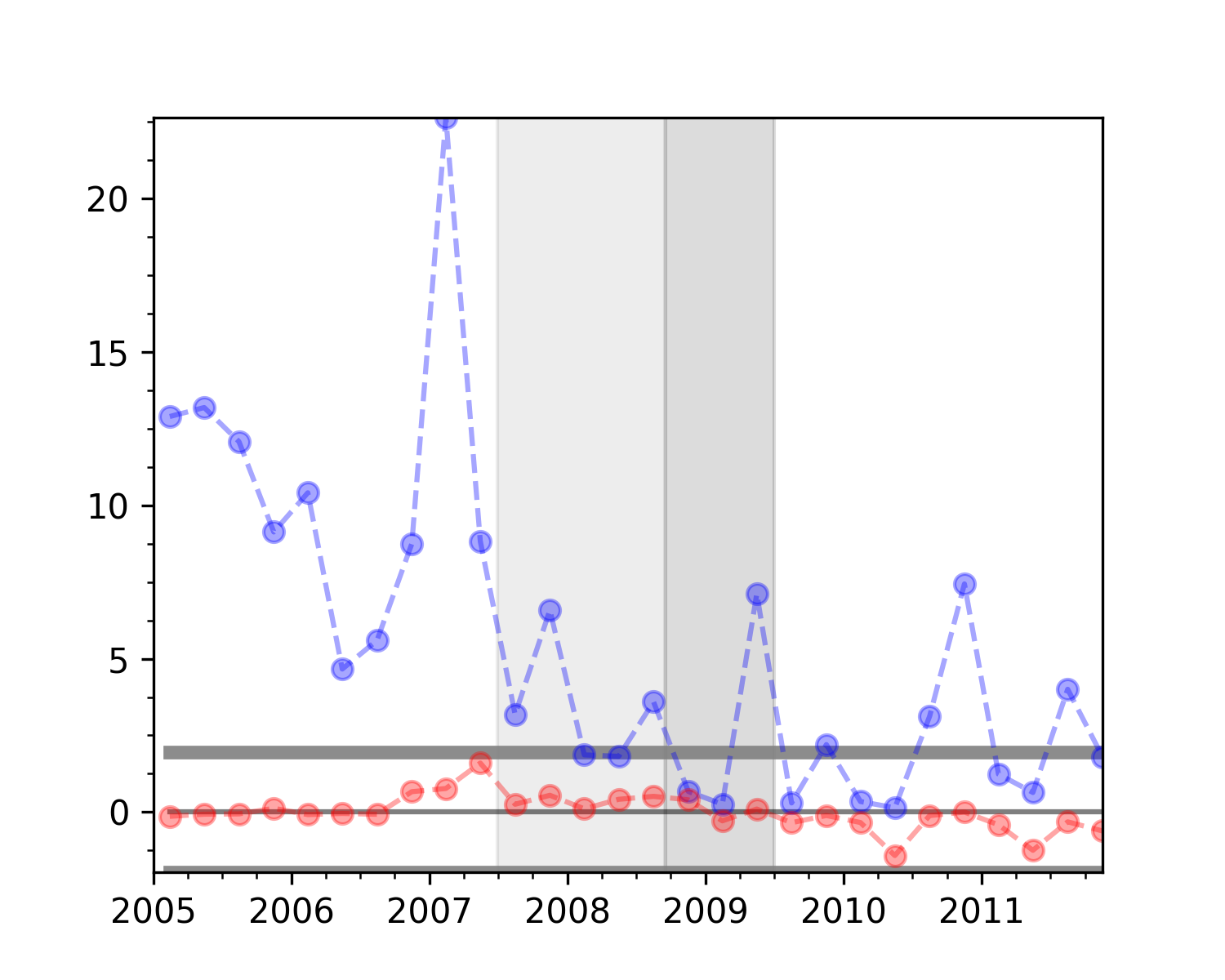}}
  \vspace{-0.4cm}
  \\
  \subfloat{\includegraphics[scale=0.3]{graphics/cpreciprocity_zscore_ts_rec/meso_cppc_unweighted_bothreciprocity.png}}
  \subfloat{\includegraphics[scale=0.3]{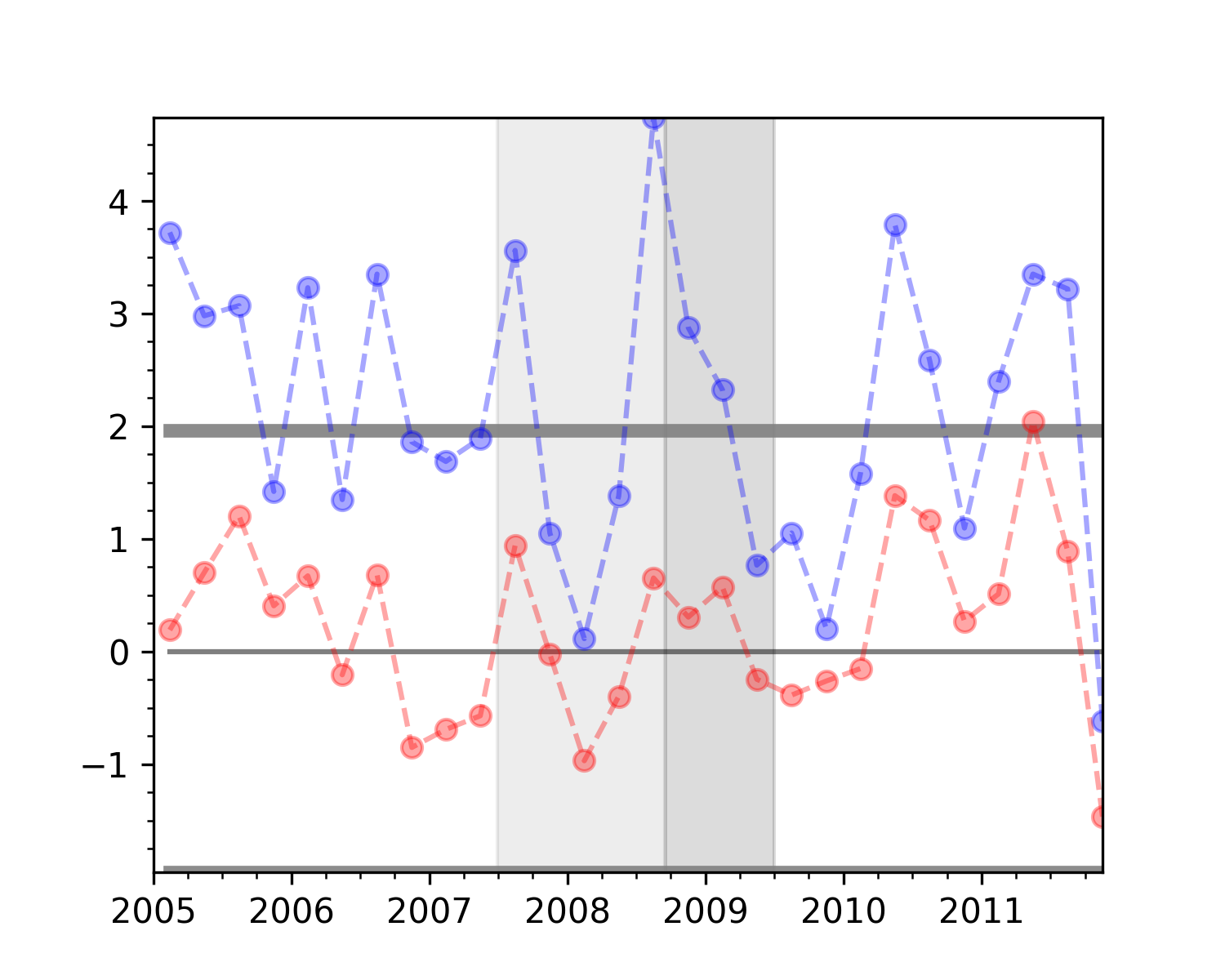}}
  \hfill\vline\hfill
  \subfloat{\includegraphics[scale=0.3]{graphics/cpreciprocity_zscore_ts_rec/meso_cppc_weighted_bothreciprocity.png}}
  \subfloat{\includegraphics[scale=0.3]{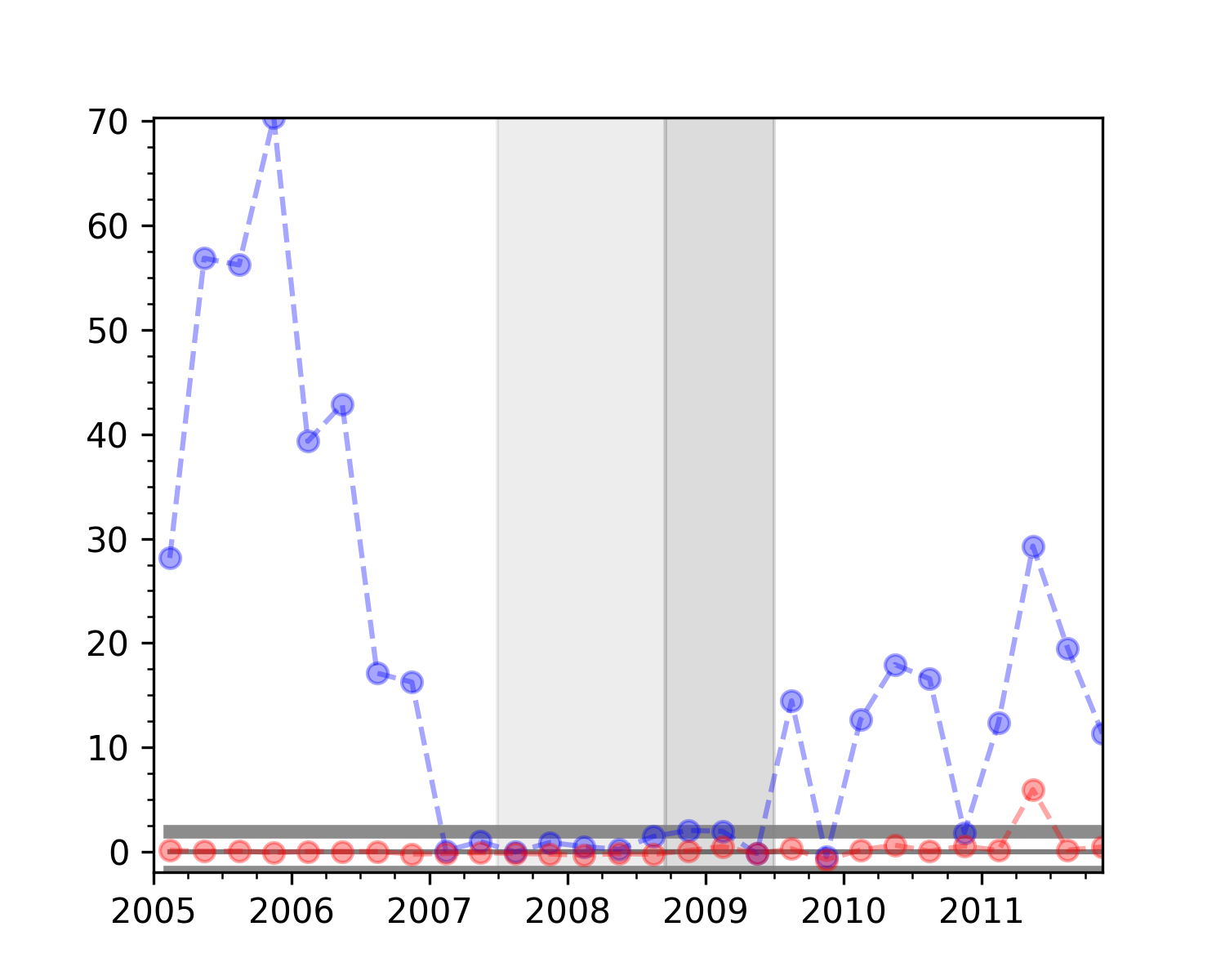}}
  \caption{\textit{Core-Periphery Decomposition (z-scores): Total Reciprocity - Unweighted (left) and Weighted (right).} Upper left panel: Core-Core. Upper right panel: Core-Periphery. Lower left panel: Periphery-Core. Lower right panel: Periphery-Periphery. DECM-filtered empirical series are shown in blue, RECM-filtered empirical series in red.
  }
  \label{fig:ts_meso_unweighted_reciprocity}
\end{figure}

Figure (\ref{fig:ts_meso_unweighted_reciprocity}) indeed reveals that block-based total reciprocities display sizeable deviations, however only for the DECM.
The RECM is in fact able to match not only the un+weighted reciprocity structure over each node's neighborhood by definition but also for the differently partitioned sub-neighborhoods.
While the DECM already underestimated the empirical reciprocity values on the network level, we now observe that this fact is driven by the presence of a periphery bank.
For both, unweighted and weighted total reciprocal relationships, positive DECM deviations appear exclusively in within-periphery and (symmetric) core-periphery blocks, but not in within-core blocks whose results turned out insignificant most of the time.
The latter finding however can be explained by the fact that banks in the core form a tightly connected sub-network, which allows them to quickly find alternative trading partners.
If anything, core banks may have even had a slight distaste for reciprocity in earlier periods with reciprocity values at or below the significance bounds.
As relationship lending is generally thought to be mitigating search and informational frictions \citep{brauning2017relationshiplending}, the incentive to form such long-lasting reciprocal links in the first place is strongly attenuated if there always exist a multitude of trustworthy trading partners.

The findings also the support the idea of core banks as market makers or intermediaries that facilitate transactions as a third-party.
In an ideal core periphery structure, this intermediated trade pattern would in fact be the only possibility for two smaller peripheral banks to trade funds with each other.
Since the DECM preserves the empirical core periphery structure, we can now infer that peripheral banks display a preference towards reciprocal trades beyond the general possibility of core-intermediation.

\medskip

Given the inherent nature of credit risk, (repeated) reciprocal trades may be a form of relationship banking while also a signal that both parties have a history of reliably fulfilling credit obligations.
Such historical information, however, may have become void once the crisis set in, as the ability to handle idiosyncratic risk in normal times may not be a good proxy for facing a large common and possibly cascading shock.
We in fact see the crisis-induced breakdown and subsequent slow return of global weighted reciprocity deviations from the DECM now particularly reflected in the within-periphery block.
The number of within-periphery reciprocal connections, on the other hand, has remained relatively stable at least until the Lehman default.
It is thus reciprocated volume per relationship that peripheral banks have reduced in crisis times, reverting back to higher values once counterparty risk had returned to normal levels.\footnote{
    Since quarterly aggregation of trades implies that high trading volume could equally well be the result of a few high-value trades or many low-value trades, the effect of reduced interaction as opposed to lower volume of funds per trade is not obvious.
    In a separate analysis based on trading count data we have checked however that it is largely due to the former effect:
    Banks tend to engage less often in counter-directed trades within these crisis quarters.
    This also highlights why the unweighted statistics display only minor deviations from the model-based prediction while the weighted statistics are able to capture the peripheral behavior of reducing interaction but not dissolving these relationships entirely.
}

\medskip

Large core banks in turn enjoy two benefits that peripheral banks generally do not possess:
On the lending side, they are able to better diversify their portfolio due to the larger scale of their operations.
On the funding side, they profit from a possible too-big-too-fail guarantee \citep{stern2004toobig}.
This could explain why reciprocity seemed to play only a minor role for them but more so for peripheral banks.

\subsection{3-Paths: Triadic Motifs}
For complex networks which exhibit features like small world properties or slowly decaying degree distributions, \cite{milo2002networkmotifs} suggest to focus on sub-graphs of the entire graph in order to understand these high-level structural properties.
They focus in particular on all non-isomorphic sub-network structures of three connected nodes in various networks over different fields of research.\footnote{
  Triadic motifs, in particular, have already been introduced earlier in the sociological literature \citep{holland1970amethod, holland1971transitivityin}.
}
Out of these \textit{triadic motifs} the authors found similar types of motifs to be overrepresented compared to a random null model in networks which share a common functionality, e.g. neurobiological neural networks, electrochemical gene regulation networks or technological electric circuit networks (which all deal in some sense with information processing).
Networks that served a different purpose, e.g. ecological food webs, in turn contained normal levels of said motifs while displaying higher abundance of (food)chain-like motifs.

\begin{figure}[H]
  \centering
  \includegraphics[scale=0.6]{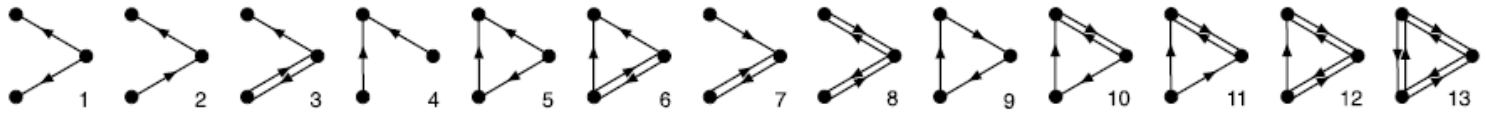}
  \caption{\textit{13 Isomorphism classes of connected Triadic Motifs. Source: \cite{squartini2013earlywarningsignals} and \cite{milo2002networkmotifs}}
  }
  \label{fig:illustration_all_motifs}
\end{figure}

Figure (\ref{fig:illustration_all_motifs}) illustrates the thirteen possible types (up to relabeling of nodes) of such connected triplets.
In the previous sections on reciprocity and core periphery partitions, the DECM turned out to be a useful benchmark model as it preserved direct connection patterns such as nodal strength and degree (given by direct connections, i.e.
paths of length one) and thus the core periphery structure.
Because of that we were able to use it in order to investigate whether reciprocity, as paths of length two, were just a mere reflection of these lower-order properties which turned out not to be the case.
In this section we are following \cite{milo2002networkmotifs} in analyzing triadic motifs, i.e. paths of length three. As our RECM preserves the DECM properties in conjunction with reciprocity, we have the advantage of being able to control for the lower-order paths in the analysis of these three-node motifs.

\medskip

In this respect, \cite{squartini2013earlywarningsignals} performed a similar analysis with interbank networks on the more opaque Dutch Over-the-Counter market.
They apply an unweighted version of our RECM as null model and found a significant under-representation of the circular motif 9 in quarters prior to the Lehman default.
Using our dataset on the electronic market for interbank deposits in Italy, we are going to extend their analysis in three ways.
First, our sample encompasses the entire crisis period as well as the later recovery quarters afterwards.
Second, we apply a configuration model (RECM) that is able to control for volume effects which in turn, as the third point, allows us to investigate the interbank network as a weighted instead of just a binary network.

\begin{center}
  \begin{table}[H]
    \begin{tabular}{ c | c | c}
      $x$ &Unweighted Motif $x$ &Weighted Motif $x$\\
      \hline\hline
          1
          &$\sum_{i \neq j \neq k} a_{ij}^{\leftarrow} a_{jk}^{\rightarrow} a_{ki}^{\not\leftrightarrow}$
          &$\sum_{i \neq j \neq k} \sqrt[2]{ w_{ij}^{\leftarrow} w_{jk}^{\rightarrow} w_{ki}^{\not\leftrightarrow} }$
          \\
          2
          &$\sum_{i \neq j \neq k} a_{ij}^{\rightarrow} a_{jk}^{\rightarrow} a_{ki}^{\not\leftrightarrow}$
          &$\sum_{i \neq j \neq k} \sqrt[2]{ w_{ij}^{\rightarrow} w_{jk}^{\rightarrow} w_{ki}^{\not\leftrightarrow} }$
          \\
          3
          &$\sum_{i \neq j \neq k} a_{ij}^{\leftrightarrow} a_{jk}^{\rightarrow} a_{ki}^{\not\leftrightarrow}$
          &$\sum_{i \neq j \neq k} \sqrt[2]{ w_{ij}^{\leftrightarrow} w_{jk}^{\rightarrow} w_{ki}^{\not\leftrightarrow} }$
          \\
          4
          &$\sum_{i \neq j \neq k} a_{ij}^{\not\leftrightarrow} a_{jk}^{\rightarrow} a_{ki}^{\leftarrow}$
          &$\sum_{i \neq j \neq k} \sqrt[2]{ w_{ij}^{\not\leftrightarrow} w_{jk}^{\rightarrow} w_{ki}^{\leftarrow} }$
          \\
          5
          &$\sum_{i \neq j \neq k} a_{ij}^{\leftarrow} a_{jk}^{\rightarrow} a_{ki}^{\leftarrow}$
          &$\sum_{i \neq j \neq k} \sqrt[3]{ w_{ij}^{\leftarrow} w_{jk}^{\rightarrow} w_{ki}^{\leftarrow} }$
          \\
          6
          &$\sum_{i \neq j \neq k} a_{ij}^{\leftrightarrow} a_{jk}^{\rightarrow} a_{ki}^{\leftarrow}$
          &$\sum_{i \neq j \neq k} \sqrt[3]{ w_{ij}^{\leftrightarrow} w_{jk}^{\rightarrow} w_{ki}^{\leftarrow} }$
          \\
          7
          &$\sum_{i \neq j \neq k} a_{ij}^{\leftrightarrow} a_{jk}^{\leftarrow} a_{ki}^{\not\leftrightarrow}$
          &$\sum_{i \neq j \neq k} \sqrt[2]{ w_{ij}^{\leftrightarrow} w_{jk}^{\leftarrow} w_{ki}^{\not\leftrightarrow} }$
          \\
          8
          &$\sum_{i \neq j \neq k} a_{ij}^{\leftrightarrow} a_{jk}^{\leftrightarrow} a_{ki}^{\not\leftrightarrow}$
          &$\sum_{i \neq j \neq k} \sqrt[2]{ w_{ij}^{\leftrightarrow} w_{jk}^{\leftrightarrow} w_{ki}^{\not\leftrightarrow} }$
          \\
          9
          &$\sum_{i \neq j \neq k} a_{ij}^{\leftarrow} a_{jk}^{\leftarrow} a_{ki}^{\leftarrow}$
          &$\sum_{i \neq j \neq k} \sqrt[3]{ w_{ij}^{\leftarrow} w_{jk}^{\leftarrow} w_{ki}^{\leftarrow} }$
          \\
          10
          &$\sum_{i \neq j \neq k} a_{ij}^{\leftarrow} a_{jk}^{\leftrightarrow} a_{ki}^{\leftarrow}$
          &$\sum_{i \neq j \neq k} \sqrt[3]{ w_{ij}^{\leftarrow} w_{jk}^{\leftrightarrow} w_{ki}^{\leftarrow} }$
          \\
          11
          &$\sum_{i \neq j \neq k} a_{ij}^{\rightarrow} a_{jk}^{\leftrightarrow} a_{ki}^{\leftarrow}$
          &$\sum_{i \neq j \neq k} \sqrt[3]{ w_{ij}^{\rightarrow} w_{jk}^{\leftrightarrow} w_{ki}^{\leftarrow} }$
          \\
          12
          &$\sum_{i \neq j \neq k} a_{ij}^{\leftrightarrow} a_{jk}^{\leftrightarrow} a_{ki}^{\leftarrow}$
          &$\sum_{i \neq j \neq k} \sqrt[3]{ w_{ij}^{\leftrightarrow} w_{jk}^{\leftrightarrow} w_{ki}^{\leftarrow} }$
          \\
          13
          &$\sum_{i \neq j \neq k} a_{ij}^{\leftrightarrow} a_{jk}^{\leftrightarrow} a_{ki}^{\leftrightarrow}$
          &$\sum_{i \neq j \neq k} \sqrt[3]{ w_{ij}^{\leftrightarrow} w_{jk}^{\leftrightarrow} w_{ki}^{\leftrightarrow} }$
    \end{tabular}
    \caption{Motif Abundance.}
    \label{tab:calculation_13_motifs}
  \end{table}
\end{center}

The last point, in particular, allows us to formulate a generalization from these unweighted to weighted motifs.
Table (\ref{tab:calculation_13_motifs}) shows in the first column the formulae of \cite{squartini2013earlywarningsignals} for calculating the abundance of unweighted motifs using the notation for unweighted (non-)reciprocated links from the previous sections.
We generalize this idea by making use of a general formula by \cite{onnela2005intensityand} for enumerating sub-graphs and combine it with the concepts of uni- and bidirectional links from the triadic motifs of \cite{squartini2013earlywarningsignals}.
While in the first column we are counting how many paths of a given motif can be found in the network, in the second column we are going to add up (volume)-weighted paths.
As we are multiplying trading volumes along a single path, we normalize dimensionality by taking an appropriate root (square if two uni- or bidirectional links are involved, and cubic in case of three such links).
This allows proper scaling units in the summation while still guaranteeing equivalence with the unweighted notion in case all edge weights ($w_{ij} \in \{0,1\}$) and therefore all (non-)reciprocated weights happen to be binary.

\medskip

Both definitions are counting paths only up to a combinatorial factor that depends on the symmetries of the respective motive. Fortunately, our z-scores are unaffected by this because this double counting happens both in the empirical as well as in the synthetic networks so that the effects cancel out after normalization and standardization: If we denote by $c$ the combinatorial factor of motif $m$, and by $m^e$ and $m^s$ its empirical and (model-)synthetic number of unique occurrences respectively, then the z-score $z = (cm^e - \overline{cm^s}) / \sqrt{ \sigma(cm^s)^2} = (m^e - \overline{m^s}) / \sigma(m^s)$ with model standard deviation $c\sigma(m^s)$ remains unaffected by this scaling coefficient.

\medskip

\begin{figure}[H]
  \centering
  \includegraphics[scale=1.0, trim = 0cm -1cm 0cm 0cm, clip]{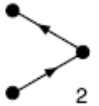}
  \includegraphics[scale=0.4]{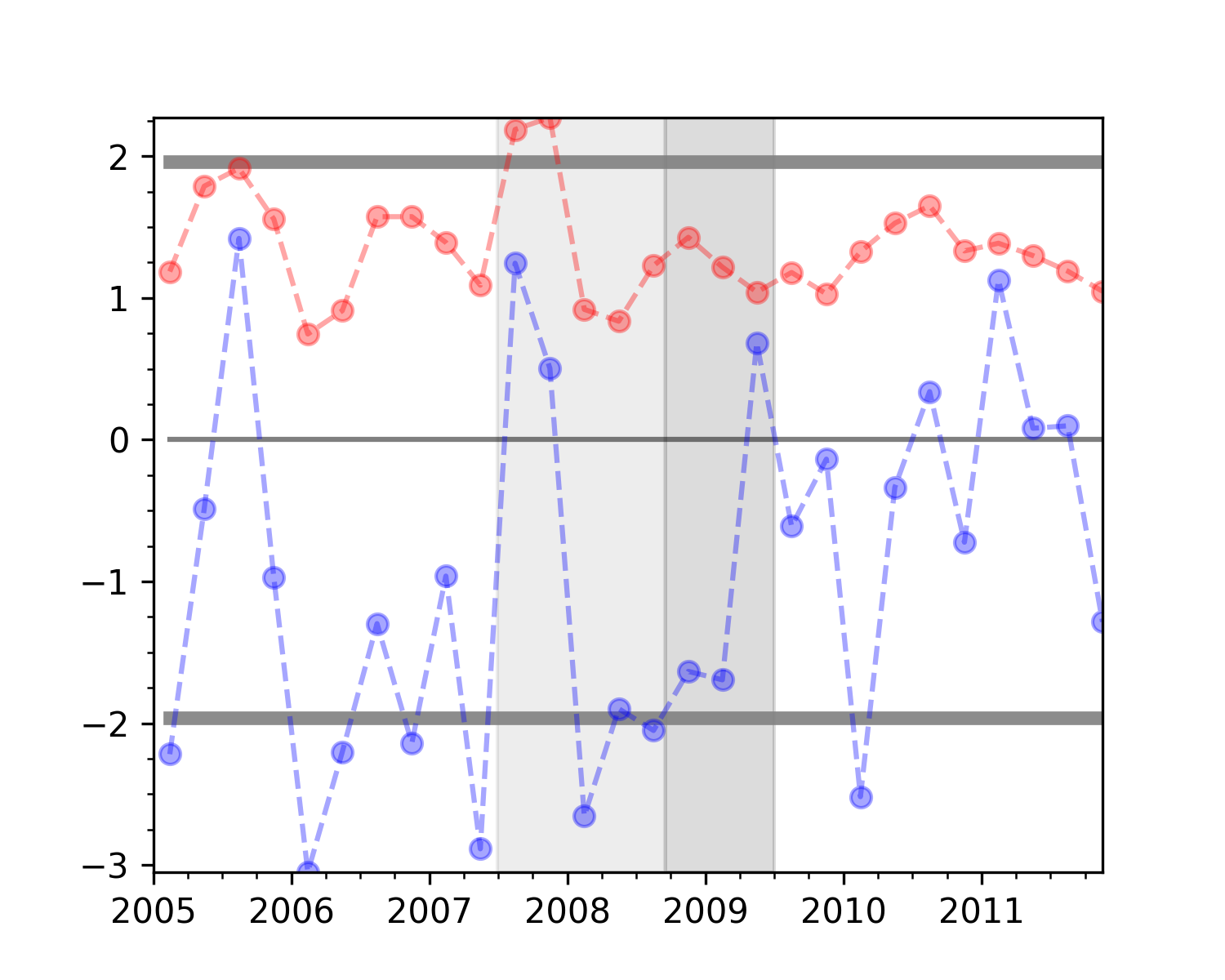}
  \includegraphics[scale=0.4]{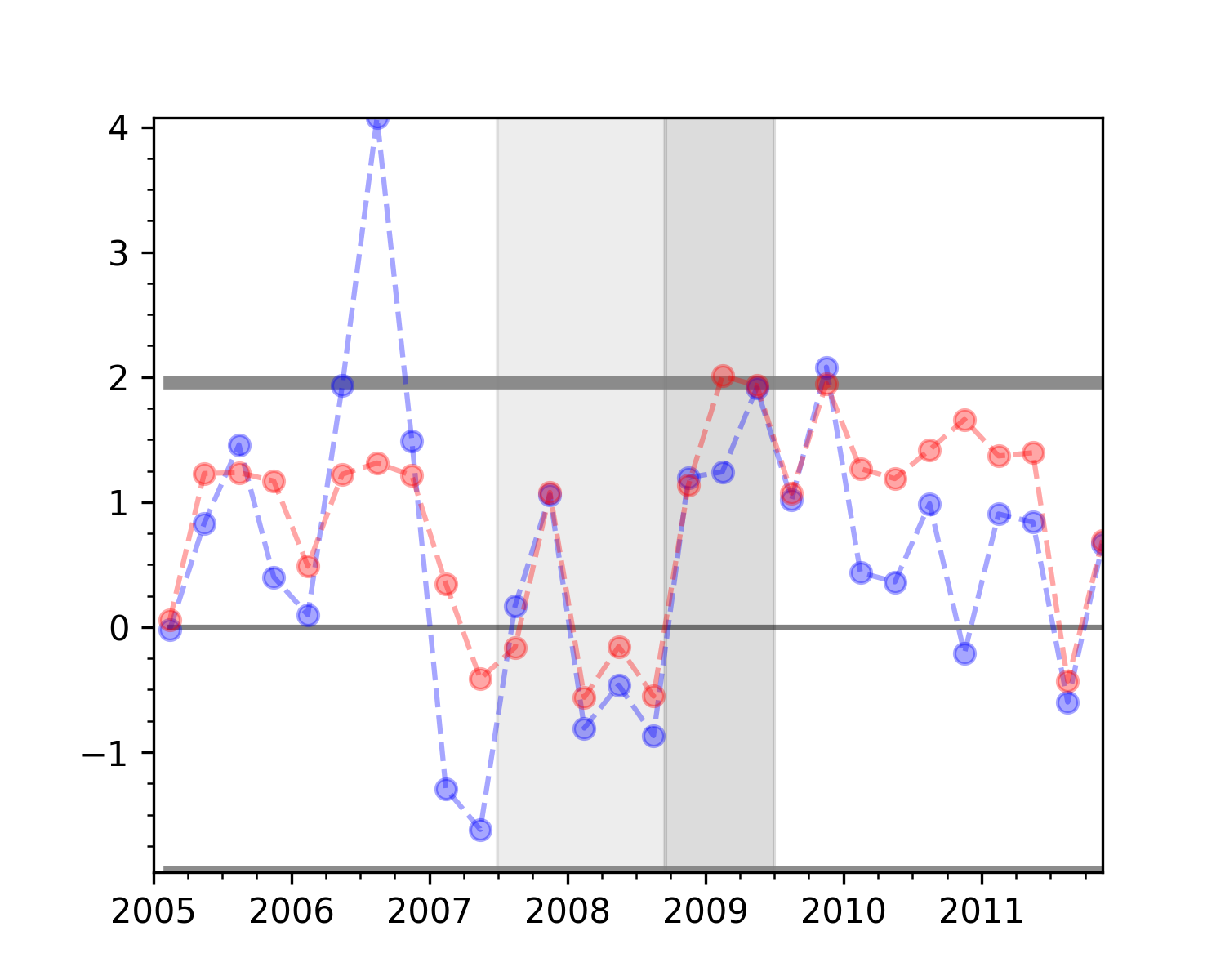}
  \\
  \includegraphics[scale=1.0, trim = 0cm -1cm 0cm 0cm, clip]{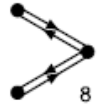}
  \includegraphics[scale=0.4]{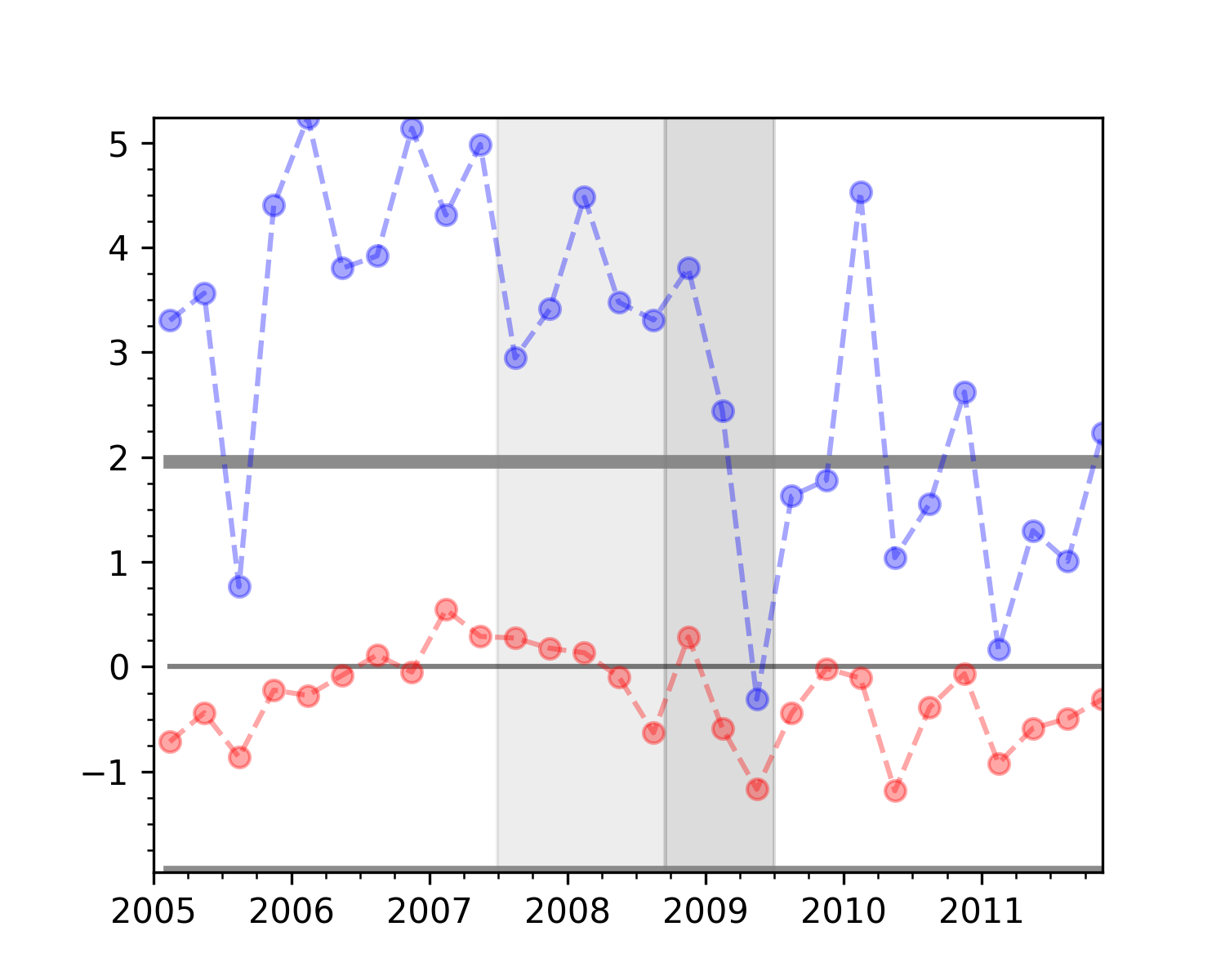}
  \includegraphics[scale=0.4]{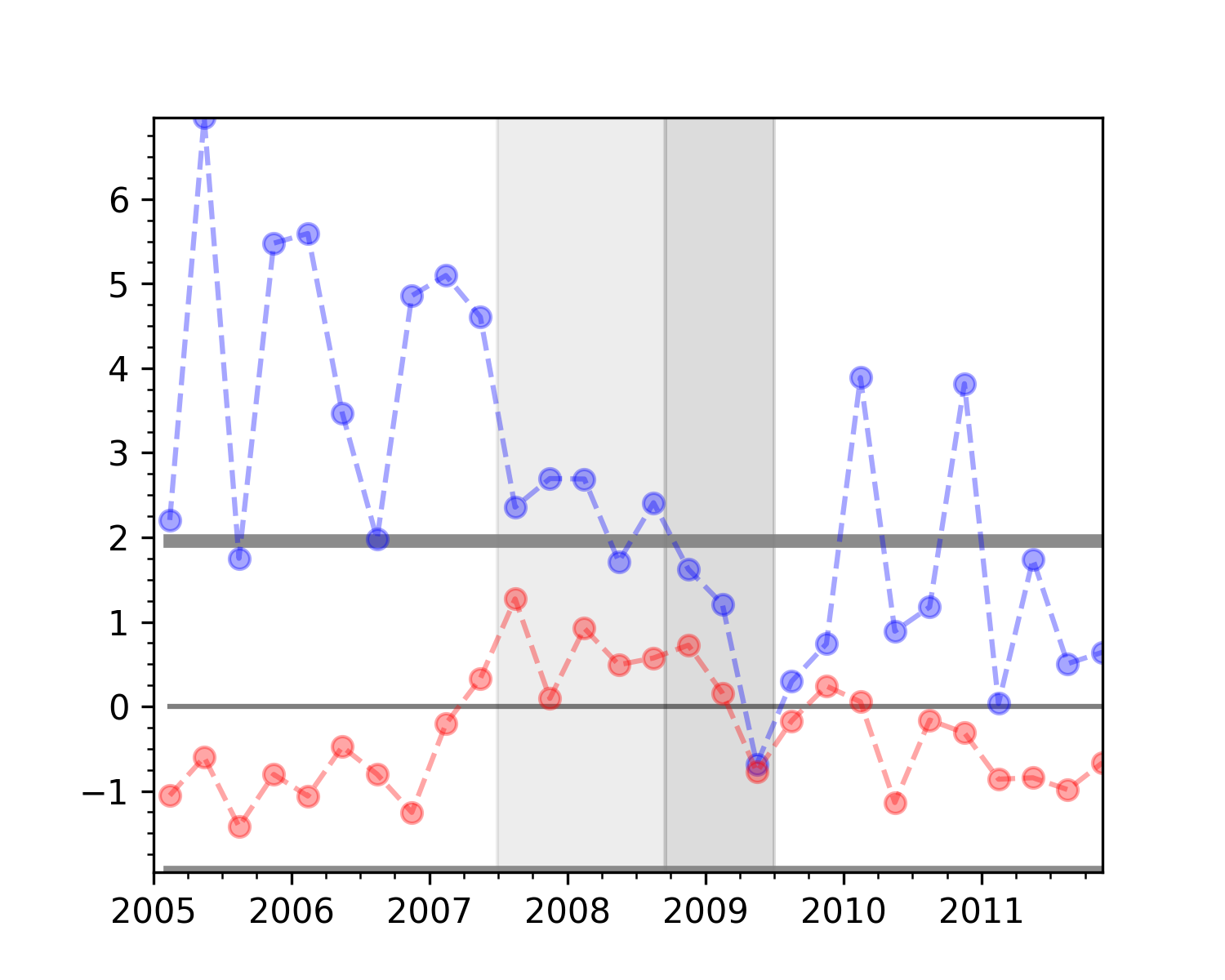}
  \caption{\textit{Abundance z-score - Motif 2\&8. Left panel: Unweighted. Right panel: Weighted.} DECM-filtered (RECM-filtered) empirical series are shown in blue (red).
  }
  \label{fig:ts_motif_2_8}
\end{figure}

The top panel of Figure (\ref{fig:ts_motif_2_8}) illustrates the z-scores of DECM and RECM for the second motif.
This intermediation motif, $X \rightarrow Y \rightarrow Z$, is mostly driven in a core periphery structure by the constellation of one core bank $Y$ intermediating trades between two peripheral banks $X,Z$.
As can be seen in the top right graphic, the weighted version is well matched by both type of models.
While the unweighted version also happens to be decently represented, the empirical data seems to display slightly fewer of these intermediation chains than predicted by the DECM in the first half of the sample.
This is somewhat surprising as one would expect intermediation a central property in such hierarchical networks and thus well captured by the core-periphery-preserving DECM. We should stress however that the magnitude of these deviations remains fairly small and irregular.

Looking at this finding from a different perspective, if one takes a look at the bottom of the figure, one finds motif 8 to be overrepresented in the empirical data compared to the DECM for both unweighted and weighted versions.
This motif can be constructed as a superposition of two converse second motifs, and should be equally well in line with a core periphery structure because a bidirectional chain $P_1 \leftrightarrow C_1 \leftrightarrow P_2$ of two periphery banks $P_1, P_2$ trading via one core bank $C_1$ does not increase the error score in the core periphery optimization.
For any other constellation of core and periphery nodes in this (and the previous) motif the error score would obviously increase.
Indeed most of the empirical trades are occurring in the core-intermediation constellation.
Given that most periphery banks in the empirical e-MID data are in fact net suppliers of funds with core banks being in net demand, both motifs however cannot be considered ideal representations for that fact.
Consequently it seems plausible that the motif with too much symmetry (8) and the motif with too little symmetry (2) are in line with the model predictions if analyzed as joint average, but differ in opposite directions if analyzed separately.

\medskip

Although the RECM preserves for each bank its total number and weight of reciprocal trades, which is based on directly reciprocated trading relationships between two banks, it does neither account for unweighted nor weighted reciprocated trading paths of length larger than one. The empirical count of Motif 8, i.e. second-order reciprocal trading chains, however is precisely captured by the model as opposed to the DECM results. Since the RECM is constructed as enhancement of this core-periphery-preserving DECM with reciprocity structure, we can now infer that second-order reciprocal intermediation chains can be retraced to the combination of (i) the core-periphery hierarchy and (ii) and first-order (or direct) reciprocity.

\medskip

\begin{figure}[H]
  \centering
  \includegraphics[scale=1.0, trim = 0cm -1cm 0cm 0cm, clip]{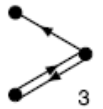}
  \includegraphics[scale=0.4]{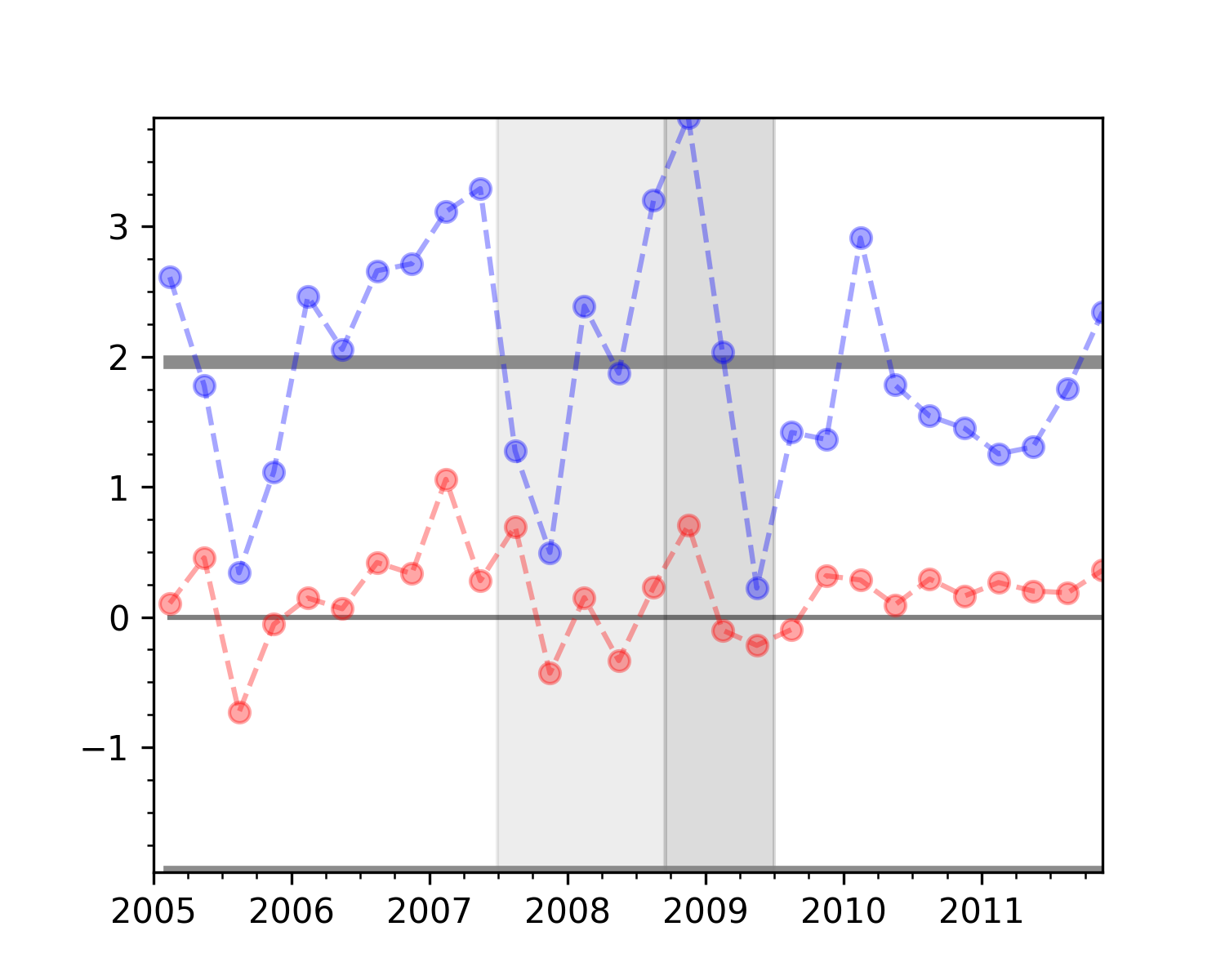}
  \includegraphics[scale=0.4]{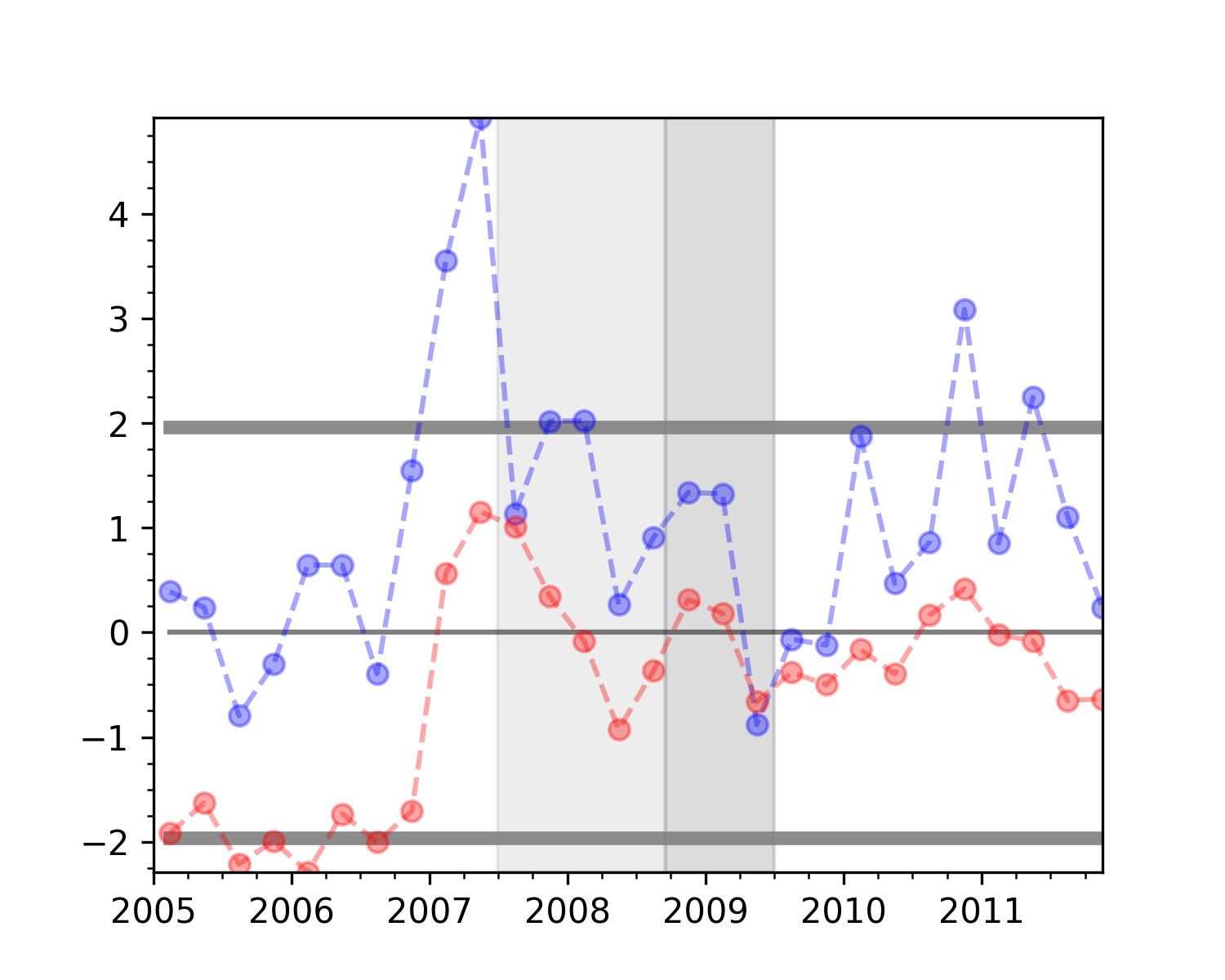}
  \\
  \includegraphics[scale=1.0, trim = 0cm -1cm 0cm 0cm, clip]{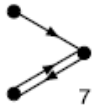}
  \includegraphics[scale=0.4]{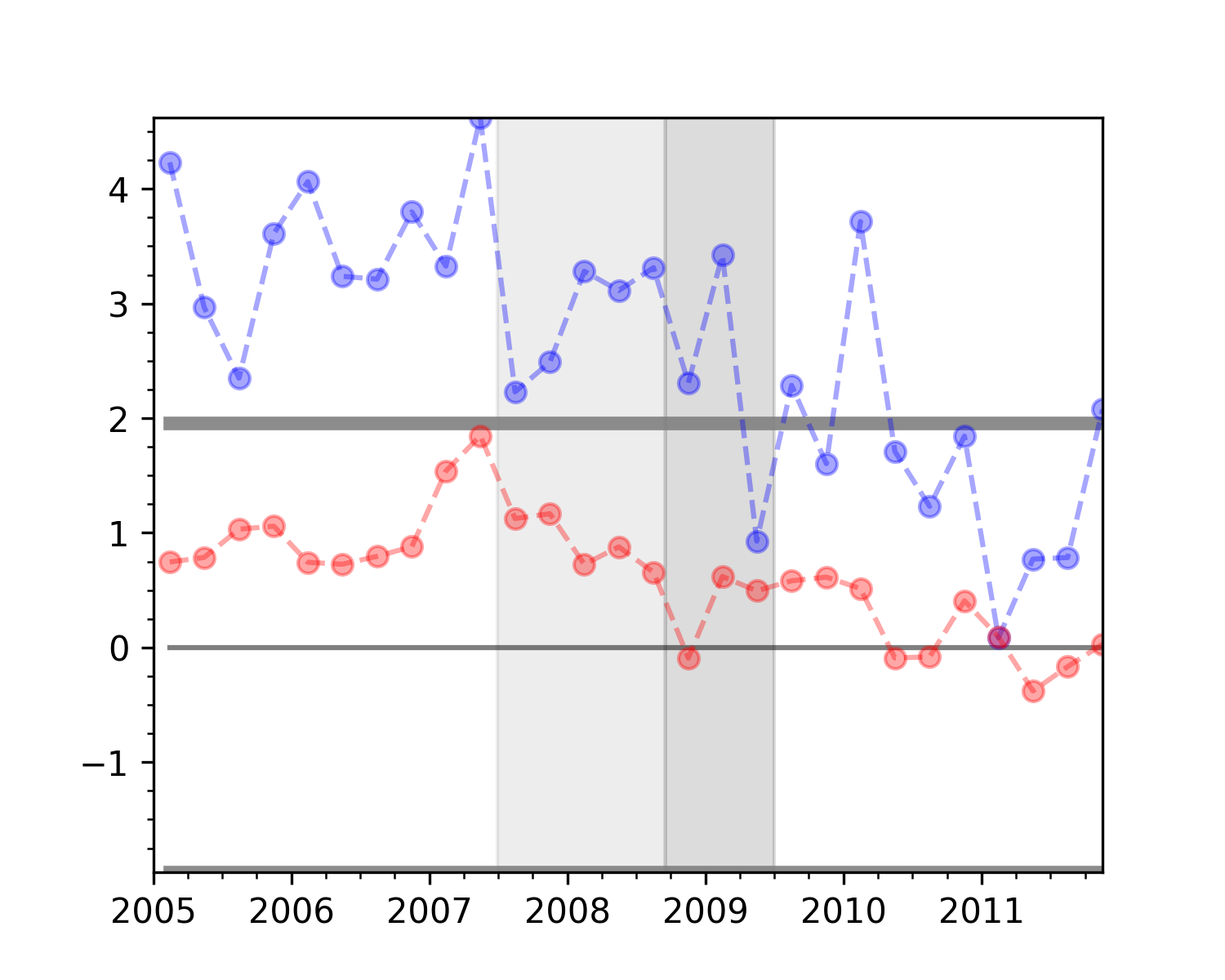}
  \includegraphics[scale=0.4]{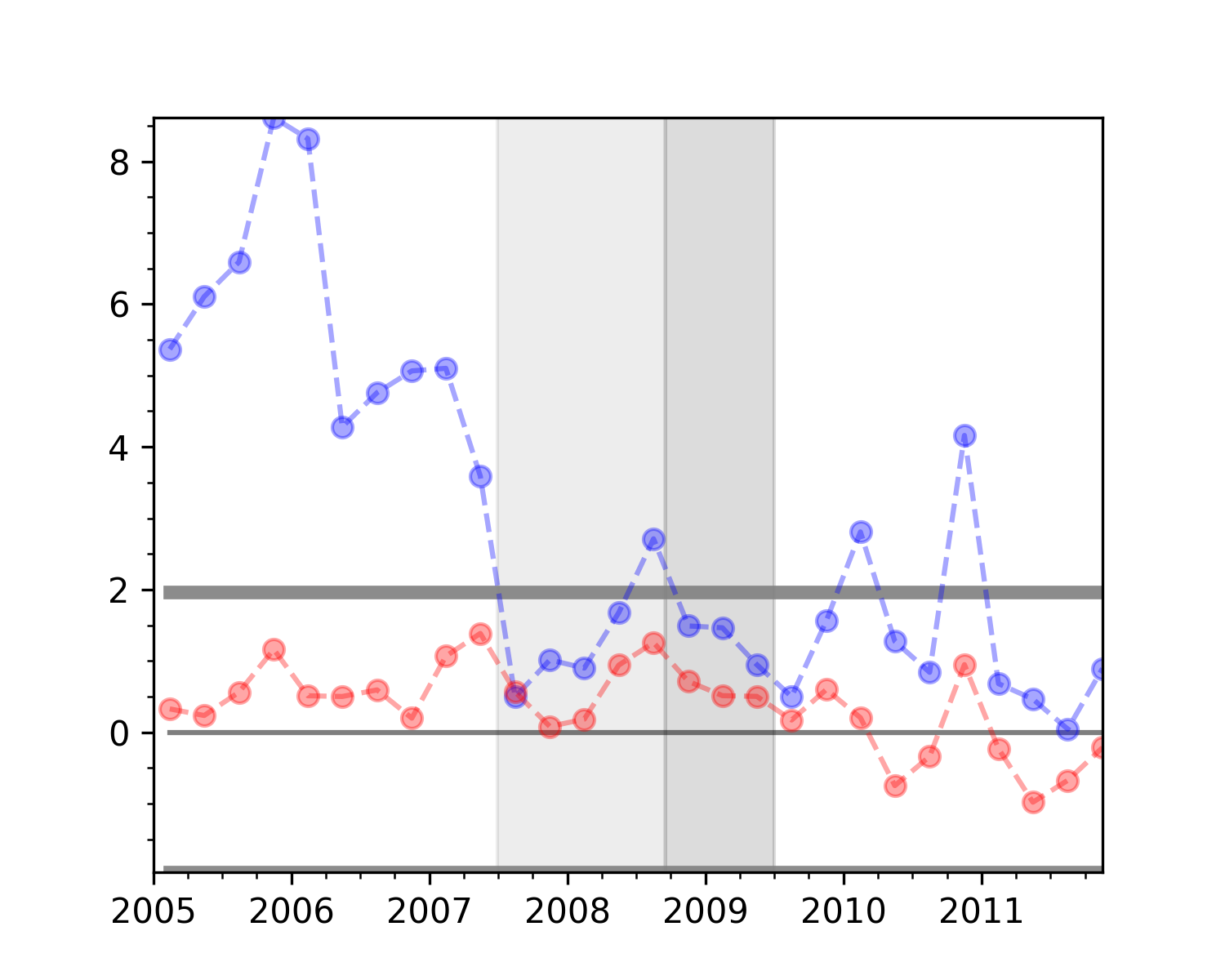}
  \caption{\textit{Abundance z-score - Motif 3\&7. Left panel: Unweighted. Right panel: Weighted.} DECM-filtered (RECM-filtered) empirical series are shown in blue (red).
  }
  \label{fig:ts_motif_7_3}
\end{figure}

Investigating the difference between these two motifs in more detail, in Figure (\ref{fig:ts_motif_7_3}), we deconstruct motif 8 into motifs 3 and 7.
Starting from the pure intermediation motif 2, $X \leftarrow Y \leftarrow Z$, the seventh motif adds a reciprocal link $X \Rightarrow Y$ from the receiving node, whereas in the third motif the intermediating node reciprocates the sending node, $Y \Rightarrow Z$.
This increase in symmetry in turn leads the DECM to underestimate the empirical count of both unweighted motifs similar to the result for the eighth motif.
This effect is more pronounced for motif 7 which also translates into an underestimation of its weighted counterpart which, this time, is not the case for motif 3.

In particular, the time-varying nature of the DECM deviations for motifs 8 and 7 reflect similar dynamics which are reminiscent of what we found for (un+)weighted reciprocity earlier.
Indeed, turning to the reciprocity-preserving RECM, we see that the occurrence of these motifs are in concordance with the model.
Like in motif 8 ($X \leftrightarrow Y \leftrightarrow Z$), for the decision to reciprocate a direct relationship, the indirect relationship is either irrelevant or, if we focus on core-intermediation in motif 7 ($P_1 \leftrightarrow C_1 \leftarrow P_2$) and 3 ($P_1 \leftarrow C_1 \leftrightarrow P_2$), already captured by the core-periphery structure which we control for with the degree sequences.

\medskip

Based on the intermediation motif 2 ($X \leftarrow Y \leftarrow Z$) we could have equally moved towards less hierarchical motifs by giving up the central role of node $Y$.
We constructed motif 7 out of motif 2 by letting $X$ reciprocate its existing trading relationship with $Y$ in case $X$ is in excess supply of funds.
Another viable option for $X$ would have certainly been the establishment of an entirely new direct trading relationship, e.g. with $Z$.
Adding to the second motif the link $X \Rightarrow Z$, instead of $X \Rightarrow Y$ in case of motif 7, yields the cyclical motif 9.
In this motif ($X \leftarrow Y \leftarrow Z \leftarrow X$) each bank is an intermediary in the flow of funds.
From a topological perspective, neither relationship is reciprocated nor severed so that the hierarchical order is completely flat.

\begin{figure}[H]
  \centering
  \includegraphics[scale=1.0, trim = 0cm -1cm 0cm 0cm, clip]{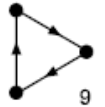}
  \includegraphics[scale=0.4]{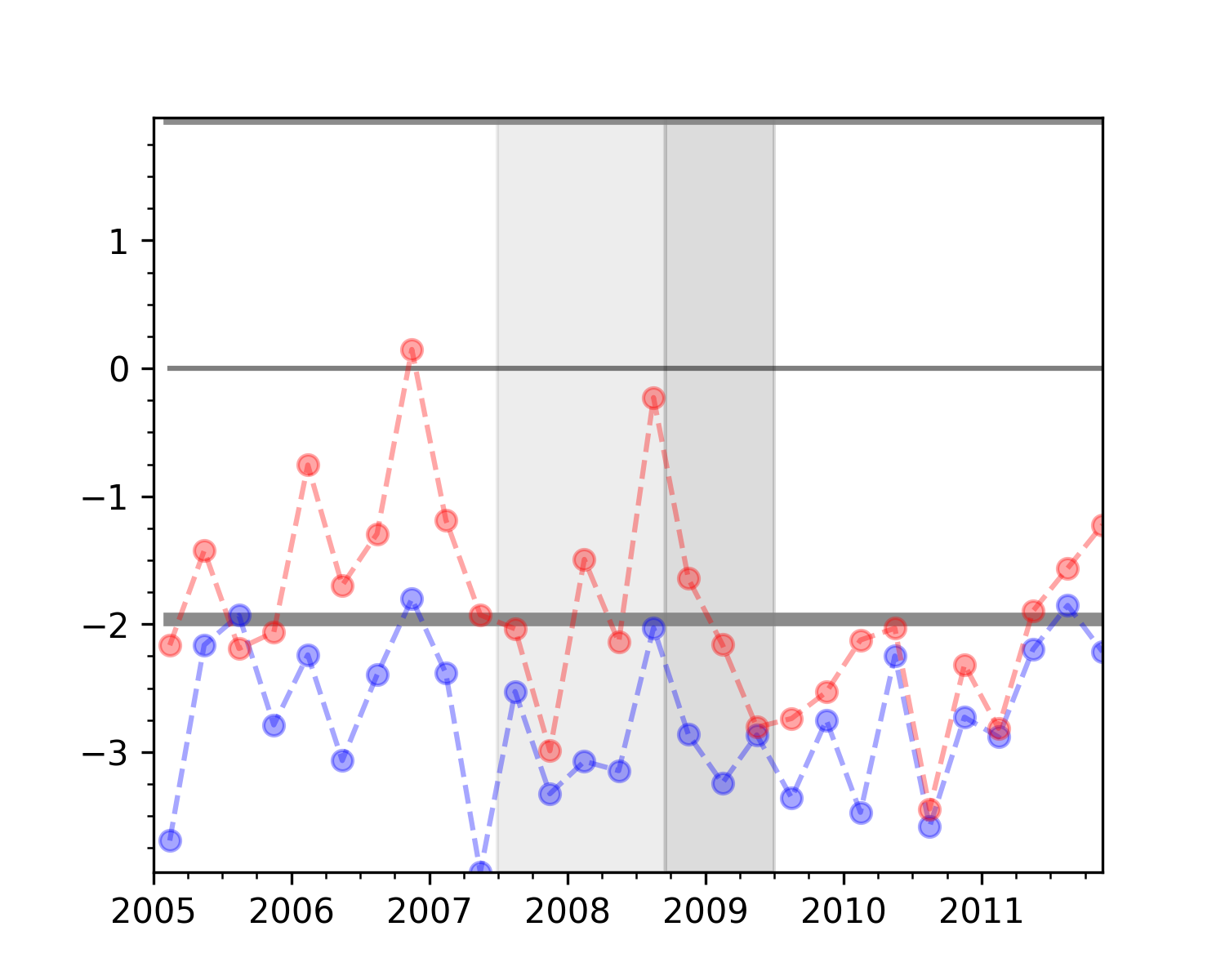}
  \includegraphics[scale=0.4]{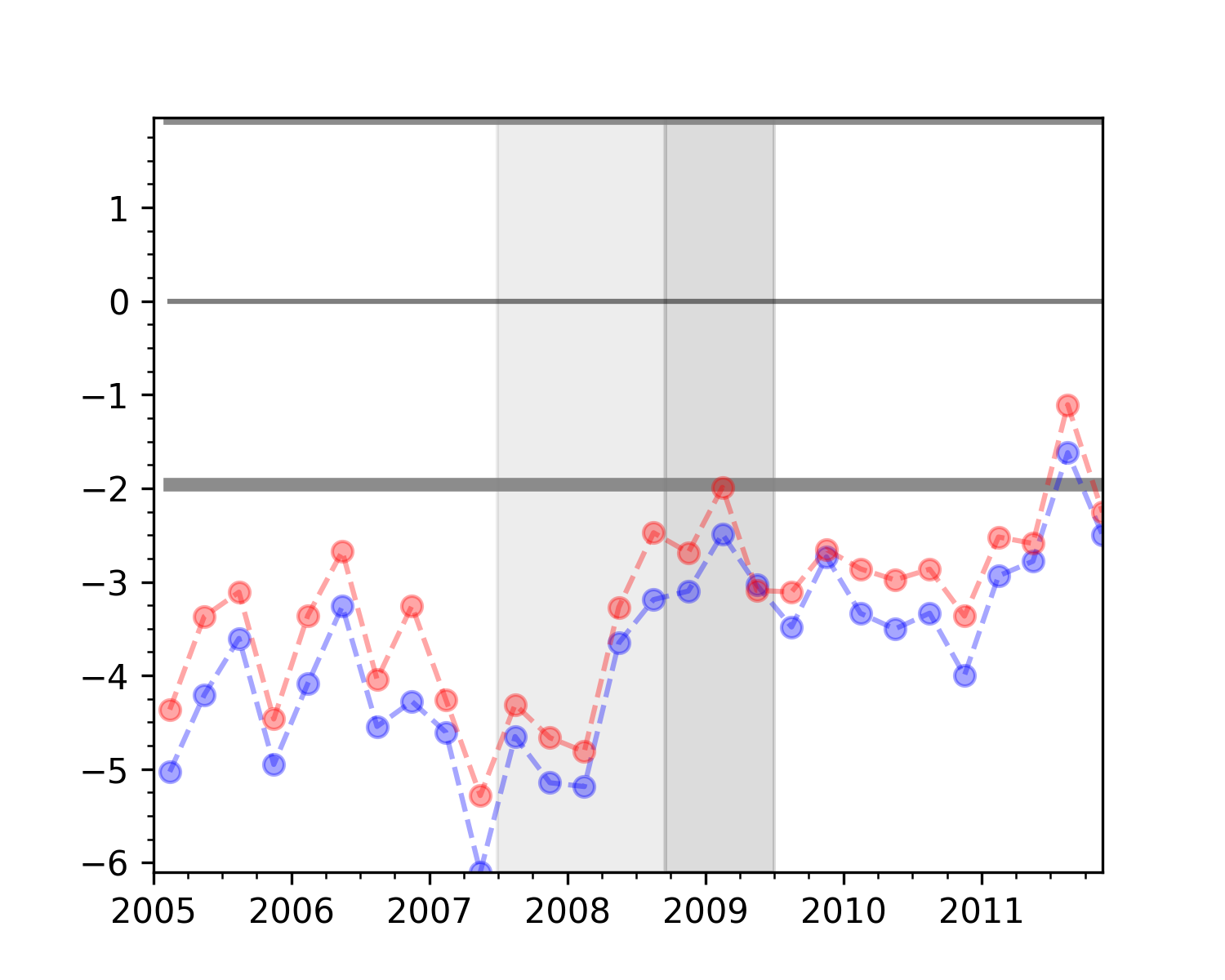}
  \caption{\textit{Abundance z-score - Motif 9. Left panel: Unweighted. Right panel: Weighted.} DECM-filtered (RECM-filtered) empirical series are shown in blue (red).
  }
  \label{fig:ts_motif_9}
\end{figure}

This requirement, however, appears to be too strict from a theoretical perspective for a hierarchical network such as an interbank market driven by a small group of large dealer banks.
Indeed such a configuration would contribute to the core periphery error score regardless of the exact partition of these three nodes into core or peripheral nodes.
It is nevertheless interesting to see that the DECM, which preserves the empirical core periphery structure, still overestimates the number of these cyclical relationships in Figure (\ref{fig:ts_motif_9}).
Banks on the e-MID hence tend to avoid (deliberately or unknowingly) these constellations above and beyond the main topological impact of the core periphery hierarchy.
Controlling for reciprocal relationships in the network weakens this effect but does not completely offset it, as can be seen for the RECM especially in the aftermath of the crisis.
\cite{squartini2013earlywarningsignals} find the same effect for a similar data period in the Dutch interbank market.
We can thus confirm their results for a different dataset.
As both datasets have in common an impartial core periphery structure, we may conjecture that this structure could bring along the general tendency to fewer cyclical trading relationships.

\medskip

Since the topological requirement for motif 9 seems relatively strict for many types of empirical networks, it can be useful to allow for small topological deviations from the cyclical nature e.g. some counter-directed links.
Weighted motifs in general are more robust in this respect as they focus on the asymmetry of weights in bilateral relationship.
This would allow, for instance, the identification of the ninth motif even in a completely connected graph.
In our case of e-MID networks in Figure (\ref{fig:ts_motif_9}), the analysis of weighted motifs now clearly confirms the unweighted tendency of banks' reluctance to engage in cyclical trading relationships.
This motif, in particular, is the first motif whose deviations from model predictions cannot be remedied by controlling for empirical reciprocity structure.

\medskip

Note that the value of weighted motifs in general is more heavily influenced by nodes with high-value relationships.
In case of motif 9, it should thus capture triadic asymmetric cycles mostly among (two or three) core banks.
This interpretation is also in line with the regime shift of the series that happens at the transition from the first to the second phase of the crisis as this is also the time when the number of core connections drastically falls leading to a smaller core with a higher error score.

\begin{figure}[H]
  \centering
  \includegraphics[scale=1.0, trim = 0cm -1cm 0cm 0cm, clip]{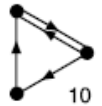}
  \includegraphics[scale=0.4]{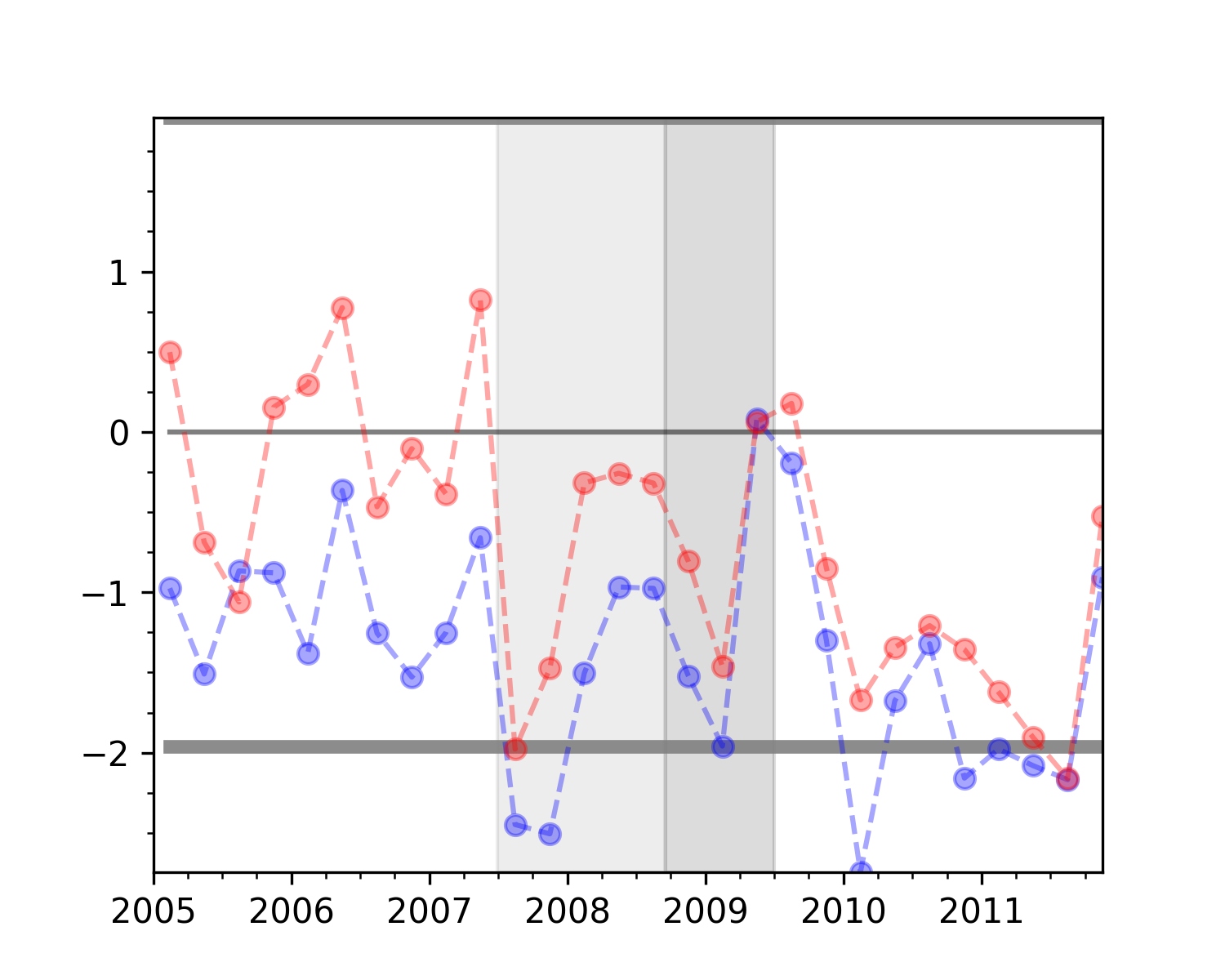}
  \includegraphics[scale=0.4]{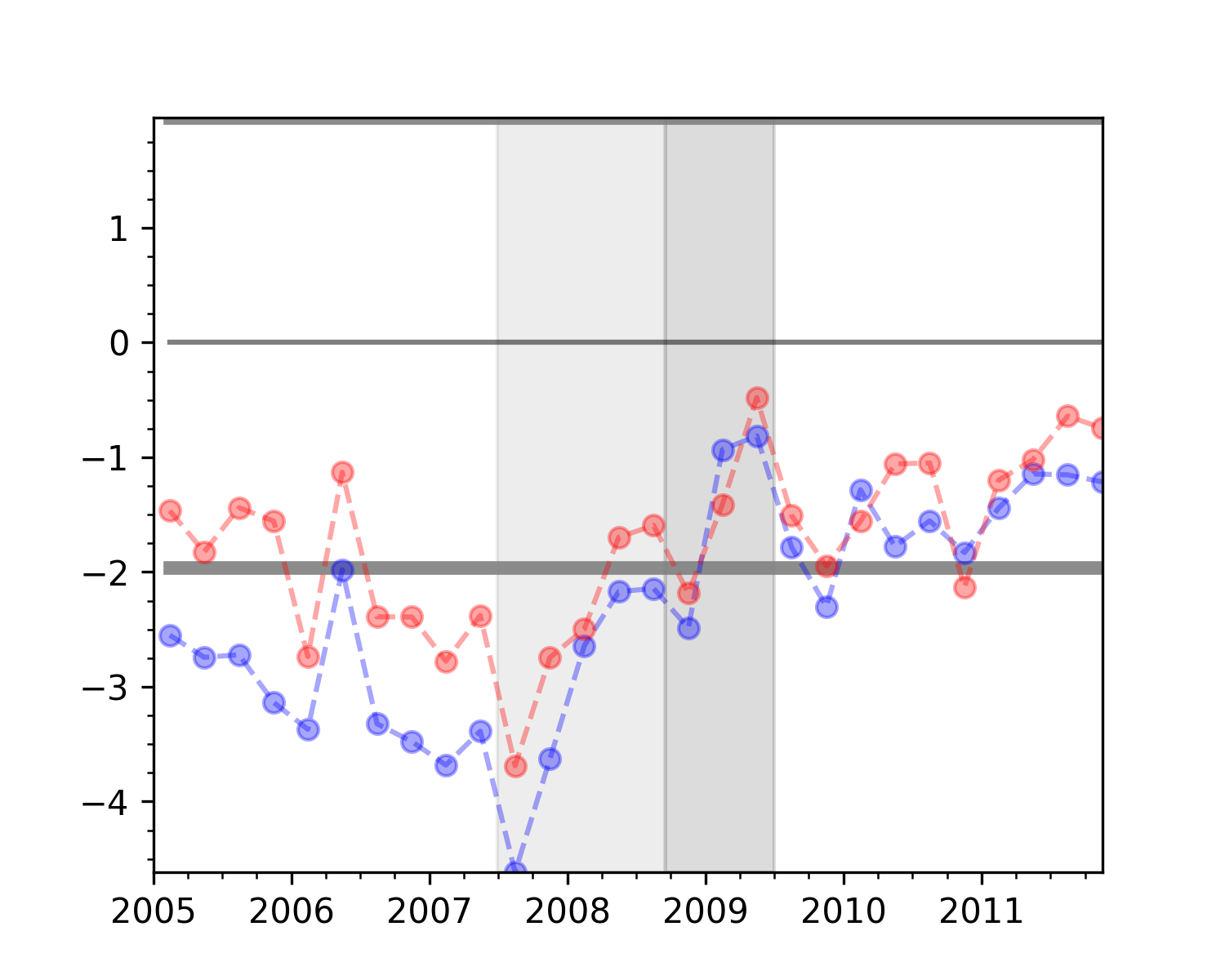}
  \\
  \includegraphics[scale=1.0, trim = 0cm -1cm 0cm 0cm, clip]{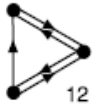}
  \includegraphics[scale=0.4]{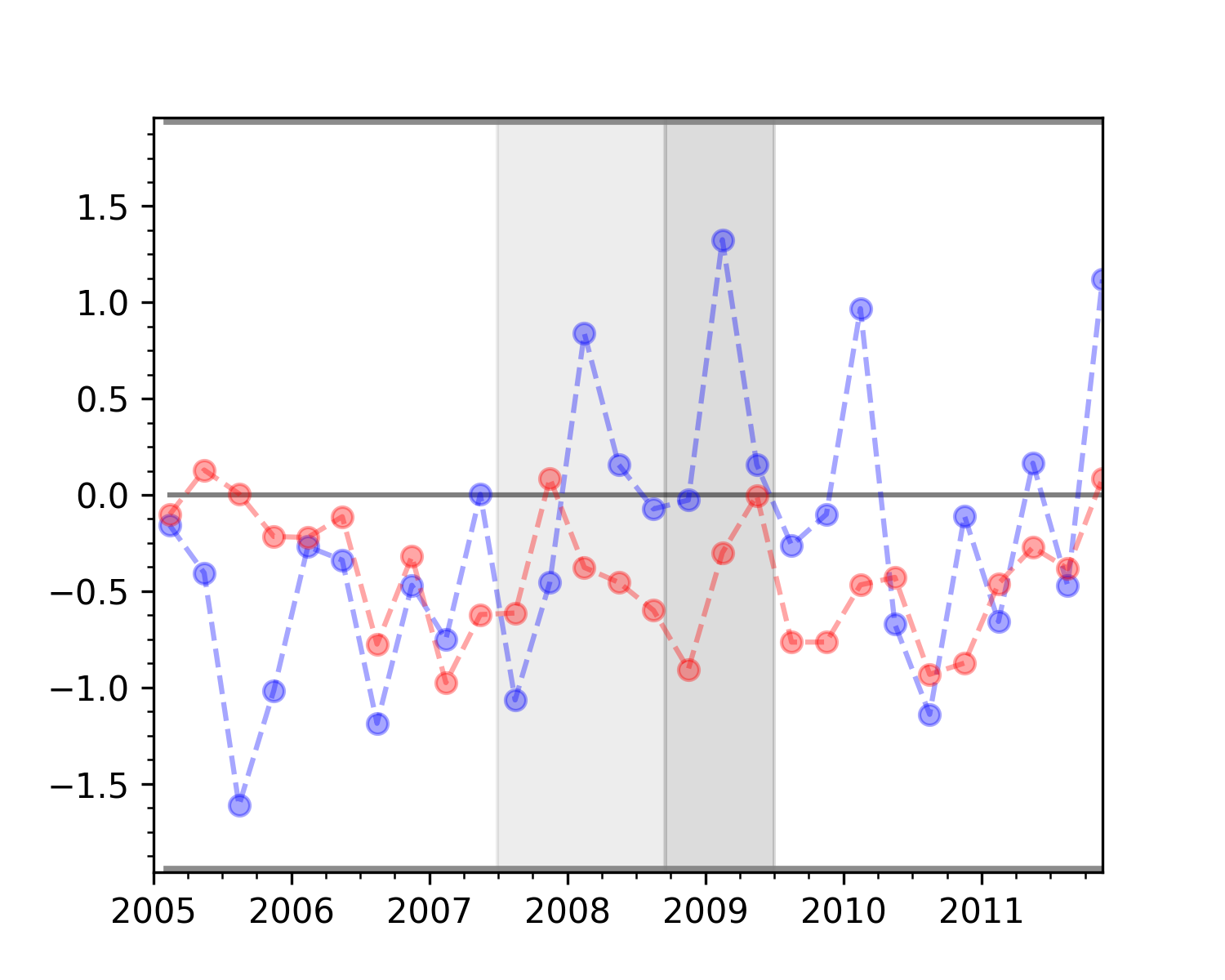}
  \includegraphics[scale=0.4]{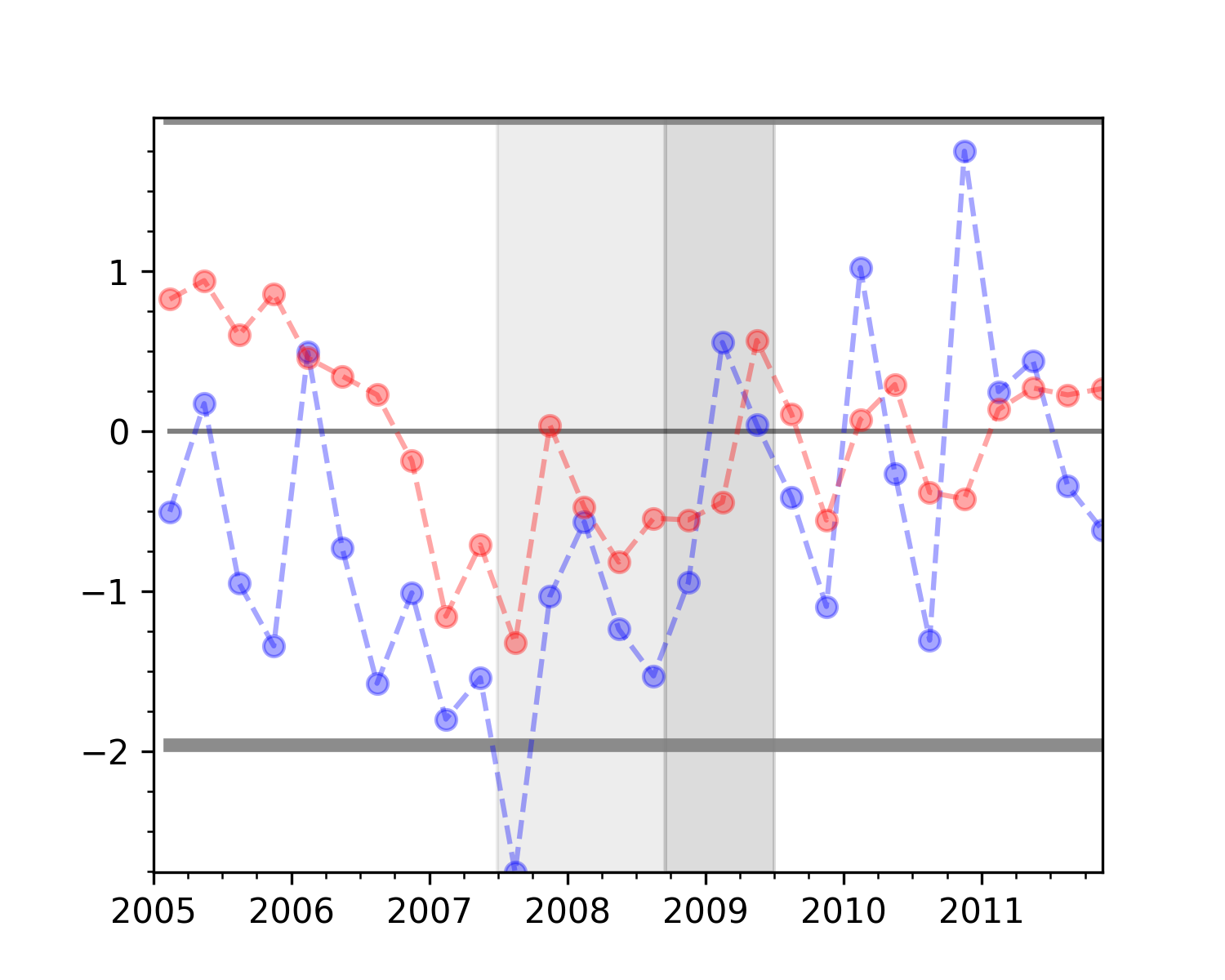}
  \caption{\textit{Abundance z-score - Motif 10\&12. Left panel: Unweighted. Right panel: Weighted.} DECM-filtered (RECM-filtered) empirical series are shown in blue (red).
  }
  \label{fig:ts_motif_10_12}
\end{figure}

Adding to this topologically cyclical motif 9 ($X \leftarrow Y \leftarrow Z \leftarrow X$) one reciprocal connection somewhere along the cycle, e.g. $Y \Rightarrow Z$, re-introduces some hierarchical tendencies that are now perfectly compatible with the theoretical notion of a core periphery structure, $P_1 \leftarrow C_1 \leftrightarrow C_2 \leftarrow P_1$, consisting of two fully connected core banks $C_1,C_2$ and a single peripheral bank, $P_1$, lending to and borrowing from the core.
This ideal structure nevertheless has been shown to only hold approximately in the empirical data.
Hence, the empirical tendency for periphery banks to be mostly on the lending side as well as their preference for reciprocal trades makes other configurations within this motif 10 also realistic.
Interestingly, the inclusion of this reciprocal link seems to improve the model fit and induces a level shift upwards in all series in the upper panel of Figure (\ref{fig:ts_motif_10_12}).
While this is not enough to fully explain the (still) small empirical values of the weighted version, the addition of another reciprocal link in motif 12 is finally able to reconcile model prediction with empirical motif values.
The time series of the twelfth motif ($X \leftrightarrow Y \leftrightarrow Z \leftarrow X$) in the bottom panel of the figure illustrates this fact for both unweighted and weighted versions.

\medskip

Almost all of the previously presented motifs ($2$,\,$3$,\,$7-10$ \& $12$) displayed deviations from the predictions of the core-periphery-preserving DECM in unweighted or weighted versions of motif abundance.
These motifs have one thing in common: They are all intransitive.
A relation $\sim$ on an underlying set $S$ is called transitive if $s_1 \sim s_2$ and $s_2 \sim s_3$ implies $s_1 \sim s_3$ for all $s_1,s_2,s_3 \in S$.
\cite{holland1971transitivityin} use this idea to define the notion of a transitive graph, i.e., a network that consists of a set of nodes, $S$, with directed edge relation, $\Rightarrow$, such that the existence of two edges, $s_1 \Rightarrow s_2 $ and $s_2 \Rightarrow s_3$ implies $s_1 \Rightarrow s_3$.\footnote{
  Note that we still use the notation from the theory section in which we defined $\Rightarrow$ as a general directed edge, which may or may not have an oppositely directed edge relationship $\Leftarrow$, as opposed to a strictly unilateral edge $\rightarrow$ which precludes the existence of a converse unilateral edge $\leftarrow$ for a given pair of nodes.
}
This implies in particular that in such a graph any two nodes that are connected via a path of edges, $s_1 \Rightarrow s_2 \Rightarrow \dots \Rightarrow s_k$, with $k \leq |S|$, also happen to be directly connected by an edge $s_1 \Rightarrow s_k$.
Put differently, the natural tendency of a node, i.e. sender or receiver, gets reinforced with each indirect relation which in turn intensifies the hierarchical structure of the network.
While most empirical networks do not fulfill the strict definition of a transitive graph, they may still be composed of a large number of transitively closed sub-graphs.
In this regard we have concentrated on (isomorphism classes of) sub-graphs of three connected nodes, i.e. triadic motifs.
Up to this point, we have exclusively focused on those motifs that do not fulfill the criteria of a transitive graph.
These intransitive motifs may in fact contain single transitive relations, yet contain at least one intransitive relation, thus contradicting the definition on the (sub-)graph level.

\medskip

For the classes of motif 2 ($X \leftarrow Y \leftarrow Z$), 3 ($X \leftarrow Y \leftrightarrow Z$), 7 ($X \leftrightarrow Y \leftarrow Z$) and 8 ($X \leftrightarrow Y \leftrightarrow Z$) it is readily apparent that there is no transitive closure of the path ($Z \Rightarrow Y \Rightarrow X$) from Z to X, by means of a directed edge $Z \Rightarrow X$.
Nor has this closure link been established in motif 9 ($X \leftarrow Y \leftarrow Z \leftarrow X $).
Instead, compared to the second motif, the converse link $X \Rightarrow Z$ in the ninth motif has even opened up two new intransitive paths, $Y \Rightarrow X \Rightarrow Z$ and $X \Rightarrow Z \Rightarrow Y$, i.e. without respective closure links $Y \Rightarrow Z$ and $X \Rightarrow Y$, which has smeared any distinction of predominantly sending nodes from receiving nodes.
By contrast, the further addition of $Y \Rightarrow Z$ to motif 9 in motif 10 ($X \leftarrow Y \leftrightarrow Z \leftarrow X$) closes one of these two newly generated intransitivities $Y \Rightarrow X \Rightarrow Z$, while the addition of $X \Rightarrow Y$ to motif 10 in motif 12 ($X \leftrightarrow Y \leftrightarrow Z \leftarrow X$) closes the second intransitivity $X \Rightarrow Z \Rightarrow Y$.
Needless to say, the missing reciprocated link $Z \Rightarrow X$ leaves the indirect relation $Z \Rightarrow Y \Rightarrow X$ and thus the entire sub-graph in both cases intransitive.
Nevertheless we clearly observe the addition of transitive edges leading to better concordance with the core periphery structure as proxied by DECM deviations.

\medskip

\begin{figure}[H]
  \vspace{-3cm}
  \centering
  \includegraphics[scale=1.0, trim = 0cm -1cm 0cm 0cm, clip]{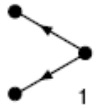}
  \includegraphics[scale=0.4]{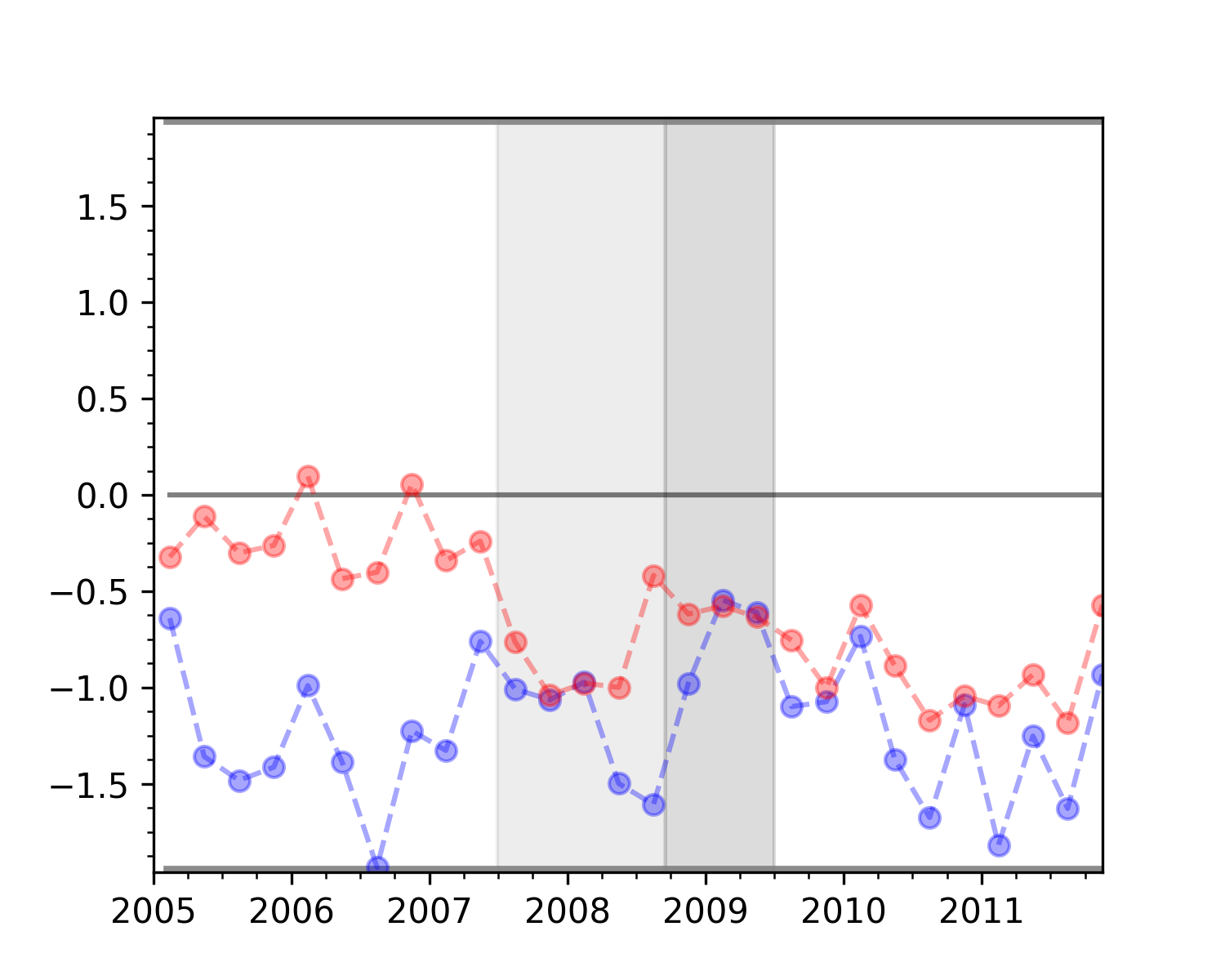}
  \includegraphics[scale=0.4]{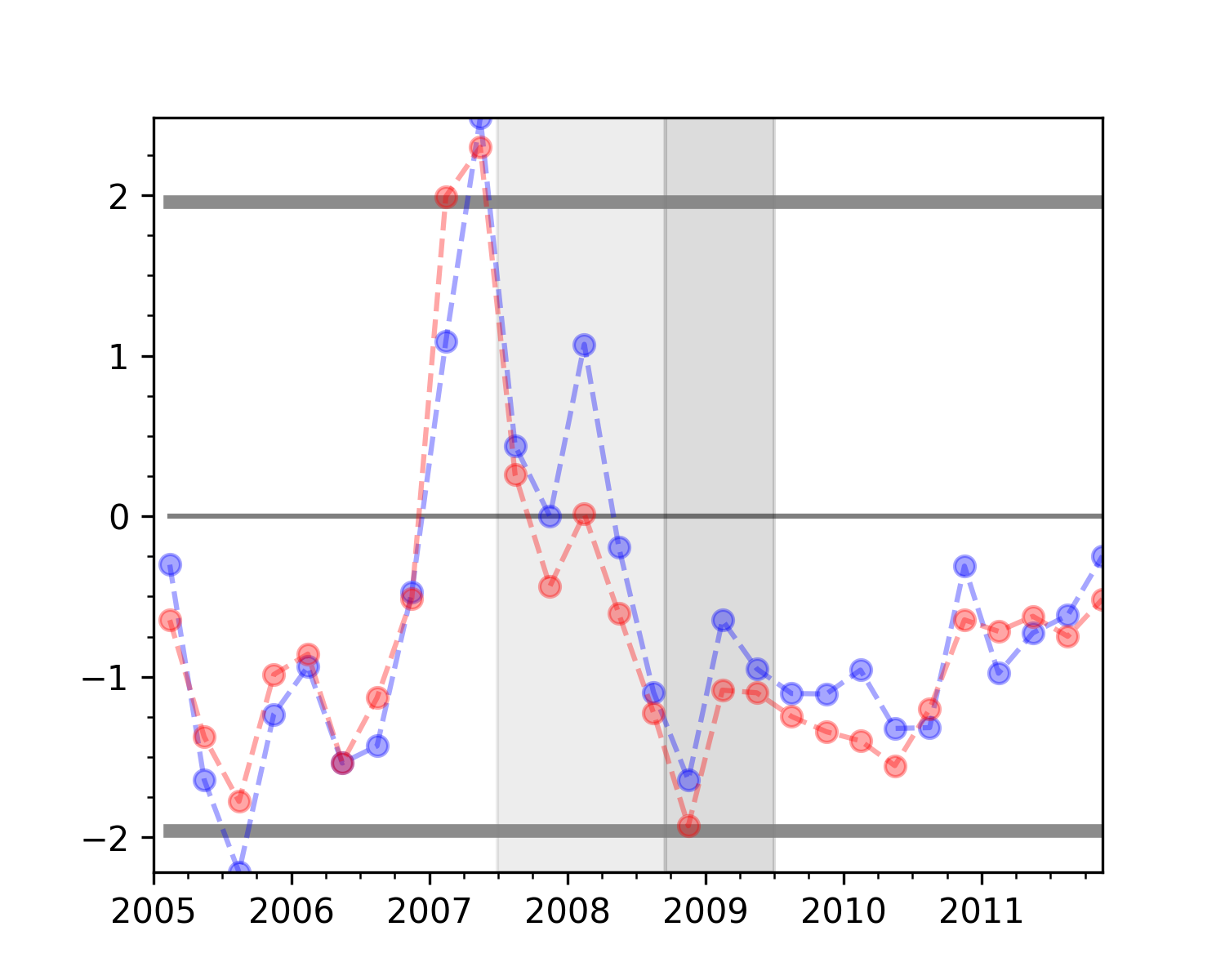}
  \\
  \includegraphics[scale=1.0, trim = 0cm -1cm 0cm 0cm, clip]{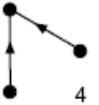}
  \includegraphics[scale=0.4]{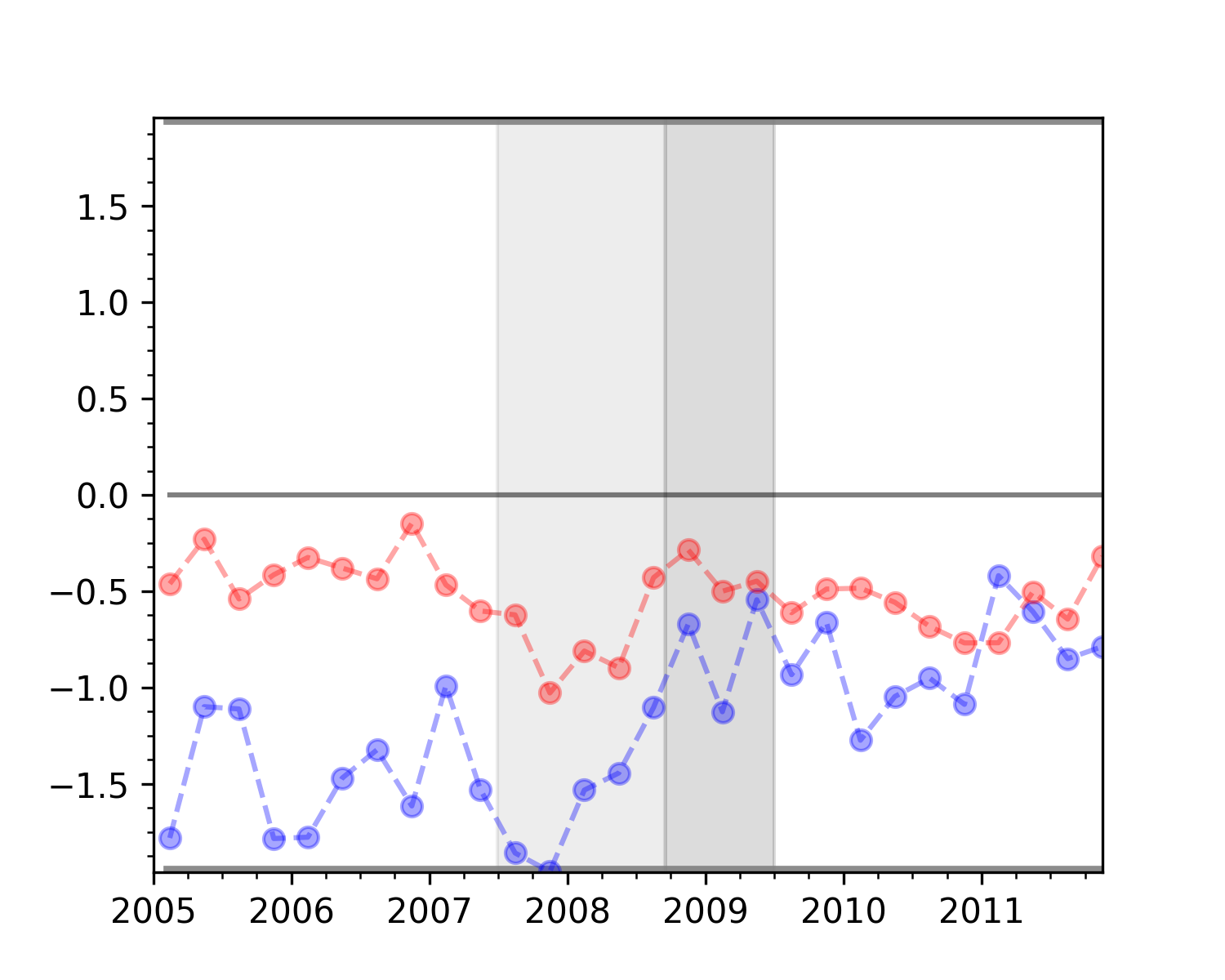}
  \includegraphics[scale=0.4]{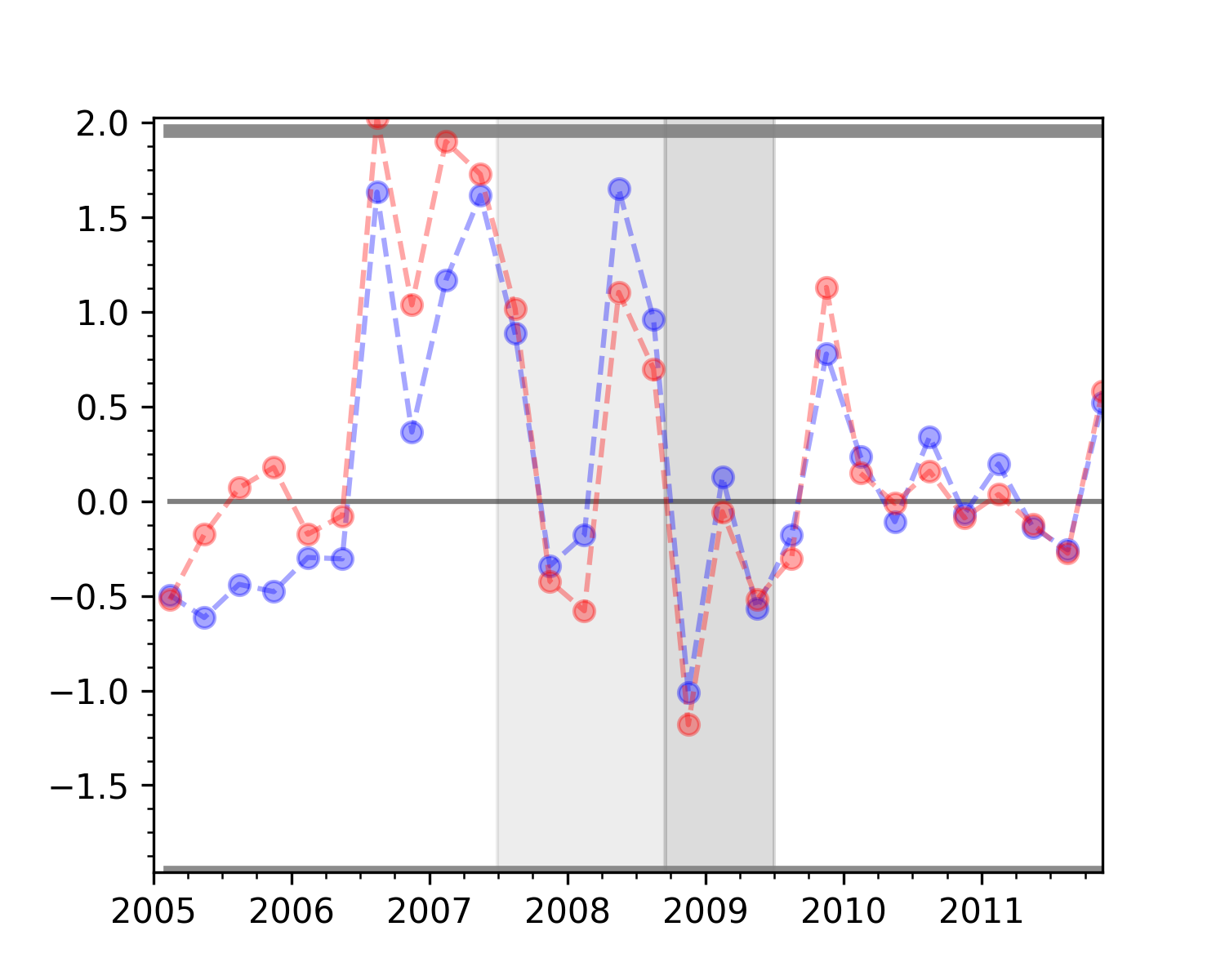}
  \\
  \includegraphics[scale=1.0, trim = 0cm -1cm 0cm 0cm, clip]{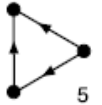}
  \includegraphics[scale=0.4]{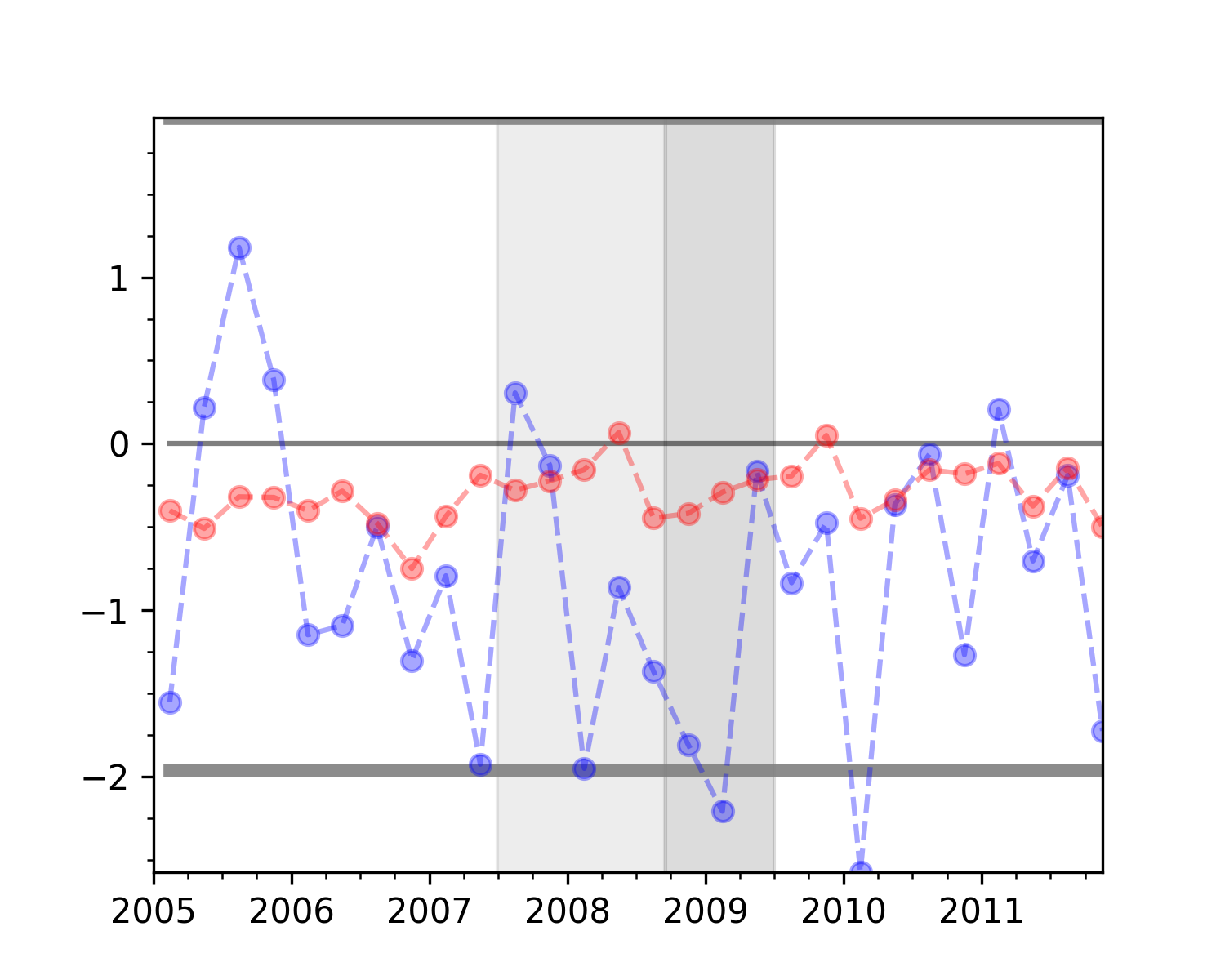}
  \includegraphics[scale=0.4]{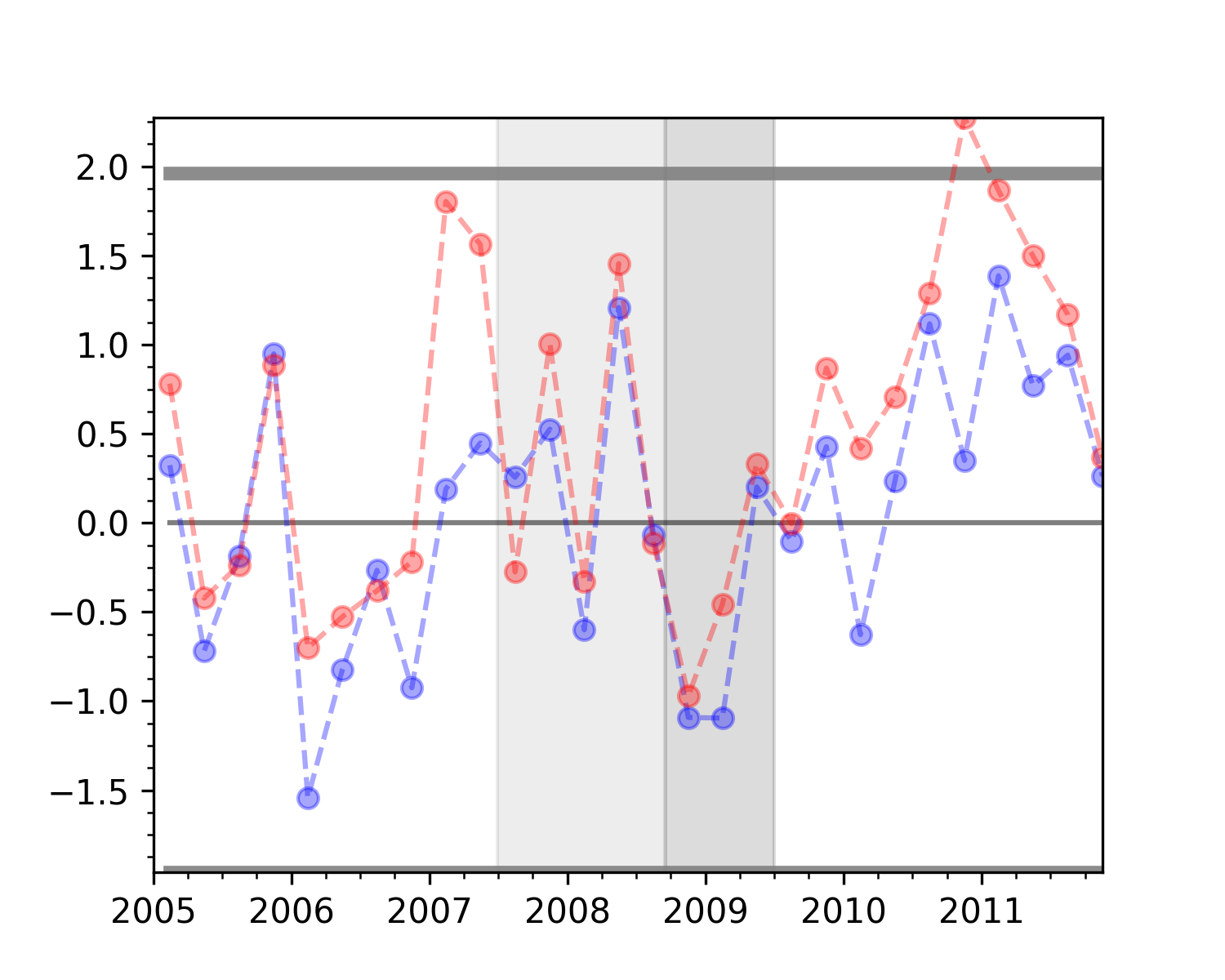}
  \\
  \includegraphics[scale=1.0, trim = 0cm -1cm 0cm 0cm, clip]{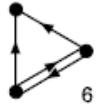}
  \includegraphics[scale=0.4]{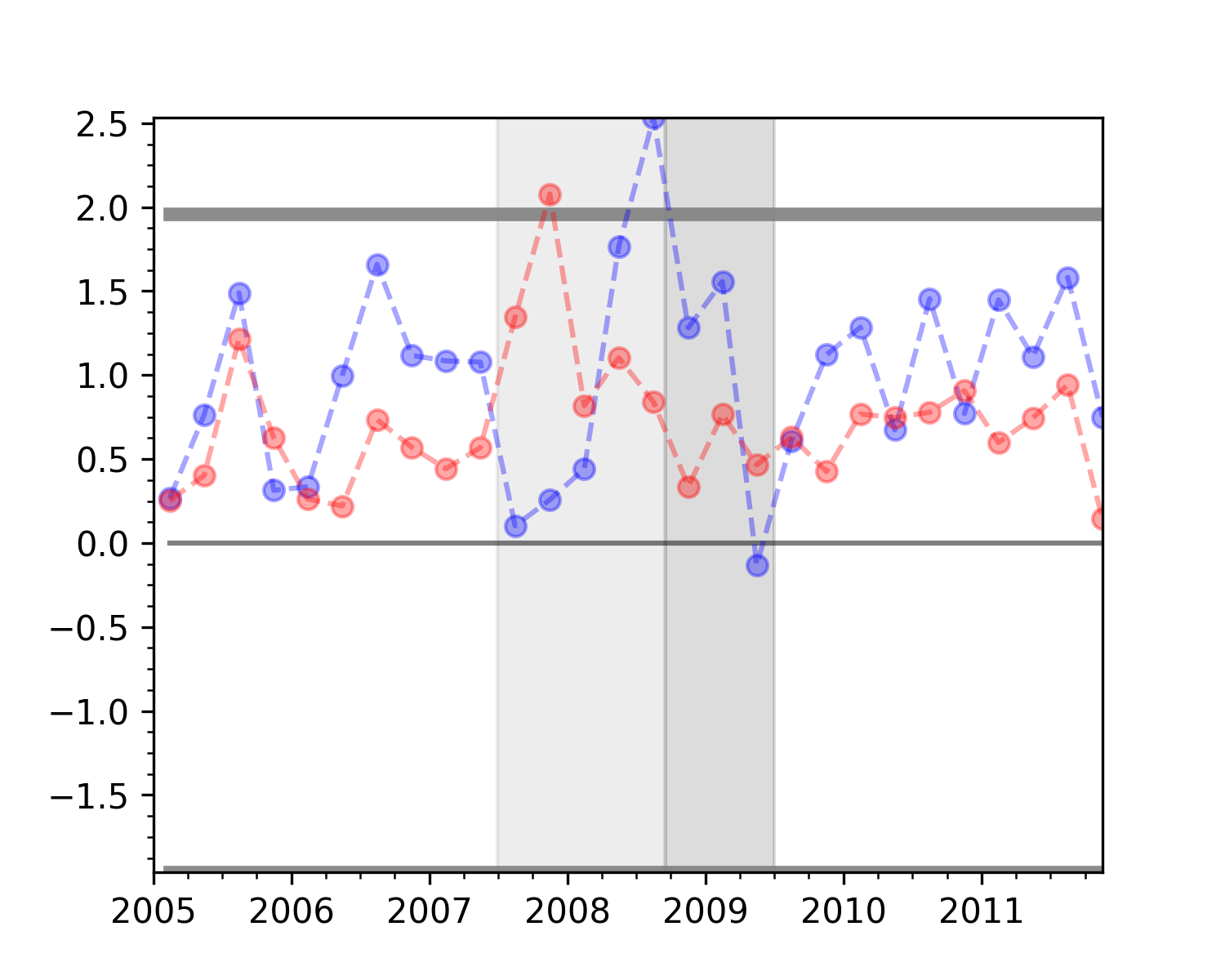}
  \includegraphics[scale=0.4]{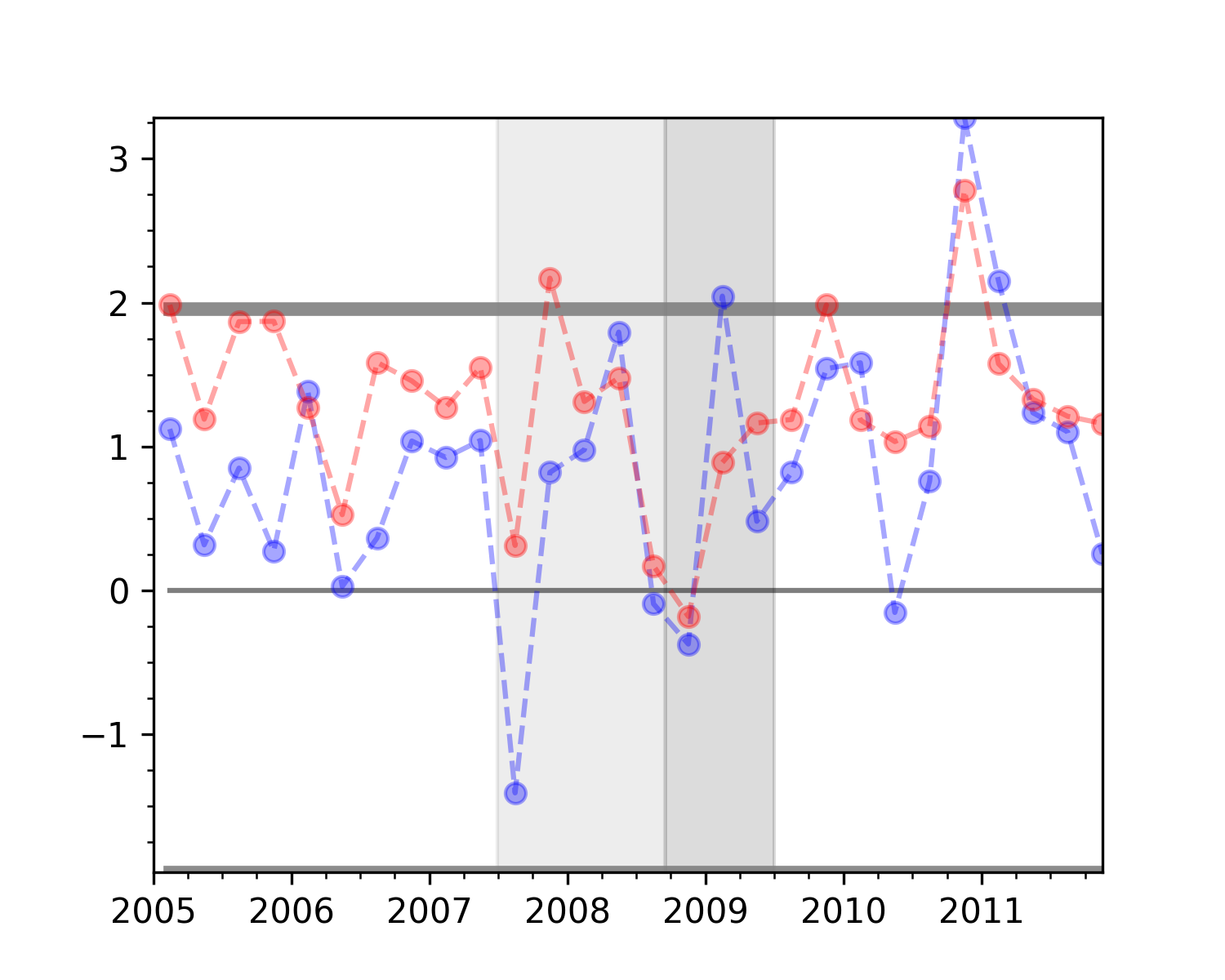}
  \\
  \includegraphics[scale=1.0, trim = 0cm -1cm 0cm 0cm, clip]{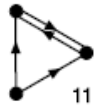}
  \includegraphics[scale=0.4]{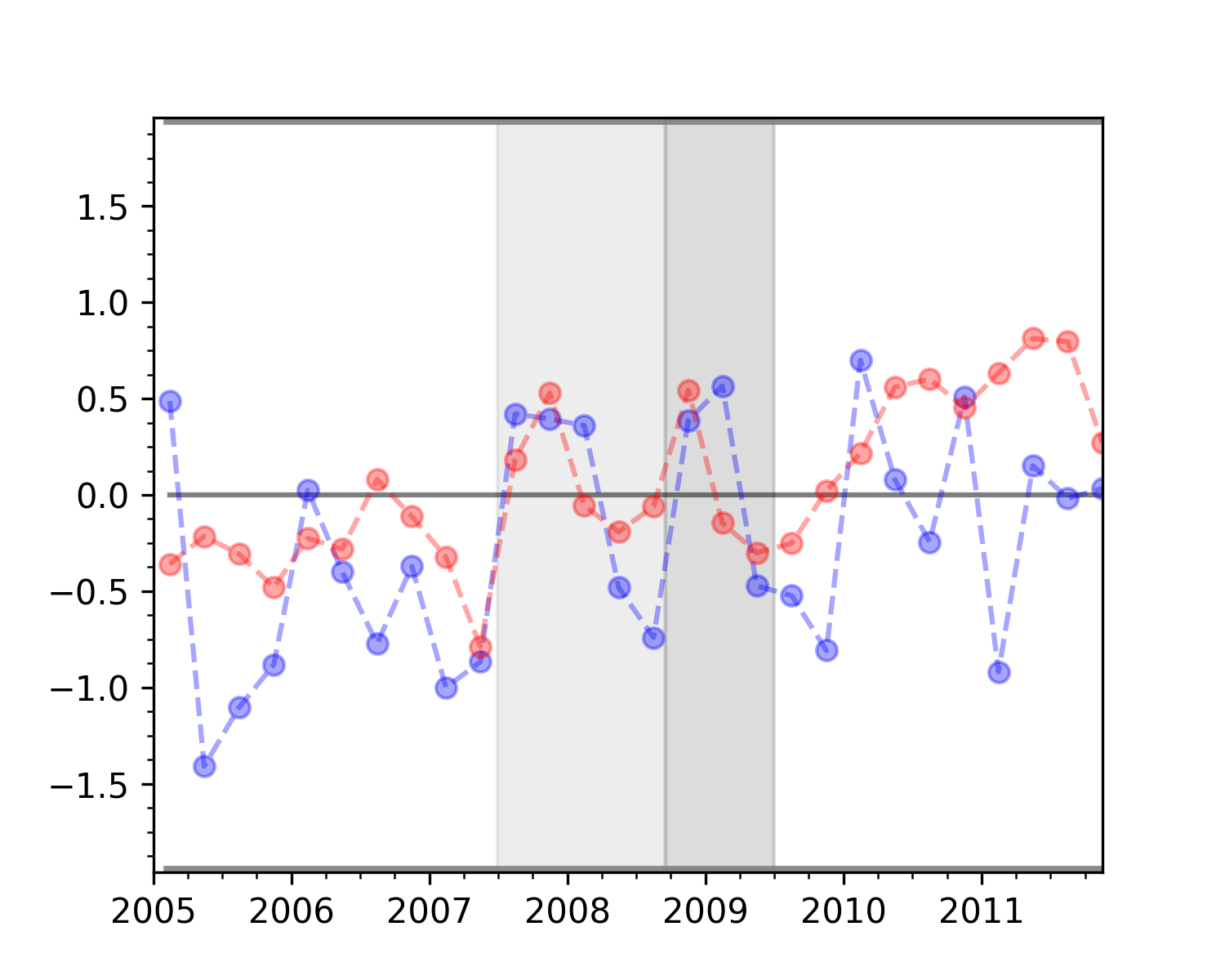}
  \includegraphics[scale=0.4]{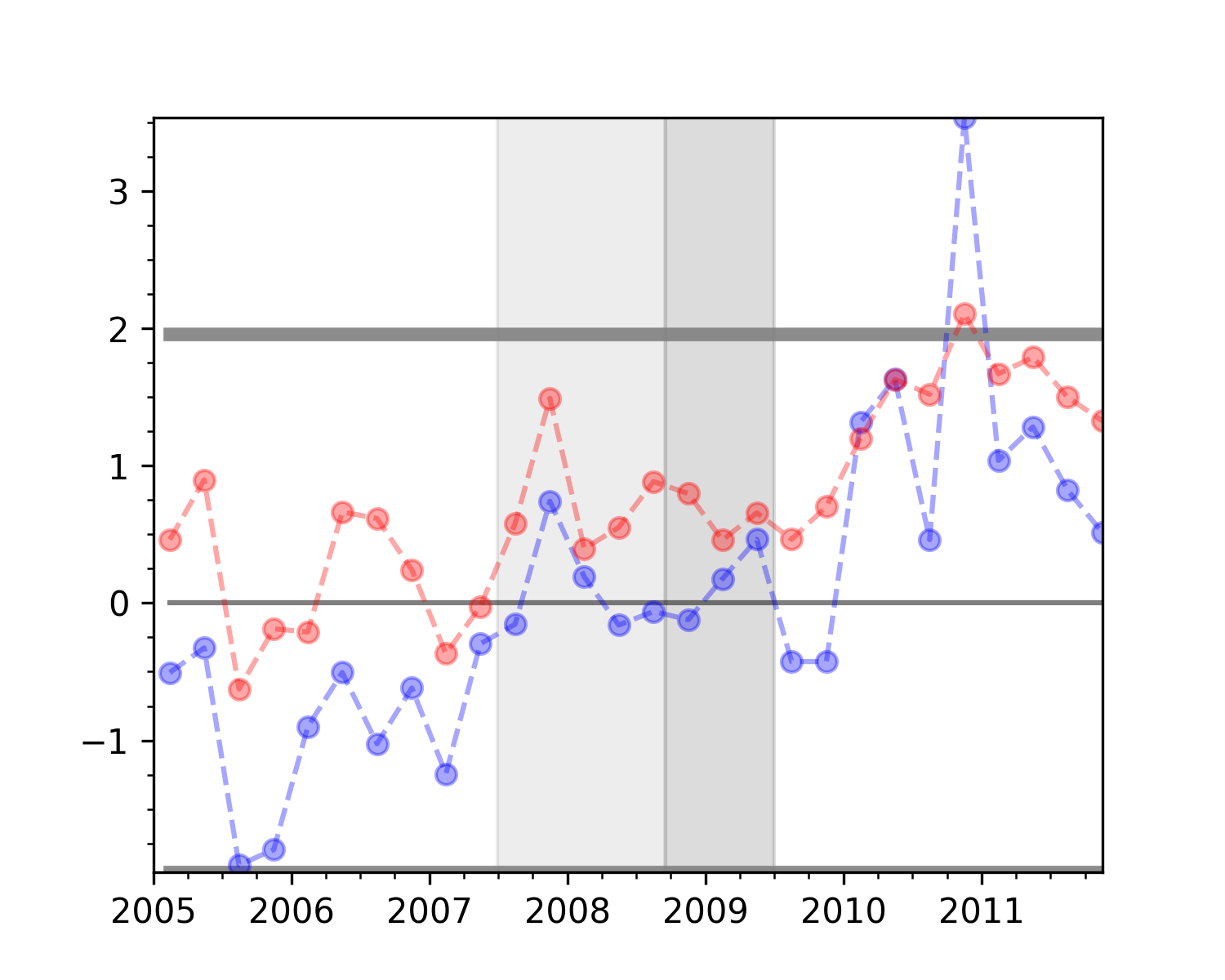}
  \\
  \includegraphics[scale=1.0, trim = 0cm -1cm 0cm 0cm, clip]{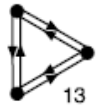}
  \includegraphics[scale=0.4]{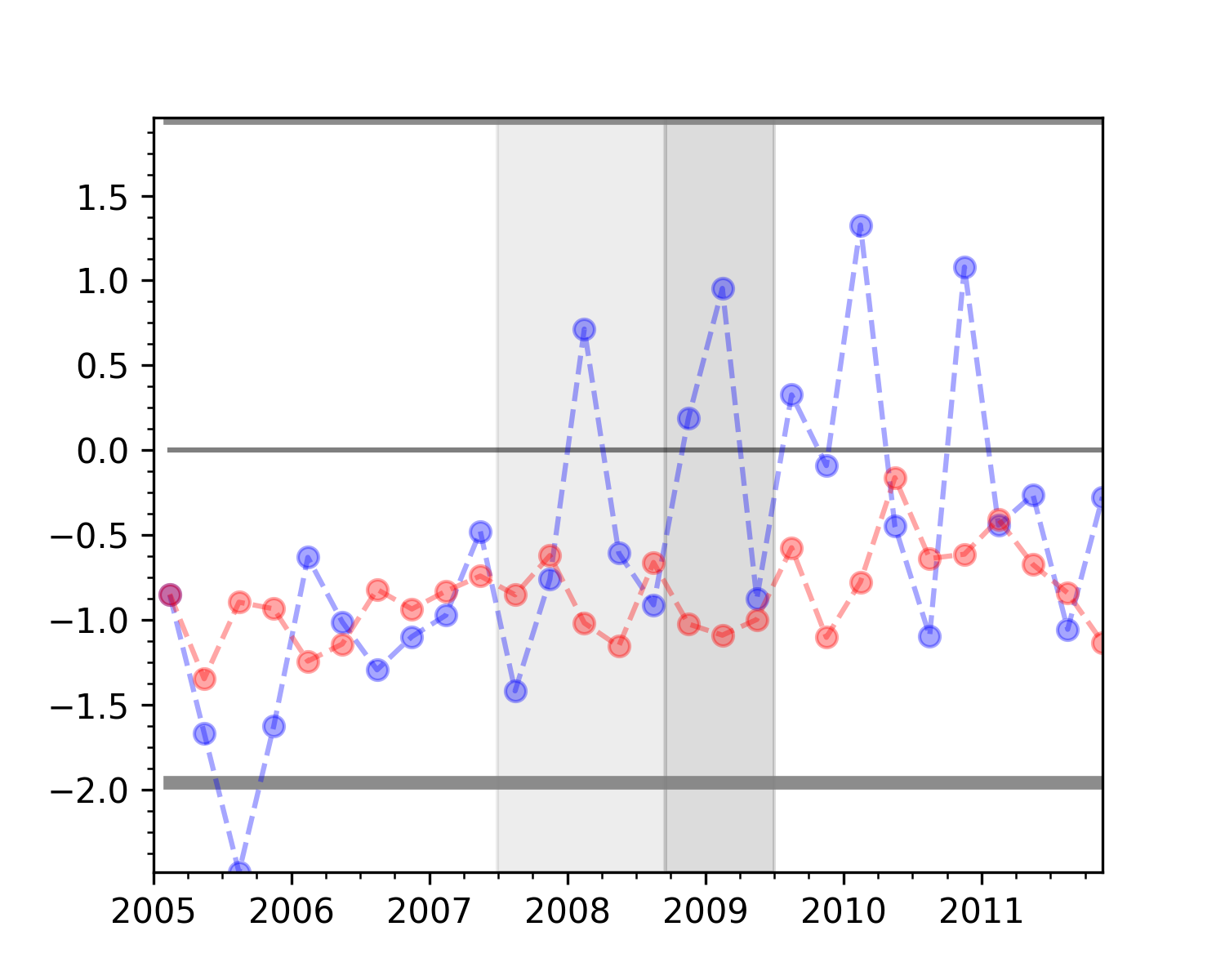}
  \includegraphics[scale=0.4]{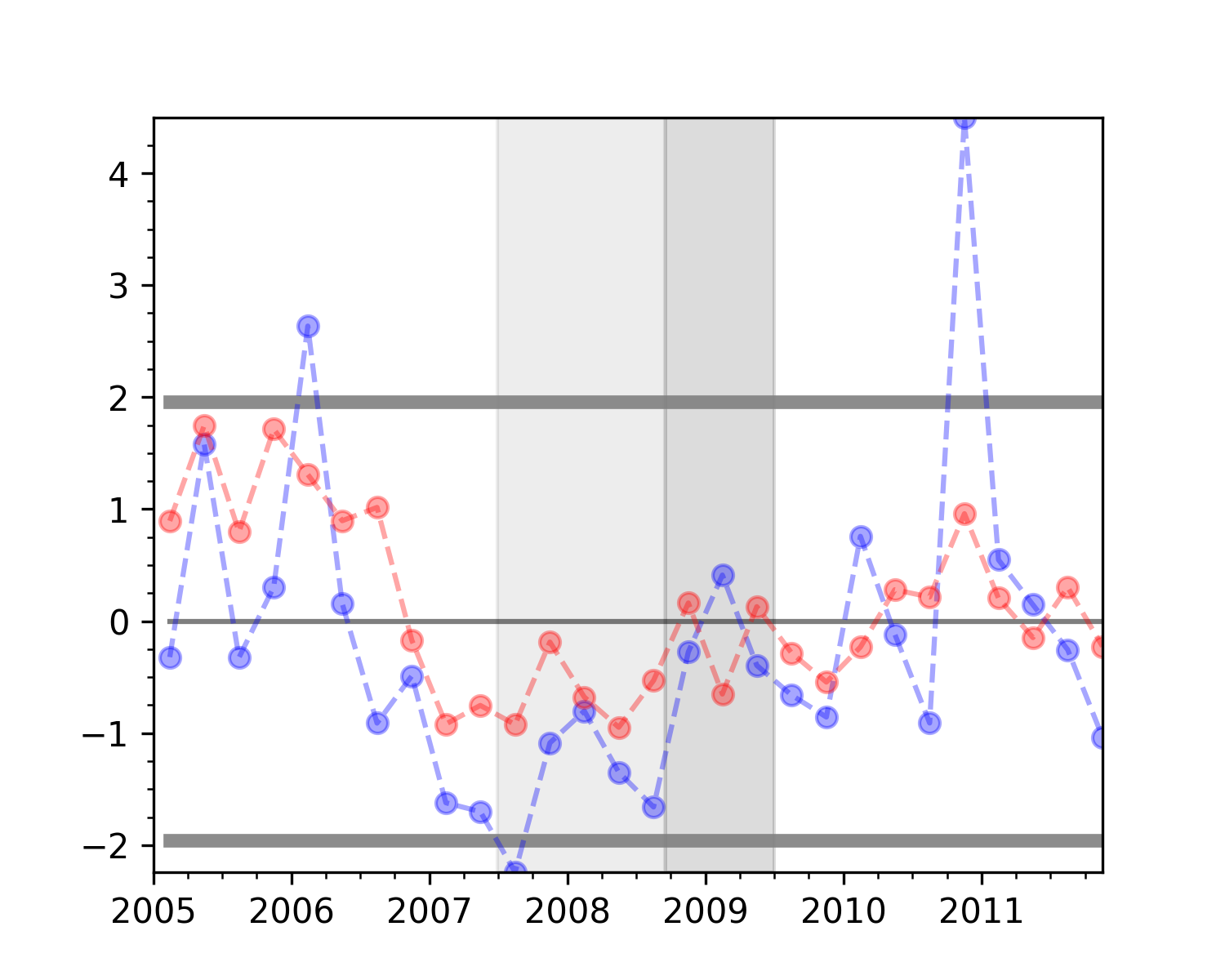}
  \caption{\textit{Abundance z-score - Transitive Motifs. Left panel: Unweighted. Right panel: Weighted.} DECM-filtered (RECM-filtered) empirical series are shown in blue (red).
  }
  \label{fig:transitive_motifs}
\end{figure}

Once we add the remaining link $Z \Rightarrow X$ to the intransitive motif 12 ($X \leftrightarrow Y \leftrightarrow Z \leftarrow X$) we have constructed the transitive cycle, motif 13 ($X \leftrightarrow Y \leftrightarrow Z \leftrightarrow X$), which theoretically coincides with an ideal within-core structure while being completely incompatible with a within-periphery structure. This thirteenth motif, in fact, also happens to be well-represented by both DECM and RECM ensembles.
In this regard, Figure (\ref{fig:transitive_motifs}) provides a summary of all transitive motifs and temporal model deviations for weighted and unweighted versions over the sample period.
It turns out that the transitive cycle is not an exception but instead follows a general tendency of transitive motifs to be in line with the core-periphery structure as measured by the small deviations from the DECM predictions.

In turn, the intransitive cycle, motif 9 ($X \leftarrow Y \leftarrow Z \leftarrow X$) consisting exclusively of asymmetric relationships, was found to be significantly underrepresented in the empirical interbank data.
Flipping one of these asymmetric relationships, for instance $Y \rightarrow Z$ instead of $Y \leftarrow Z$, however rendered these deviations insignificant.
The resulting motif 5 ($X \leftarrow Y \rightarrow Z \leftarrow X$) reinstantiated some hierarchical structure with an exclusive lender ($Y$), an exclusive borrower ($Z$) and an intermediary ($X$).
The seemingly small change of a single link can thus have large consequences for the structure and functioning of a network.

\medskip

In a similar vein, \cite{finger2013networkanalysis} investigate normalized versions of the unweighted motifs 5 and 9 in the form of clustering coefficients and also find the cyclical motif 9 to be underrepresented in the e-MID data, while the transitive motif 5 is overrepresented compared to a (homogeneous) Erd\"{o}s-Renyi null model.
Our randomized graph ensembles are generalizations of the Erd\"{o}s-Renyi model that take into account the observed core-periphery structure while the RECM is furthermore able to preserve the (reciprocal) normalization factors of the clustering coefficients.
We can therefore support the findings of the authors that banks on the e-MID disfavor asymmetric cyclical triads, while we add to their findings that the apparent preference for these so-called transitive triads (Motif 5) may just be a natural correlate of the hierarchy displayed in the empirical core-periphery network structure.

\medskip

Through the lens of a core periphery structure, with core banks $C_1, C_2$ and periphery banks $P_1,P_2$, we can regard the remaining motifs (1,\,4,\,6 \& 11) as mirror-images of inter-group financing.
Given that core banks are predominantly borrowing on the unsecured interbank market, the core-financing Motif 1 ($P_1 \leftarrow C_1 \rightarrow P_2$) occurs much less frequently than the periphery-financing Motif 4 ($P_1 \rightarrow C_1 \leftarrow P_2$).
Both motifs however are in line with a bipartition of nodes into a set of lending nodes and a set of borrowing nodes.
Hence, while motif 4 may be the more relevant sub-graph for interbank networks, both triads are theoretically compatible with a core periphery structure, which translates into good fits of the random null models.
Motifs 11 and 6 can be constructed out of Motifs 1 and 4, respectively, by replacing the non-existent link with a reciprocal trading relationship.
In an ideal core-periphery structure a non-existent link points towards the appearance of two peripheral nodes, while two core nodes would rather establish a reciprocal link.
Hence the financing side, i.e. core or periphery, is likely to switch in the transition from Motif 1 ($P_1 \leftarrow C_1 \rightarrow P_2$) to Motif 11 ($P_1 \rightarrow C_1 \leftrightarrow C_2 \leftarrow P_1$), and from Motif 4 ($P_1 \rightarrow C_1 \leftarrow P_2$) to Motif 6 ($P_1 \leftarrow C_1 \leftrightarrow C_2 \rightarrow P_1$).
Thus, it is now Motif 11 that should occur more frequently than Motif 6 while both are again perfectly consistent with a core periphery structure, displaying only random fluctuations in the null models.

\medskip

All in all, the consistency of the DECM with \textit{all} transitive triplets indicates that both types contain the same informational content, i.e. the heterogeneous in- and out-degree sequences that give rise to the core periphery structure in the market.
While some intransitive triplets gave rise to substantial deviations from the DECM, most of these deviations from a core-periphery structure could be remedied once reciprocal relationships were controlled for in the RECM.
Only the reluctance to form intransitive cycles appears to be a genuine effect on the triadic level which goes beyond the explanatory power of the hierarchical structure on the interbank market.

\newpage
\section{Conclusion}
Financial crises are turbulent times, and the Italian Market for Interbank Deposits (e-MID) is no exception in this regard.
Using a detailed dataset on bilateral trading relationships among Italian banks, we documented the development of financial networks formed on this trading platform from the early build-up phase of the recent financial crisis to its peak and immediate aftermath.
In particular, weighted reciprocity has shown a marked transition over the different phases.
Compared to a statistical null model for directed and weighted networks controlling for trading partners and volume of each bank, reciprocal trading volume has been significantly over-expressed in the pre-crisis quarters.
Once the markets showed early signs of heightened counterparty risk, this gap immediately closed and remained in line with the model-implied values for the entire crisis period only to slowly widen up afterwards again.

\medskip

Decomposing the network into a set of large core and small periphery banks, we find that this effect is mostly driven by the periphery engaging in less intensive reciprocal trading relationships in times of crisis.
The network however has not only undergone changes in this dyadic pattern, which we may interpret as early-warning signal of financial distress, but also in its triadic patterns.
Using a null model which can additionally control for these dyadic reciprocity effects, we found most triadic patterns to be a random outcome of these lower order effects.
The only exception to this finding is the tendency to form non-hierarchical triadic trading cycles which are significantly less represented in the empirical data than the models suggest.
Taking into account the trading volume inside these triadic cycles, we see a shift towards the model-implied values once the peak of the crisis sets in.
Nonetheless, the difference remains significantly negative so that we may conclude that while banks display a preference towards reciprocal trading relationships (in which funds as well as risk immediately loop back), they exhibit a distaste for higher-order fund and risk cycles beyond the hierarchical core-periphery structure.

\medskip

In this paper, we reviewed the general literature on interbank networks and empirically investigated the hierarchical structure of a specific interbank network.
The main contribution however is methodological:
We developed a novel exponential random graph model, the Reciprocal Enhanced Configuration Model (RECM), in closed form which controls for degree, strength and reciprocity simultaneously.
The model controls for all three features for each node in both their weighted and unweighted versions at the same time.
This is important as the literature has shown that models exclusively controlling for weighted features miss the relevant topological information to distribute sufficient probability mass on sparse network configurations (which we also confirmed in our application).
To generate networks from this ensemble we furthermore decompose the probability distribution and derive an efficient sampling scheme.

\medskip

As the model nests many existing models which could only control for a subset of these features simultaneously, it provides a useful tool to distill out the effects of reciprocity on higher-order network features.
In particular, this allowed us to formulate and immediately investigate weighted versions of the traditional unweighted triadic motifs.
As we enumerate over all directed (integer-)weighted networks, thus allowing for sparse and dense configurations, our model is very generally applicable in other fields as well.
One way of application could be the usage as a statistical null model for empirical networks, another way could be to use it as a generative model for creating networks based on specific degree-, strength- and reciprocity distributions (e.g. counterfactual stress tests of financial networks based on specific distributional assumptions) or to simply use it for compact representations of temporally aggregated unweighted networks.

\newpage
\bibliographystyle{apalike}
\bibliography{sections/diss.bib} 

\end{document}